\documentclass[a4paper,11pt]{article}
\usepackage{jheppub} % for details on the use of the package, please see the JINST-author-manual
\usepackage{draft}
\usepackage{amsthm}
\newcommand{\beq}{\begin{eqnarray}}
\newcommand{\eeq}{\end{eqnarray}}
\usepackage{xcolor}
\usepackage{mathrsfs}
\usepackage{tikz}
\usetikzlibrary{arrows.meta, positioning}
\usepackage{comment}
\usetikzlibrary{decorations.pathmorphing,calc,shapes.geometric}

\usepackage{tabularx}
\usepackage{booktabs}
\usepackage{makecell}

\newcommand{\Endvect}{\text{End}_{\textrm{v}}}

\newcommand{\Obj}[1]{\text{Obj}(#1)}
\newcommand{\Hom}[3]{\text{Hom}_{#1}(#2,#3)}
\newcommand{\norm}[1]{\left\lVert #1 \right\rVert}
\usepackage{tikz}
\usepackage{amsmath}

\usetikzlibrary{cd}

\usetikzlibrary{decorations.markings}

\tikzset{line/.style={line width=0.25mm},
curve/.style={line,smooth,tension=1},
->-/.style={decoration={
  markings,
  mark=at position #1 with {\arrow[>=stealth]{>}}},postaction={decorate}},
-<-/.style={decoration={
  markings,
  mark=at position #1 with {\arrow[>=stealth]{<}}},postaction={decorate}},
}

\tikzset{bg/.style={opacity=.3}}
\def\LeftEnd{-6}
\def\RightEnd{0}
\def\GapCenter{-3}
\def\GapHalf{0.40}

\usepackage{dsfont}
\def\id{\mathds{1}}

\title{\boldmath The Many Faces of Non-invertible Symmetries }
\author{Shadi Ali Ahmad${}^{1}$,}
\affiliation{$^{1}$Center for Cosmology and Particle Physics, New York University, New York, NY 10003, USA}
\author{Marc S.~Klinger${}^{2,}$${}^3$,}
\affiliation{${}^{2}$Walter Burke Institute for Theoretical Physics, California Institute of Technology, Pasadena, CA 91125}
\affiliation{${}^{3}$Illinois Center for Advanced Studies of the Universe \& Department of Physics,~University of Illinois,~1110 West Green St.,~Urbana IL 61801, U.S.A.}
\author{and Yifan Wang${}^{1}$}
\emailAdd{shadiraliahmad@gmail.com,klingerm@caltech.edu,yw6417@nyu.edu}

\abstract{We investigate the interplay between algebraic and categorical notions of non-invertible symmetries. In particular, a fusion categorical symmetry $\mathcal{C}$ is shown to induce an algebraic symmetry encoded in a weak Hopf algebra $H$ which is Tannaka-Krein dual to $\mathcal{C}$ in the sense that $\mathcal{C} = \text{Rep}(H^*)$. The latter duality is not unique, and consequently the algebraic symmetry acts on an extended system relative to the categorical one. 
We present an approach to analyzing the symmetry breaking patterns of weak Hopf algebraic non-invertible symmetries. The central ingredient is a certain conditional expectation, which serves as the analog of a group averaging map for a non-invertible symmetry. The index of this conditional expectation emerges as a quantum information theoretic quantity that determines the extent to which the underlying symmetry can be broken. 
Ambiguities which ensue from the non-uniqueness of the categorical reconstruction lead to distinct properties of symmetry breaking compared to the invertible case. Finally, we exemplify our approach through topological and conformal quantum field theories in which non-invertible symmetries are naturally interpreted as defect operators in the presence of boundary conditions.
}

\begin{document}
\maketitle
\flushbottom

\section{Introduction}

\subsection{Motivation and Summary}
Symmetry is a powerful and universal guiding principle in explaining the complicated dynamics of many-body quantum systems such as lattice models and quantum field theories (QFT). In recent years, the study of symmetry and its applications has seen rapid development, owing to the important insight that symmetries can be thought of as general topological defect operators, which has led to the notion of generalized symmetries \cite{Gaiotto:2014kfa}. As is the case with ordinary global symmetries described by groups, there is an abstract mathematical framework that captures all intrinsic properties of generalized symmetries (irrespective of their realization in specific systems) and it is given in general by a fusion $n$-category ($e.g.$ $n=d-1$ in a general $d$-dimensional QFT).  The fusion $n$-category  generalizes the familiar group by allowing for more general compositions of symmetry operations that are non-invertible and subject to stringent constraints that can be thought of as generalizations of familiar Ward identities for ordinary symmetries. Interestingly, even though the study of fusion categories is a relatively young subject \cite{etingof2017tensor} (even more so for the higher $n$-category generalizations \cite{lurie2009higher,Johnson-Freyd:2020usu}), it was recently realized that such symmetries are ubiquitous in both lattice systems and QFTs, and they play an important role in dynamical questions such as the fate of renormalization group flows, the infra-red phase structure and transitions in relation to symmetry breaking, and the interplay with dualities (see e.g. \cite{Schafer-Nameki:2023jdn,Brennan:2023mmt,Bhardwaj:2023kri,Shao:2023gho} for recent reviews). 
In these investigations, it has become clear that there is typically a lot of freedom in realizing a given categorical symmetry, even in a fixed bulk system, for example by choosing boundary conditions or inserting impurities or defects. Such choices are naturally encoded in the representation of the symmetry acting on the relevant operator algebra. An immediate challenge is to find universal properties of the symmetries that may be refined by the choice of (higher) representation (for recent works see e.g. \cite{Bartsch:2022mpm,Bartsch:2022ytj,Bhardwaj:2023wzd,Bartsch:2023pzl,Bartsch:2023wvv,Cordova:2024vsq,Copetti:2024onh,Cordova:2024iti,Cordova:2024nux,Choi:2024tri,Bhardwaj:2024igy,Gagliano:2025gwr}).

There is a complementary approach to symmetries that is in a sense more general and more concrete at the same time, based on endomorphism actions on a closed operator algebra, for which non-invertibility is clearly not a surprise. This appears to be a natural approach to tackle the challenge raised above. The algebraic approach and the study of generalized symmetries in this way have a long history (even though the terminology was different, see \cite{Bischoff:2014xea} for a review). Important results include the axiomatic formulation of local QFT due to Haag–Kastler by attaching C$^*$-algebras to space-time regions subjected to consistency conditions and the Doplicher-Haag-Roberts (DHR) superselection theory that describes the local realization of symmetries (endomorphisms) in the algebraic approach \cite{Haag:1963dh,Doplicher:1971wk,Doplicher:1973at,doplicher_new_1989,Longo:1989tt,Longo:1990zp,Longo:1994xe,Guido:1997gi}. One can argue that fusion category symmetries in QFT were first discovered in this context via the category of such local endomorphisms up to natural equivalence relations in the algebra formalism. The algebraic approach nicely connects the abstract, completely theory-independent notion of symmetries to central ingredients of quantum systems such as Hilbert space, observables, and entanglement. 
Despite its complexity, this framework of operator algebras is surprisingly robust in distilling nontrivial universal properties of QFT, as is exemplified by various inequalities for entropic quantities \cite{Casini:2022rlv}.
It is then natural to explore the question of generalized symmetries (as defined above) and their universal physical signatures in this framework. 

In fact, important progress was made in \cite{Casini:2019kex,Furuya:2020wxf,Casini:2020rgj,Casini:2021tax,Casini:2021zgr,Magan:2021myk,Ares:2022koq,Benedetti:2024dku,Benedetti:2024utz}, where entropic order parameters for certain generalized symmetries and their universal signatures were proposed and analyzed.  
Compared to conventional symmetry order parameters, the entropic version has several advantages. It is universally defined for general algebras by the relative entropy $S(\psi|\psi_{\rm sym})$ between a state $\psi$ and its symmetrization $\psi_{\rm sym}$ (known as entanglement asymmetry in \cite{Ares:2022koq} in a more specific setting).\footnote{Further generalizations include the one-parameter family of Renyi relative entropies.} It provides a faithful diagnostic for symmetry breaking since the relative entropy is non-negative and zero only when the state is symmetric.
It also quantifies the amount of symmetry breaking and appears to obey an upper bound which only depends on universal data of the symmetry category, which is a consequence of a stronger entropic equality known as the \textit{entropic certainty principle} studied in \cite{Magan:2020ake,Hollands:2020owv,Longo:2022lod}. For example, for a discrete group symmetry $G$,
\ie 
0\leq S(\psi|\psi_{\rm sym}) \leq \log |G|\,.
\label{groupbound}
\fe
The saturation of the upper bound has interesting implications, for example in the case of a thermal state at high temperature this leads to the universal `equipartiton' in the density of high energy states also studied in \cite{Harlow:2021trr,Lin:2022dhv,Pal:2020wwd} and a similar consideration explains features in the symmetry-resolved entanglement entropy recently studied in \cite{Xavier:2018kqb,Murciano:2020vgh,DiGiulio:2022jjd,Kusuki:2023bsp,Saura-Bastida:2024yye,Choi:2024wfm}.
Furthermore, the entropic order parameters are applicable to more general transitions, such as those seen in mixed-states and out-of-equilibrium scenarios including symmetry restoration in the quantum Mpemba effect \cite{Ares:2024nkh,Benini:2024xjv,Kusuki:2024gss,Fossati:2024ekt,DiGiulio:2025ems,Ares:2025onj,Fujimura:2025rnm}. Finally, while the entropic order parameters are generally difficult to calculate in the continuum theory, it can be accessed on the lattice with tensor network techniques such as density matrix renormalization group (DMRG) (e.g. in \cite{Capizzi:2023xaf,Khor:2023xar}). 

Despite these exciting developments, the bridge between the categorical and algebraic approaches to symmetries is still under-developed\footnote{Though see Ref.~\cite{Evans:2025msy} for an algebraic framework for fusion categorical symmetries acting on the lattice without passing to an algebraic avatar of the symmetry category as we do here. This is made possible by a lattice version of the DHR superselection theory that applies to abstract spin systems \cite{Jones:2023ptg,Hataishi:2025pxu}. In this context, the fusion categorical symmetries act via localizable (DHR) bimodules with respect to the ``observable'' algebra. } and many subtleties remain to be addressed. 
%In particular, the entropic order parameters, as introduced in the algebraic approach \cite{Casini:2020rgj}, are naturally
%defined using the symmetry algebra, instead of the symmetry category, acting on the observable algebra. 
The pressing task which we aim to undertake here is to clarify the definition of entropic order parameters detecting non-invertible symmetry breaking given the concise categorical input and incorporate the effects of choices therein on the values and bounds of such order parameters.
As we will see, due to the the necessity of making choices on the algebraic side to match the categorical input, there is no unique entropic order parameter for a symmetry category. Nonetheless, all such quantities share the same desired properties as a faithful diagnostic for symmetry breaking and obey universal bounds. We will also see how anomalies for symmetries enter in these bounds in an interesting way. Roughly speaking, anomalous symmetries produce more diversity in symmetry breaking, as is measured by the relative entropy $S(\psi|\psi_{\rm sym})$.

 In $d=2$, it is well-known that the algebraic actions of a fusion category $\cal C$ are encoded by strip algebras $\text{Str}_{\mathcal{C}}({\cal M})$ where $\cal M$ is a left $\cal C$-module category \cite{Ostrik:2001xnt,Kitaev:2011dxc}. The strip algebra is also known as annular algebra in \cite{GHOSH20161537} that generalizes Ocneanu’s tube algebra and ladder category in \cite{Barter:2018hjs} (see reviews in recent works e.g. \cite{Jia:2024rzr,Cordova:2024iti}). Such algebras naturally arise when studying the 2d QFT with $\cal C$-symmetry in the presence of boundaries where $\cal M$ encodes possible superselection sectors \cite{Cordova:2024vsq,Cordova:2024iti}. They also play a role in defining the entanglement entropy by factorizing the full Hilbert space \cite{Ohmori:2014eia,Choi:2024tri,Choi:2024wfm}. Mathematically, these strip algebras are C$^*$ weak Hopf algebras (WHAs) \cite{B_hm_1996},\footnote{WHAs are generalizations of Hopf algebras by relaxing certain compatibility conditions involving the counit and the coproduct. Familiar examples of Hopf algebras are quantum groups which were first studied in integrable systems \cite{Drinfeld:1985rx,Drinfeld:1986in,Jimbo:1985vd} that can be thought of quantization of Lie groups \cite{Faddeev:1987ih}. In modern terms, quantum group and Hopf algebra are often used interchangeably, similarly for quantum groupoid and WHA. See \cite{nikshych2002finite} for a review in the finite setting which will be the main focus here. The reason behind term groupoid will become clear shortly (see also Appendix~\ref{app: strip}).
} which are \textit{reconstructed} from the fusion category via the Tannaka–Krein theorem \cite{Ostrik:2001xnt,etingof2017tensor},\footnote{Here  $H^*$ is the \emph{dual WHA}. The dual of a WHA $H$ is itself a WHA modeled on the dual space $H^*$. Its algebraic and co-algebraic operations are inherited from $H$ by dualization. For example, the product of $\hat{h}_1,\hat{h}_2 \in H^*$ is defined by $\mu_{H^*}(\hat{h}_1,\hat{h}_2)(h) = \hat{h}_1 \otimes \hat{h}_2 \circ \Gamma_H(h)$ for each $h \in H$. Likewise, the coproduct is given by $\Gamma_{H^*}(\hat{h})(h_1 \otimes h_2) = \hat{h} \circ \mu_H(h_1,h_2)$.} 
\ie 
\cC={\rm Rep}(\text{Str}_{\mathcal{C}}({\cal M})^*)\,.
\label{TKrec}
\fe 
This relation provides a natural connection from the fusion category to the algebraic framework for symmetries in this restricted setting (e.g., involving only topological degrees of freedom). The role of WHAs as symmetries of 2d conformal field theories (CFTs) was also studied a long time ago in \cite{Roche:1990hs,Mack:1991by,Mack:1991tg,Fuchs:1994ya,Mack:1994bb,Rehren:1996ku}\footnote{To be precise, the class of Hopf-like symmetries considered in \cite{Roche:1990hs,Mack:1991by,Mack:1991tg} are weak quasitriangular quasi-Hopf algebras. The ``quasitriangular'' structure on such an algebra $H$ is specified by an invertible element in $H \otimes H$, which endows the category of $H$-modules with a braiding. The ``weak'' condition is as in weak Hopf algebras (WHA), e.g.\ $\Delta(\id) \neq \id \otimes \id$, while the ``quasi'' condition relaxes coassociativity of the coproduct through the introduction of a (possibly nontrivial) coassociator.
It is worth emphasizing that reconstruction theorems for categorical symmetries apply equally to weak quasi-Hopf algebras and to weak Hopf algebras \cite{etingof2017tensor}. In this work we follow \cite{B_hm_1996} and restrict to algebraic symmetries realized by WHAs, thereby retaining strict coassociativity and the resulting favorable self-duality properties.
} and revisited recently in \cite{Gabai:2024puk,
Gabai:2024qum}. In higher dimensions, the layered structure of the higher fusion category that describe generalized symmetries thereof leads to higher (weak) Hopf algebras that are only recently starting to be investigated \cite{green2023tannakakreinreconstructionfusion2categories,Gagliano:2025gwr}.

In this work, we start with a general operator algebra $M$ and symmetry fusion category $\cal C$ and identify the choices of algebraic actions on $M$ in terms of WHAs $H$ reconstructed from $\cal C$. We then construct, explicitly, the symmetrization map on the algebra (and dually on the states) and establish bounds on the entropic order parameter refined by these choices using index theory \cite{Jones1983,Pimsner1986,Kosaki1991,Watatani1990,Teruya1992Index}. Generalizing the group case in \eqref{groupbound}, we find that the entropic order parameter is bounded as: 
\ie 
0\leq S(\psi|\psi_{\rm sym}) \leq \log \left( r|\cC| \right)\,,
\label{genbound}
\fe
where the upper bound depends on the total quantum dimension $|\cC|$ and $r$ is the number of irreducible idempotents that make up the unit (identity) in $H$.\footnote{As we explain later, for WHA symmetry realized via a strip algebra ${\rm Str}_\cC(\cM)$ with a $\cC$-module category $\cM$, $r={\rm Irr}(\cM)$ counts the number of simple objects in $\cM$ (i.e. the rank of $\cM$).} 
We emphasize that the weak feature of WHA is tied to $r>1$ and for anomalous fusion category symmetries only a genuine WHA (i.e. not Hopf) realization is possible  \cite{Thorngren:2019iar}. Thus, the bound allows for a larger range for the entropic order parameter. 
If the non-anomalous symmetry category $\cC$ acts on the system via a Hopf algebra, the inequality is strengthened to
\ie 
0\leq S(\psi|\psi_{\rm sym}) \leq \log \left( |\cC| \right)\,.
\label{genboundna}
\fe
The previously studied group case \eqref{groupbound} only applies for an invertible non-anomalous symmetry and is a special case of \eqref{genboundna}.

Throughout the discussion, we will intentionally minimize the assumptions on the algebra involved and so the results apply to general algebras that arise in QFT (e.g.  the full operator algebra or the local operator algebra of a region). There is also no assumption on the spacetime dimension, although we limit ourselves to the top layer of the higher WHA that is only sensitive to the representation of codimension-1 topological defect operators.\footnote{For example, the coproduct of WHA is crucial for defining the monoidal (tensor or fusion) product on the category side.} We give explicit examples to illustrate the bound \eqref{genbound} in simple quantum mechanical models and 2d CFT. 
We expect that the incorporation of locality and causality axioms via the nets of algebras will uncover further universal properties of generalized symmetries which we plan to address in future work. We also expect the upper bound in \eqref{genbound} may be further strengthened depending on the properties one assumes for the state (we will see evidence for this in Section~\ref{sec: examples}).

\subsection{Outline of the Paper}
The rest of the paper is organized as follows. In Section~\ref{sec: Prelims} we establish our point of view on categorical and algebraic non-invertible symmetries and their interrelation. In Section~\ref{sec: algebras from categories}, we demonstrate how to induce from a categorical symmetry action $\Phi: \mathcal{C} = \text{Rep}(H_{\mathcal{M}}^*) \rightarrow \text{End}(M)$, an associated weak Hopf algebraic symmetry action $\rho^{\mathcal{C},\mathcal{M}}: H_{\mathcal{M}} \rightarrow \Endvect(M_{\mathcal{M}})$. The correspondence $\mathcal{C} = \text{Rep}(H^*_{\mathcal{M}})$ is an instance of the Tannaka-Krein duality which states that any multi-fusion category is equivalent to the category of representations of a WHA. This correspondence is not unique and depends upon the choice $\mathcal{M}$ of a left indecomposable module category of $\mathcal{C}$. 

The original categorical symmetry action is a functor $\Phi: \mathcal{C} \rightarrow \text{End}(M)$, where $M$ is an operator algebra representing the observable degrees of freedom of the `bare' system and $\text{End}(M)$ is its category of endomorphisms. More generally, one could consider a functor from $\mathcal{C}$ to $\text{Bim}(M)$, the category of bimodules of the system algebra $M$ \cite{longo_landauers_2018,Jones:2023ptg,Evans:2025msy}.\footnote{$\text{End}(M)$ is a subcategory of $\text{Bim}(M)$. In the event that $M$ is properly infinite, it is possible to identify an endomorphism representative (up to unitary equivalence) for each bimodule, and thus the category $\text{End}(M)$ is sufficient for describing most symmetries \cite{longo_landauers_2018}. Historically, it has been common to further restrict to a subcategory of $\text{End}(M)$ consisting of endomorphisms which are `localizable' and `transportable' \cite{Doplicher:1971wk, Doplicher:1973at, dop_why}. We emphasize, however, that not all symmetries satisfy these requirements.}
One can interpret $\text{Bim}(M)$ as a categorification of the unital completely positive maps acting on $M$ \cite{longo_landauers_2018}.\footnote{Non-invertible symmetries in two-dimensional conformal field theory were formulated in terms of these unital completely positive maps (also known as quantum channels) and more precisely as quantum operations on the conformal net \cite{Bischoff:2016jmy,Bischoff:2016rpu,Bischoff:2020iac,Bischoff:2021snn,Bischoff:2022fxf} (see also the recent review in \cite{Okada:2024qmk}).} Roughly, we can think of this categorical symmetry action as describing how the different representations of $M$ are shuffled around; we shall establish this point more concisely in Section~\ref{sec: examples}. The associated algebraic symmetry constructs a non-invertible action of the WHA $H_{\mathcal{M}}$ on an \emph{extended} system $M_{\mathcal{M}}$. In the physical context, these additional operators have a natural interpretation as coming from boundary conditions. In the map $\rho^{\mathcal{C},\mathcal{M}}: H_{\mathcal{M}} \rightarrow \Endvect(M_{\mathcal{M}})$, $\Endvect(M_{\mathcal{M}})$ is simply the set of maps from $M_{\mathcal{M}}$ to itself which preserve the vector space structure of $M_{\mathcal{M}}$.\footnote{It is possible to derive from $\rho^{\mathcal{C},\mathcal{M}}$ an action $\tilde{\rho}^{\mathcal{C},\mathcal{M}}: H_{\mathcal{M}} \rightarrow \text{End}(M_{\mathcal{M}})$, where $\text{End}(M_{\mathcal{M}})$ is the set of maps from $M_{\mathcal{M}}$ to itself which preserve the algebraic structure of $M_{\mathcal{M}}$ e.g. its product and involution. Briefly, this can be accomplished by constructing the charged intertwiners of the Q-system associated with the invariantizing conditional expectation of the action $\rho^{\mathcal{C},\mathcal{M}}$. Actually, constructing this conditional expectation is one of the main objects of the present note. Thus, we have chosen to emphasize the action $\rho^{\mathcal{C},\mathcal{M}}$ rather than $\tilde{\rho}^{\mathcal{C},\mathcal{M}}$.} 

In Section~\ref{sec: order parameters}, we provide a construction of entropic order parameters detecting the breaking of weak Hopf algebraic non-invertible symmetries. We take as our starting point an algebraic action $\rho: H \rightarrow \Endvect(A)$ (e.g. $A$ could be an extended system algebra $M_\cM$ from above). The entropic order parameter in this context is obtained by first constructing a weak Hopf algebraic analog of a group averaging map which symmetrizes operators in the system algebra $A$ under the action of $H$. Algebraically, this map defines a conditional expectation, and we provide a comprehensive discussion of the important role played by index theory in the study of symmetry breaking. In particular, we establish the inequality \eqref{Bound for WHA} constraining the range of the entropic order parameter for general WHA symmetry. 
In the case that the action $\rho$ arises from a categorical symmetry by following the procedure outlined in Section~\ref{sec: algebras from categories}, our work provides an algorithm for constructing order parameters for fusion categorical symmetry breaking, and leads to the inequality \eqref{genbound}. We emphasize that the resulting order parameter is not uniquely determined by the original categorical symmetry alone. Following from the non-uniqueness of the correspondence $\mathcal{C} = \text{Rep}(H_{\mathcal{M}}^*)$, the entropic order parameter is determined by the categorical symmetry action $\Phi$ along with the choice of faithful left module category $\mathcal{M}$. 

In Section~\ref{sec: examples} we exhibit the abstract findings of Sections \ref{sec: algebras from categories} and \ref{sec: order parameters} through explicit examples. In Section~\ref{sec: Fib Cat} we consider a toy quantum mechanical system admitting a categorical symmetry with $\mathcal{C} = \text{Fib}$. Here, $\text{Fib}$ is the Fibonacci category, and we promote this categorical symmetry to an algebraic symmetry using the unique indecomposable left module category which is equal to $\text{Fib}$ itself. The resulting non-invertible symmetry is encoded in the strip algebra $\text{Str}_{\text{Fib}}(\text{Fib})$ which is a $13$-dimensional semisimple, non-degenerate WHA. The extended system algebra is equivalent to the dual of $\text{Str}_{\text{Fib}}(\text{Fib})$ which is isomorphic to itself. We compute the symmetrizing conditional expectation, its index, and identify a state which maximizes the associated entropic order parameter. This state saturates a \textit{strengthened} bound valid for finite dimensional systems and its entropic order parameter is the log of the index of the symmetrizing conditional expectation computed via the Pimsner-Popa bound \cite{pimsner1986entropy}. In Section~\ref{sec: TQFT Fus} we use the fact that indecomposable 2d TQFTs are in correspondence with indecomposable $\mathcal{C}$-module categories to generalize the preceding example. This allows us to regard the observable degrees of freedom of the TQFT as encoded in a unital Frobenius algebra, which is acted upon by a categorical symmetry $\mathcal{C}$. We also consider a generalization of the conditional expectation applied to fusion algebras that may not be WHA. Finally, in Section \ref{sec: ex CFT} we apply our framework to general non-chiral 2d CFTs with prescribed conformal boundary conditions and provide nontrivial checks on the entropic equalities we derived in this work.

We conclude in Section~\ref{sec: discussion} with a series of suggestions for future work. These suggestions center largely around the application of our general toolkit to the problem of symmetry breaking and gauging for continuum field theories and quantum gravity. Especially,   the formalism we have developed could be useful for understanding the problem of black hole radiation and its relation to replica symmetry breaking. Pursuing this direction should shed light on the physical interpretation of the index \cite{Leutheusser:2025zvp}, and the role of non-invertible symmetries in modifying notions of locality and geometricity \cite{Shao:2025mfj,Harlow:2025cqc}. We are also aware of the need to move beyond the assumption of finite index in order to analyze systems that admit non-invertible symmetries of `infinite' size. We remark on a possible approach to this problem. 

In addition to the content of the main text, this note also contains many comprehensive appendices which cover more technical details. While the main text can be read without referring to these appendices, we feel that they contain important information that round out a growing understanding of non-invertible symmetries. To encourage the reader to delve into these we include a detailed description of each appendix here. 

In Appendix~\ref{app: gloss}, we provide a glossary of frequently used symbols appearing in the main text for the convenience of the reader. 

In Appendix~\ref{app: FusCat}, fusion categories are briefly defined for completeness. In Appendix~\ref{app: WHA}, we provide an introduction to WHAs with a special focus on the theory of integrals. Integral elements of WHAs generalize the more familiar notion of measures on groups. As compact groups admit a special bi-invariant measure called the Haar measure, any finite, semisimple, non-degenerate WHA admits a special integral element called the Haar integral. This element is appropriately bi-invariant and can be used for many of the same purposes as one would use a Haar measure. In particular, the Haar integral can be used to define a symmetrizing conditional expectation given any non-invertible symmetry action of the WHA. We provide (albeit in a somewhat abstract form) explicit formulas for the Haar integral element of any finite, semisimple, non-degenerate WHA and of the index of the conditional expectation that it induces when the WHA acts canonically on its dual. This index can be used to bound the index of the symmetrization conditional expectation for any action of the same WHA, and to quantify this index exactly in the case that the action in question is standard and properly outer. 

In Appendix~\ref{app: strip} we specialize our analysis to the class of WHAs which are obtained from Tannaka-Krein duality using a `canonical' choice of weak fiber functor for a given module category. These are the strip algebras, $\text{Str}_{\mathcal{C}}(\mathcal{M})$, which have appeared extensively in the study of TQFTs and CFTs (see e.g. \cite{Cordova:2024iti} and the references therein). We provide an explicit account of the strip algebra, its elements, and its weak Hopf structure maps in terms of a diagrammatic basis which makes the connection to topological defects and boundary conditions manifest. We derive an expression for the Haar integral of the strip algebra in this basis and compute directly the index of the conditional expectation it induces. This leads to the remarkably intuitive result that the index of the Haar conditional expectation for a strip algebra $\text{Str}_{\mathcal{C}}(\mathcal{M})$ is equal to the product of the quantum dimension of the symmetry category $\mathcal{C}$ and the rank of the module category $\mathcal{M}$. The appearance of the module category in this quantity reflects the important role of the non-uniqueness of Tannaka-Krein duality for general non-invertible symmetries. 

In Appendix~\ref{app: ECR} we explore the entropic order parameter from a fully algebraic point of view for general inclusions, $N \subset M$, of von Neumann algebras with non-trivial (but finite-dimensional) centers. We derive the so-called entropic certainty relation in this context by employing Kosaki's spatial definition of the index. The content of the present note can then be understood as a special case of this general analysis applied to inclusions of `depth-2'. For our purposes, an inclusion of depth-2 is defined by the fact that its Jones' basic construction is isomorphic to the sequence of inclusions $M^{\rm sym}_{\rho} \subset M \subset M \times_{\rho} H$. Here, $M$ is a system algebra which admits a non-invertible symmetry action $\rho: H \rightarrow \Endvect(M)$, where $H$ is in general a WHA. $M^{\rm sym}_{\rho}$ is the symmetric subalgebra of $M$ under the action $\rho$, and $M \times_{\rho} H$ is the crossed product of $M$ by the action of $\rho$. The algebra $M \times_{\rho} H$ admits an action by the dual WHA, $\hat{\rho}: \hat{H} \rightarrow \Endvect(M \times_{\rho} H)$, and $M$ itself can be viewed as the symmetric subalgebra of $M \times_{\rho} H$ with respect to this action. These observations allow us to interpret the entropic certainty equation as a relation between the amount of symmetry breaking for $H$ and the amount of symmetry breaking for $\hat{H}$. The sum of the entropic order parameters for either case must always be equal to the expectation value of the log of the index of the Haar conditional expectation of $H$. This presents an approach to computing and interpreting the index, and finding states that saturate the entropic order parameter bound by symmetrizing with respect to a dual non-invertible symmetry.

\textbf{Note added:} We have coordinated the submission of this work with the authors of~\cite{Benini:2025lav}, who also study the entanglement asymmetry as an entropic order parameter in systems with higher and non-invertible symmetries. We thank the authors of~\cite{Benini:2025lav} for communication and coordination.

\section{From Invertible to Non-invertible Symmetries}  \label{sec: Prelims}

It is well known that order parameters for the breaking of group-like symmetries may be constructed by comparing the information content of a state and its symmetric part~\cite{Casini:2020rgj,Ares:2022koq}. For example, one can consider a theory with a symmetry given by a finite\footnote{The Haar average is also well defined for compact groups in which \eqref{Group Averaging Intro} is replaced by an analogous integral over $G$ with respect to the Haar measure. If $G$ is locally compact, but not compact, this integral can still be defined but does not preserve the normalization of the state.} group $G$ in a state described by a normalized density operator $\rho$. Using the Haar average, one can symmetrize the state to obtain
\begin{equation} \label{Group Averaging Intro}
    \rho_{\rm G} = \frac{1}{|G|}\sum_{g} U(g) \rho U(g)^{\dagger}, \quad U(g) \rho_{\rm G}U(g)^{\dagger} = \rho_{\rm G}~\forall g\in  G\,,
\end{equation}
where $U$ is a unitary representation of $G$ on the Hilbert space of the theory and $|G|$ is the order of the group. The relative entropy between $\rho$ and $\rho_{\rm G}$ serves as an order parameter for the breaking of the $G$ symmetry\footnote{It is interesting to note that this relative entropy may be obtained from a more general quantity called the $n$-th Renyi relative entropy
\begin{equation}
    \Delta_G S^{(n)}(\rho) := \frac{1}{1-n} \log \left[ \frac{\text{tr} \rho^{n}_{\rm G}}{\text{tr} \rho^{n}} \right]\,,
\end{equation}
in the limit $n \to 1$. For each $n$ though, $\Delta_G S^{(n)}(\rho)$ can likewise serve as order parameters. Any such order parameter shares the property that the symmetry is broken in the state $\rho$ if and only if $\Delta_G S^{(n)}(\rho) \neq 0$, but for specific $n$, the features of the symmetry breaking pattern one accesses may be different.}
\begin{equation}
    \Delta_G S(\rho) \equiv  S(\rho| \rho_{\rm G}) = \text{tr}\left[\rho \log \rho - \rho \log \rho_{\rm G} \right]\,,
\end{equation}
which obeys the bound \eqref{groupbound}.

In this note, we provide a framework for adapting the above approach to entropic order parameters for a broader notion of symmetries. As a necessary prerequisite to that discussion, we shall now provide a description of what is meant here by general symmetries (from both categorical and algebraic perspectives).

We take the point of view that our quantum system is described by a subalgebra of observables acting on a Hilbert space $\mathscr{H}$, $M \subset B(\mathscr{H})$. For example, in the algebraic approach to Quantum Field Theory (QFT), $M$ could coincide with the set of observables localized in a particular spacetime subregion \cite{Haag:1963dh,Doplicher:1971wk,Doplicher:1973at,doplicher_new_1989,Longo:1989tt,Longo:1990zp,Longo:1994xe,Guido:1997gi}.\footnote{In future work, we plan to describe a more complete approach in which we consider a collection of algebras coinciding with the observables in any open subset of a spacetime. In this context, extra conditions will be placed on the compatibility between algebras which encode the local and causal properties of the theory. } Given a unitary representation $U: G \rightarrow U(\mathscr{H})$ we obtain an \emph{automorphism} action of $G$ on the algebra $M$:
\beq\label{groupconj}
	\alpha: G \rightarrow \text{Aut}(M), \qquad \alpha_g(x) = U(g)^{\dagger} x U(g)\,.
\eeq
The set $\text{Aut}(M)$ consists of all invertible, algebra preserving maps from $M$ to itself. That is, $\alpha \in \text{Aut}(M)$ is a bijection $\alpha: M \rightarrow M$ which is compatible with the product and the star involution on $M$
\beq \label{Automorphism Action}
	\alpha(xy) = \alpha(x)\alpha(y), \qquad \alpha(x^*) = \alpha(x)^*\,.
\eeq 
In this respect, we say that a group $G$ acts on $M$ as an \emph{invertible symmetry} if there exists a homomorphism $\alpha: G \rightarrow \text{Aut}(M)$. 

A natural generalization of an invertible symmetry is a \emph{non-invertible symmetry}. To define a non-invertible symmetry, we generalize from the set $\text{Aut}(M)$ of bijective algebra preserving maps to the set $\Endvect(M)$ of vector space preserving, but not necessarily invertible maps.
%Elements $\Endvect(M) \ni \rho: M \rightarrow M$ are called \emph{endomorphisms} of the system $M$. 
Given a set $S$, we say that $S$ acts on $M$ as a \emph{non-invertible symmetry} if there exists a map $\rho: S \rightarrow \Endvect(M)$. More specifically, given another algebra $A$, we say that $A$ acts on $M$ as an \emph{algebraic non-invertible symmetry} if there exists a homomorphism $\rho: A \rightarrow \Endvect(M)$ (i.e. such that $\rho_{a_1} \circ \rho_{a_2} = \rho_{a_1 a_2}$). In other words, $M$ is an $A$-module. In the Hilbert space picture, such an action may be implemented by a representation $V: A \rightarrow B(\mathscr{H})$ such that
\ie  \label{Endomorphism Action}
	V(a) x = \rho_{(a_{(1)})}(x) V(a_{(2)})\,,
\fe 
where we have adopted the standard Sweedler notation for the coproduct on the RHS (see standard textbook \cite{Majid:1996kd}).
This generalizes \eqref{groupconj} for the group symmetries. We refer to $V(a)$ as `charged intertwiners' (generalizing the notions in \cite{Longo:1994xe}) and for point-localized endomorphisms (obtained from a limiting procedure \cite{Buchholz:1988ae}), they correspond to twisted sector operators (e.g. attached to the topological defect network whose action on $B(\cH)$ is represented by $\rho$).\footnote{Since $\text{Aut}(M) \subset \Endvect(M)$, the notion of non-invertible symmetry subsumes also invertible symmetries. When we discuss a non-invertible symmetry we leave open the possibly that some of the symmetry generators may be invertible e.g. belong to $\text{Aut}(M)$.}

If, in addition to preserving $M$ as a set, $\rho \in \Endvect(M)$ also preserves the product and involution on $M$, it is called an algebra endomorphism of $M$. We denote the set of algebra endomorphisms simply by $\text{End}(M)$. One can regard $\rho \in \text{End}(M)$ as a (not necessarily faithful) representation of the algebra $M$. Often, a general endomorphism can be decomposed into a direct sum of irreducible elements, as in the representation theory of a group. More precisely, the \emph{sector} associated with a given endomorphism may decompose as a direct sum. Here, by sector we mean the inner unitary equivalence class of an endomorphism
\beq
\label{sectors}
	[\rho] \equiv \{\text{Ad}_u \circ \rho \; | \; u \in U(M) \}\,.
\eeq
Irreducible sectors would coincide with superselection sectors in the physical sense.\footnote{The existence of such endomorphisms is guaranteed for a properly infinite algebra $M$, given a categorical symmetry action implemented by unital completely positive maps, via a generalized Stinespring dilation theorem \cite{longo_landauers_2018} (possibly after enlarging the underlying Hilbert space $\cH$). However, obtaining explicit realizations of these endomorphisms as operators in $U(\cH)$ remains highly nontrivial. Most known constructions are confined to free (bosonic or fermionic) theories, where the presence of canonical (anti)commutation relations (CCR/CAR algebras) enables a concrete description of the endomorphisms through their action on free field modes \cite{Buchholz:1988ae,Mack:1989uz,Fuchs:1991qv,Bockenhauer:1994tr,Bockenhauer:1996gt}.} 

To formalize this notion, it becomes useful to \emph{categorify} our analysis of $\text{End}(M)$. As a category, $\text{End}(M)$ has objects given by endomorphisms of the algebra $M$. For any pair $\rho,\sigma \in \text{End}(M)$ the hom-set $\text{Hom}(\rho,\sigma)$ is given by the collection of \emph{inner} intertwining elements $t \in M$ such that
\beq
	t \rho(x) = \sigma(x) t, \qquad \forall x \in M\,.
    \label{intertwiner}
\eeq
We say that $\rho \in \text{End}(M)$ is \emph{irreducible} if it is irreducible as a representation (i.e. provided $\rho(M) \subset B(\mathscr{H})$ admits no non-trivial invariant subspaces). In the Hilbert space setting, this is equivalent to saying that the set of all operators in $M$ which commute with $\rho(M)$ are complex multiples of the identity. The category $\text{End}(M)$ is called \emph{semisimple} if every $[\rho]$ can be decomposed as a finite direct sum of irreducible sectors. The irreducible sectors carry \textit{charges}, and the properties of the category translate to their fusion rules, conjugation, and multiplicities. In particular, the   sectors (of equivalence classes of endomorphisms) in \eqref{sectors} correspond to the topological defects in the fusion category and the intertwiners \eqref{intertwiner} correspond to topological junction vectors between the defects.  

The categorical point of view on the set of endomorphisms of $M$ unlocks yet another generalization of the notion of symmetry which we refer to as a \emph{categorical symmetry}. Given a category $\mathcal{C}$, we say that $\mathcal{C}$ acts on $M$ as a \emph{non-invertible categorical symmetry} if there exists a functor $\Phi: \mathcal{C} \rightarrow \text{End}(M)$. In this respect, a categorical symmetry can be thought of as a special kind of non-invertible symmetry in which the objects of a category act on the system $M$ by endomorphisms which intertwine between different sectors. 

A useful example that is dual to a standard group action is the notion of a categorical symmetry action by the category $\text{Rep}(G)$ of finite dimensional representations of a (typically compact) group $G$. The analog of irreducible sectors in the context of the category $\text{Rep}(G)$ are, predictably, irreducible unitary representations of the group. Thus, one can think of the action of $\text{Rep}(G)$ as being implemented on the algebra $M$ by (direct sums of) charged intertwiners \eqref{Endomorphism Action} which carry labels given by irreps of $G$. A useful way to access all of these physical representations is by referring to the regular representation $L^{2}(G)$, which decomposes into a direct sum of irreducibles with multiplicity given by their dimension.

The irreducible representations of $G$ also happen to form a basis for the Abelian algebra of functions on $G$. This can be seen by recognizing that the representative functions
\beq
	g \mapsto \phi_{Uij}(g) = \bra{e_i^{(U)}} U(g) \ket{e_j^{(U)}}\,,
\eeq	
are dense in the algebra $L^{\infty}(G)$. Here, $U: G \rightarrow U(\mathscr{H}_U)$ is a unitary irrep with $\ket{e^{(U)}_i}$ an orthonormal basis for $\mathscr{H}_U$. The product of two such representative functions is given by
\begin{flalign}
	\phi_{Uij}(g) \phi_{Vkl}(g) &= \bra{e_i^{(U)}} \otimes \bra{e_k^{(V)}} U \otimes V(g) \ket{e_j^{(U)}} \otimes \ket{e_l^{(V)}} \nonumber \\
    &= \sum_{W \in G^{\vee}} \sum_{m,n = 1}^{\text{dim}(\mathscr{H}_W)} N^{Wmn}_{Uij,Vkl} \phi_{Wmn}(g)\,,
\end{flalign}
where $G^{\vee}$ is the collection of (equivalence classes of) unitary irreps of $G$ and $N^{Wmn}_{Uij,Vkl}$ are the fusion coefficients governing the tensor product of irreps. From this observation, it is possible to show that the \emph{categorical} symmetry action $\Phi: \text{Rep}(G) \rightarrow \text{End}(M)$ equivalently gives rise to an \emph{algebraic} symmetry action $\rho: L^{\infty}(G) \rightarrow \text{End}(M)$. 

The correspondence between categorical and algebraic symmetries can be further generalized to the context of \emph{fusion categories} and \emph{finite connected WHAs} \cite{calaque2008lecturestensorcategories,hayashi1999canonicaltannakadualityfinite,etingof2017fusioncategories,muger2006abstract,MUGER2003159,Halvorson:2006wj}. A fusion category (over $\mathbb{C}$) is a $\mathbb{C}$-linear, semisimple, monoidal, rigid category with finitely many (isomorphism classes of) simple objects. A WHA is a weak bialgebra $(H,\mu,\eta,\Gamma,\epsilon)$ along with a linear map $S: H \rightarrow H$ called the antipode.\footnote{We review the definitions of the fusion algebra and WHA in the Appendix. However, we emphasize that these details are not necessary to follow the main line of the paper. All of the necessary details will be discussed in the main text.} Tannaka-Krein duality establishes that, given any fusion category $\mathcal{C}$, there exists a family of WHAs $\mathcal{F}$ such that $\mathcal{C} = \text{Rep}(H)$ for each $H \in \mathcal{F}$. This family can be indexed by the choice of an indecomposable module category $\mathcal{M}$ over $\mathcal{C}$ \cite{Ostrik:2001xnt,Bai:2025zze}. In the next section we expand upon this correspondence and provide a complete algorithm for cycling through the different algebraic non-invertible symmetries which can be realized for a given fusion categorical symmetry.

\section{Algebraic Symmetries from Categorical Symmetries} \label{sec: algebras from categories}

In this section, we discuss how to obtain algebraic non-invertible symmetries from categorical ones. Our starting point is a fusion category $\mathcal{C}$ acting on a system $M \subset B(\mathscr{H})$ via a functor $\Phi: \mathcal{C} \rightarrow \text{End}(M)$. We therefore think of the objects of $\mathcal{C}$ as acting on the system $M$ by endomorphisms. An overview of the algebras associated with a fusion category $\mathcal{C}$ is presented in Table \ref{Table: Fusion Cat Alg}. 

\begin{table}[h!]
\centering
\footnotesize
\begin{tabularx}{\textwidth}{>{\raggedright\arraybackslash}m{1.1cm}
                                     >{\raggedright\arraybackslash}m{3.0cm} 
                                     >{\centering\arraybackslash}m{1.6cm}
                                     >{\centering\arraybackslash}m{1.3cm}
                                     >{\centering\arraybackslash}m{2.0cm}
                                     >{\centering\arraybackslash}m{1.8cm}
                                     >{\centering\arraybackslash}m{2.0cm}}
\toprule
\textbf{Name} &
\textbf{Algebra} & 
\textbf{C$^*$ Completion} & 
\textbf{Weak Hopf} & 
\textbf{RepCat} & 
\textbf{Acts on} & 
\textbf{Canonical} \\
\midrule

Fusion &
\makecell[l]{
$\text{Fus}(\mathcal{C}) = \bigoplus_{U}$\\
$\text{Hom}_{\mathcal{C}}(U,U)$}
& Yes & No & N/A & $M$ & Yes \\

\addlinespace[0.5em]

Strip &
\makecell[l]{
$\text{Str}_{\mathcal{C}}(\mathcal{M}) = \bigoplus_{X,X',Y,Y'}$\\
$\text{Hom}_{\mathcal{C}}(\overline{[Y',Y]}, [X,X'])$}
& Yes & Yes 
& \makecell{$\text{Rep}\left(\text{Str}_{\mathcal{C}}(\mathcal{M})^*\right)$\\$ = \mathcal{C}$} 
& $\frac{\text{Str}_{\mathcal{C}}(\mathcal{M}) \otimes M}{\text{Fus}(\mathcal{C})}$ 
& \makecell{No, depends\\on $\mathcal{M}$} \\

\addlinespace[0.5em]

Tube &
\makecell[l]{
$\text{Tub}(\mathcal{C}) = \bigoplus_{U,V,W}$\\
$\text{Hom}_{\mathcal{C}}(U \otimes V, W \otimes U)$}
& Yes & No 
& \makecell{$\text{Rep}(\text{Tub}(\mathcal{C}))$\\ $= \mathcal{Z}(\mathcal{C})$} 
& $M \otimes B(L^2(\mathcal{C}))$ 
& Yes \\
\bottomrule
\end{tabularx}
\caption{Guide to the algebras associated with a fusion category $\mathcal{C}$ supporting a categorical symmetry action $\Phi: \mathcal{C} \rightarrow \text{End}(M)$. The fusion algebra $\text{Fus}(\mathcal{C})$ is formed by taking the linear extension of the fusion rules for simple objects in the category. It obtains a C$^*$ completion through its embedding in the tube algebra $\text{Tub}(\mathcal{C})$, and inherits an endomorphism action on the system $M$ directly through $\Phi$. The strip algebra $\text{Str}_{\mathcal{C}}(\mathcal{M})$ (its dual) is obtained by appealing to the Tannaka-Krein duality which realizes $\mathcal{C}$ as the representation category of a WHA. The WHA is not unique; $\mathcal{M}$ is an 
indecomposable faithful module category of $\mathcal{C}$ which indexes a subset of possible choices. The strip algebra inherits an action on the extended system algebra $M_{\mathcal{M}} \equiv (\text{Str}_{\mathcal{C}}(\mathcal{M}) \otimes M)/\text{Fus}(\mathcal{C})$ which combines the action of the category $\mathcal{C}$ on the original system $M$ with its action on the module category $\mathcal{M}$. This action is non-canonical insofar as it depends on the choice of module category, however this choice may be natural given a particular physical setting. Finally, the tube algebra can be formed purely from the category $\mathcal{C}$ and is defined by the properties that (a) it possesses a universal C$^*$ completion and (b) its representation category computes the Drinfeld center of $\mathcal{C}$. Schematically, one may regard $\text{Tub}(\mathcal{C})$ as a double of the fusion algebra (the Drinfeld center can be read as a categorical double $\mathcal{Z}(\mathcal{C}) \simeq (\mathcal{C} \boxtimes \mathcal{C}^{op})^*_{\mathcal{C}}$), and it acts on the system $M$ plus the full set of twisted sectors which form its so-called admissible representations \cite{ghosh2015annularrepresentationtheoryrigid}.} \label{Table: Fusion Cat Alg}
\end{table}

\subsection{Fusion Algebra}

As a fusion category, $\mathcal{C}$ possesses a finite set of irreducible (simple) objects which we denote by $\text{Irr}(\mathcal{C})$. Because it is semisimple, the tensor product of any pair of objects $X,Y \in \Obj{C}$ can be re-expressed as a direct sum of objects $U \in \text{Irr}(\mathcal{C})$. In particular, the tensor products of any two irreducible objects is given by
\beq
	U \otimes V = \bigoplus_{W \in \text{Irr}(\mathcal{C})} N_{UV}^W W\,,
\eeq
where $N_{UV}^W$ are called \emph{fusion coefficients}. Because it is rigid, every object $X \in \Obj{\mathcal{C}}$ possesses a dual object $\overline{X} \in \Obj{\mathcal{C}}$. Using these facts, we can fashion the set of irreducible objects into an involutive algebra called the fusion algebra, $\text{Fus}(\mathcal{C})$. The elements of $\text{Fus}(\mathcal{C})$ are complex linear combinations of simple objects with product given by the linear extension of the fusion rules, and involution given by the linear extension of object duality. In this regard, we can already think of the categorical action of $\mathcal{C}$ as giving rise to an algebraic action by $\text{Fus}(\mathcal{C})$. 

To illustrate the general structure of the fusion algebra, it will be useful to call to mind the following example. The invertible symmetry of a group $G$ acting upon an algebra $M$ by automorphism can be recast as a categorical symmetry action of the category $\text{Vec}_G$. The category $\text{Vec}_G$ has as objects $G$-graded vector spaces and as morphisms intertwining maps between these vector spaces which preserve the grading. The simple objects of this category are labeled by group elements and may therefore be identified with the group itself, $\text{Irr}(\text{Vec}_G) \simeq G$. Let us denote simple objects by $\ell_g$. The fusion rules simply encode the group product, and object duality is encoded in the group inversion:
\beq
	\ell_g \otimes \ell_h = \ell_{gh}, \qquad \overline{\ell_g} = \ell_{g^{-1}}\,.
\eeq
In this regard, the fusion algebra of $\text{Vec}_G$ is nothing but the usual group algebra $\text{Fus}(\text{Vec}_G) = \mathbb{C}[G]$. 

A nice feature of the group algebra $\mathbb{C}[G]$ is that it possesses a universal completion to a C$^*$-algebra. This completion can be obtained, for example, by constructing the left regular representation of the group algebra acting on the Hilbert space $L^2(G)$. The simple objects $\ell_g$ act on elements $\psi \in L^2(G)$ as translations:
\beq
	\bigg(\ell_g \psi\bigg)(h) \equiv \psi(g^{-1}h)\,.
\eeq
and can be shown to be unitary operators with respect to the natural inner product on $L^2(G)$. The universal C$^*$-completion of $\mathbb{C}[G]$ is its completion under the operator norm induced by this inner product. The resulting C$^*$-algebra is called the group convolutional algebra $\mathcal{L}(\text{Vec}_G) \equiv \mathcal{L}(G)$. We can extend the categorical action of $\text{Vec}_G$ to an algebraic action of $\mathcal{L}(G)$. It is well known that representations of $\mathcal{L}(G)$ are in correspondence with unitary representations of $G$ itself, and thus we can regard the resulting endomorphism action of $\mathcal{L}(G)$ as emerging from an automorphism action of the group $G$, as advertised. 

Returning to the case of a general fusion category, we would like to lift the fusion algebra $\text{Fus}(\mathcal{C})$ to a C$^*$-algebra, to determine the full algebraic action induced by the functor $\Phi$. However, a problem emerges because $\text{Fus}(\mathcal{C})$ does not come equipped with a natural, universal topology like in the case of the group algebra. In the recent literature, a trio of different approaches to overcoming this problem have been proposed \cite{Popa_2015,Neshveyev_2016,GHOSH20161537}. For the purposes of this note, we will concentrate on the approach of \cite{GHOSH20161537} because it is the most transparent algebraically. It has now been appreciated that all three approaches are equivalent, but nevertheless yield interesting insights into different aspects categorical symmetry.

The general approach of \cite{GHOSH20161537} is to arrive at the C$^*$ completion of $\text{Fus}(\mathcal{C})$ by embedding it into a larger algebra which itself possesses a universal C$^*$ norm. The appropriate algebra to consider is the tube algebra, $\text{Tub}(\mathcal{C})$. The tube algebra is defined by
\beq
	\text{Tub}(\mathcal{C}) \equiv \bigoplus_{U,V,W \in \text{Irr}(\mathcal{C})} \text{Tub}_{VW}^{U}(\mathcal{C}) = \bigoplus_{V,W \in \text{Irr}(\mathcal{C})} \text{Tub}_{VW}(\mathcal{C})\,,
\eeq
with
\beq
	\text{Tub}_{VW}^U(\mathcal{C}) \equiv \text{Hom}(U \otimes V, W \otimes U), \qquad \text{Tub}_{XY}(\mathcal{C}) \equiv \bigoplus_{U \in \text{Irr}(\mathcal{C})} \text{Tub}_{XY}^U(\mathcal{C})\,.
\eeq
For the explicit algebraic operations of $\text{Tub}(\mathcal{C})$, e.g. its product and involution, we refer the reader to \cite{GHOSH20161537}. Given these definitions, it is not difficult to show that $\text{Tub}_{\mathbb{I}\mathbb{I}}(\mathcal{C}) = \text{Fus}(\mathcal{C})$. Thus, we have an inclusion of involutive algebras, $\text{Fus}(\mathcal{C}) = \text{Tub}_{\mathbb{I} \mathbb{I}}(\mathcal{C}) \subset \text{Tub}(\mathcal{C})$.  

The tube algebra was originally studied for its relationship to the Drinfeld center \cite{MUGER2003159,Ocneanu2016ChiralityFO,etingof2017fusioncategories,Bai:2025zze}. We take as a definition the fact that the Drinfeld center $\mathcal{Z}(\mathcal{C})$ is (monoidally) equivalent to the category of $*$-representations of the tube algebra: $\mathcal{Z}(\mathcal{C}) \simeq \text{Rep}(\text{Tub}(\mathcal{C}))$. Although this was not the original definition, this correspondence provides the most constructive way to analyze $\mathcal{Z}(\mathcal{C})$. 

For our present purpose, the most important feature of the tube algebra is that it possesses a universal C$^*$ completion. In \cite{GHOSH20161537}, this completion is obtained by considering a generalization of the GNS construction for $\text{Tub}(\mathcal{C})$ centered around so-called weight $U$-annular states, for $U \in \text{Irr}(\mathcal{C})$. A universal representation for $\text{Tub}(\mathcal{C})$ is obtained by taking the direct sum of GNS Hilbert spaces of all possible weight $U$-annular states for each $U \in \text{Irr}(\mathcal{C})$. We denote this Hilbert space by $L^2(\mathcal{C})$, and note that it can be regarded as a generalization of the regular representation of the group, which is itself a direct sum of irreducible unitary representation spaces. 

The universal C$^*$-tube algebra of $\mathcal{C}$ is the completion of $\text{Tub}(\mathcal{C})$ in the operator norm of its universal representation. Upon completing $\text{Tub}(\mathcal{C})$ we likewise obtain a C$^*$ completion of $\text{Fus}(\mathcal{C})$ as a subalgebra. The resulting C$^*$-algebra $\mathcal{L}(\mathcal{C})$ is a generalization of the group convolutional algebra which inherits an endomorphism action on $M$ via the functor $\Phi$. Thus, for any fusion categorical symmetry, we can always construct a C$^*$-algebraic symmetry action. 

\subsection{Strip Algebra}

In the previous subsection, we have constructed a C$^*$-algebra, $\mathcal{L}(\mathcal{C})$, which inherits an endomorphism action on $M$ through the categorical action $\Phi: \mathcal{C} \rightarrow \text{End}(M)$. Nevertheless, it is not immediately clear whether $\mathcal{L}(\mathcal{C})$ is a WHA, and thus we cannot immediately conclude that $\mathcal{C} \simeq \text{Rep}(\mathcal{L}(\mathcal{C})^*)$. In this subsection and the next, we demonstate how to extend the action of the fusion algebra to that of a WHA whose representation category is equivalent to $\mathcal{C}$. 

To address this problem, let us first describe how one can arrive at a WHA, which we will denote by $A$ for the purposes of this construction, such that $\mathcal{C} = \text{Rep}(A)$ for a general fusion category. This follows from Tannaka-Krein duality, and is described in great detail in \cite{Bai:2025zze}. Here, we simply review the important ingredients. The central structure which is required to construct $A$ is a faithful and exact separable Frobenius functor $F: \mathcal{C} \rightarrow \text{Vec}$, also known as a weak fiber functor.\footnote{For our purposes we shall restrict our attention to a subclass of such functors which can be directly constructed. For a more detailed definition we refer the reader to \cite{Bai:2025zze}.} Given such a functor, we can construct an algebra $A_F \equiv \text{End}(F)$ consisting of natural transformations from $F$ to itself. In this case, a natural transformation is a map $\eta: \text{Obj}(\mathcal{C}) \rightarrow \text{Hom}_{\mathcal{C}}$ such that
\beq
	\eta_X \in \text{Hom}_{\mathcal{C}}(F(X),F(X)), \qquad \eta_Y \circ F(f) = F(f) \circ \eta_X, \; \forall f \in \text{Hom}_{\mathcal{C}}(X,Y)\,.
\eeq
The set $\text{End}(F)$ becomes an algebra under the composition of natural transformations. 

For any fiber functor $F$, $\mathcal{C} = \text{Rep}(\text{End}(F))$. While the preceding discussion is valid for arbitrary finite $\mathbb{C}$-linear categories, in the event that $\mathcal{C}$ is moreover a fusion category the algebra $\text{End}(F)$ can be endowed with a weak Hopf structure \cite{Bai:2025zze}.

In \cite{Bai:2025zze}, a large class of fiber functors for $\mathcal{C}$ are constructed, $F_{\mathcal{M},B}$, which are labeled explicitly by the choice of (a) a semisimple, faithful, module category $\mathcal{M}$ over $\mathcal{C}$ and (b) a Frobenius algebra $B$ such that $\mathcal{M} = \text{Rep}(B)$. The existence of such a $B$ is ensured through a theorem of Hayashi \cite{hayashi1999canonicaltannakadualityfinite}. In particular, the algebra $B_{\mathcal{M}} \equiv \mathbb{C}^{\oplus |\text{Irr}(\mathcal{M})|} = \bigoplus_{X \in \text{Irr}(\mathcal{M})} \text{Hom}_{\mathcal{M}}(X,X)^{op}$ satisfies $\mathcal{M} = \text{Rep}(B_{\mathcal{M}})$, and may always be promoted to a Frobenius algebra (provided $\mathcal{M}$ is semisimple). 

We therefore obtain a full score of WHAs, $A^{\mathcal{C}}_{\mathcal{M},B} = \text{End}(F_{\mathcal{M},B})$, such that $\mathcal{C} = \text{Rep}(A^{\mathcal{C}}_{\mathcal{M},B})$. The algebra $B$ turns out to be equivalent to the target counital base subalgebra of $A^{\mathcal{C}}_{\mathcal{M},B}$ \cite{Bai:2025zze}. For our purposes, we will be most interested in the `canonical' such algebra for which $B = B_{\mathcal{M}}$. Then, the choice of weak Hopf representative is fixed by the choice of module category $\mathcal{M}$. 

The WHA $A^{\mathcal{C}}_{\mathcal{M},B_{\mathcal{M}}}$ is closely related to another standard algebraic structure which has appeared in recent literature called the strip algebra. Adopting the notation of \cite{Cordova:2024iti}, $A^{\mathcal{C}}_{\mathcal{M},B_{\mathcal{M}}} = \text{Str}_{\mathcal{C}}(\mathcal{M})^*$. The strip algebra is based on the vector space
\beq
	\text{Str}_{\mathcal{C}}(\mathcal{M}) = \bigoplus_{X,X',Y,Y' \in \text{Irr}(\mathcal{M}), U \in \text{Irr}(\mathcal{C})} \text{Hom}_{\mathcal{M}}(Y', U \triangleright Y) \otimes \text{Hom}_{\mathcal{M}}(U \triangleright X, X')\,,
\eeq	
where here $U \triangleright X$ is the action of $\mathcal{C}$ on $\mathcal{M}$. Using the language of internal homs\footnote{The internal hom $[X,\cdot]: \mathcal{M} \rightarrow \mathcal{C}$ is defined as the right adjoint of the functor $\cdot \triangleright X: \mathcal{C} \rightarrow \mathcal{M}$ for each $X \in \text{Obj}(\mathcal{M})$. We refer to $[X,Y]$ as the internal hom between $X$ and $Y$.}, the strip algebra can be rewritten strictly in terms of homomorphims of $\mathcal{C}$ as
\beq \label{Strip in terms of internal homs}
	\text{Str}_{\mathcal{C}}(\mathcal{M}) = \bigoplus_{X,X',Y,Y' \in \text{Irr}(\mathcal{M})} \text{Hom}_{\mathcal{C}}(\overline{[Y',Y]}, [X,X'])\,.
\eeq
In Appendix~\ref{app: strip} we provide a complete description of strip algebras induced from fusion categories in terms of a diagrammatic calculus. 

As with the fusion algebra, it is useful to pause to consider the strip algebra for the category $\mathcal{C} = \text{Vec}_G$ which carries an invertible symmetry of the group $G$. Given the trivial choice of module category $\mathcal{M} = \text{Vec}$, the strip algebra is exactly equivalent to the group C$^*$-algebra, $\text{Str}_{\text{Vec}_G}(\text{Vec}) = \mathcal{L}(G)$. If we take $\mathcal{M} = \text{Vec}_G$, the strip algebra organizes itself into a direct sum of matrix algebras of dimension $|G|^2$, each labelled by an element of $g$:
\beq
    \text{Str}_{\text{Vec}_G}(\text{Vec}_G) = \bigoplus_{g \in G} \mathbb{C}^{|G| \times |G|}\,.
\eeq    
One may interpret this as an amplification of the standard group algebra in which, rather than taking the complex span of the group, we take the span of the group with a field of $|G| \times |G|$ matrices. From this point of view, we have the natural inclusion $\text{Str}_{\text{Vec}_G}(\text{Vec}) \subset \text{Str}_{\text{Vec}_G}(\text{Vec}_G)$. 

Associated with any pair $(\mathcal{C},\mathcal{M})$ we can define another category called the dual category of $\mathcal{C}$ with respect to $\mathcal{M}$ and denoted by $\mathcal{C}_{\mathcal{M}}^*$. This can be thought of as the category of endomorphisms of $\mathcal{M}$ with opposite monoidal structure so that $\mathcal{C}_{\mathcal{M}}^*$ acts on $\mathcal{M}$ from the right, while $\mathcal{C}$ itself acts from the left. In fact, the category $\mathcal{C}_{\mathcal{M}}^*$ encodes the representations of $\text{Str}_{\mathcal{C}}(\mathcal{M})$:
\beq \label{Rep of Strip}
	\text{Rep}(\text{Str}_{\mathcal{C}}(\mathcal{M})) \simeq \mathcal{C}_{\mathcal{M}}^*\,.
\eeq
The Hopf dual of the strip algebra is another WHA which in fact is equivalent to the strip algebra of the dual category $\mathcal{C}_{\mathcal{M}}^*$ with respect to $\mathcal{M}$ as a module:
\beq \label{Dual of Strip}
	\text{Str}_{\mathcal{C}}(\mathcal{M})^* = \text{Str}_{\mathcal{C}_{\mathcal{M}}^*}(\mathcal{M})\,.
\eeq
In other words, duality in the weak Hopf sense is functorially related to duality in the categorical sense (with respect to a given module category). Combining \eqref{Rep of Strip} and \eqref{Dual of Strip} provides an alternative point of view on the equality
\beq
	\text{Rep}(\text{Str}_{\mathcal{C}}(\mathcal{M})^*) = \text{Rep}(\text{Str}_{\mathcal{C}_{\mathcal{M}}^*}(\mathcal{M})) \simeq (\mathcal{C}_{\mathcal{M}}^*)_{\mathcal{M}}^* \simeq \mathcal{C}\,.
\eeq	

Eqn. \eqref{Rep of Strip} should be compared with the relationship between the tube algebra and the Drinfeld center. In fact, the Drinfeld center can be identified with the dual of the categorical double $\mathcal{C} \boxtimes \mathcal{C}^{op}$ with respect to $\mathcal{C}$ viewed as a module category: $Z(\mathcal{C}) \simeq (\mathcal{C} \boxtimes \mathcal{C}^{op})_{\mathcal{C}}^*$. Thus, from \eqref{Rep of Strip} we conclude that $\text{Str}_{\mathcal{C} \boxtimes \mathcal{C}^{op}}(\mathcal{C})$ is Morita equivalent to the tube algebra. The Drinfeld double of a WHA $H$ is a particular WHA $D(H)$ modeled on the space $H \otimes H^*$ and inheriting its product and coproduct structure for $H$, $H^*$, and their action on each other \cite{Kassel1995}. If $\mathcal{C} = \text{Rep}(H)$, the Drinfeld center is (monoidally) equivalent to the representation category of the Drinfeld double $D(H)$. So, we also conclude that the tube algebra is Morita equivalent to $D(H)$. We therefore land on the following sequence of Morita equivalences:
\beq \label{Tube, Double, Center}
	\text{Tub}(\mathcal{C}) \simeq \text{Str}_{\mathcal{C} \boxtimes \mathcal{C}^{op}}(\mathcal{C}) \simeq D(\text{Str}_{\mathcal{C}}(\mathcal{M})^*)\,.
\eeq

\subsection{Rieffel Induction from Fusion to Strip}

Unfortunately, the strip algebra appears to suffer from the opposite problem as the fusion algebra. The fusion algebra possesses a natural endomorphism action on the system, but is not clearly weak Hopf. The dual of the strip algebra is clearly weak Hopf and satisfies $\mathcal{C} \simeq \text{Rep}(H)$, but it is not clear that it gives rise to an algebraic action on the system. We can remedy this mismatch by invoking the machinery of Rieffel induction to promote the representation $\rho: \text{Fus}(\mathcal{C}) \rightarrow \text{End}_{\textrm{v}}(M)$ to a representation of $\text{Str}_{\mathcal{C}}(\mathcal{M})$ acting on an enlarged system $M_{\mathcal{M}}$. 

Let us briefly recall the procedure of Rieffel induction \cite{RIEFFEL1974176}. Let $A$ and $B$ be a pair of C$^*$-algebras. A $B$-rigged space is a Banach space $X$ admitting a right representation $r: B \rightarrow B(X)$ and a compatible $B$-valued inner product $G: X \times X \rightarrow B$ such that
\beq
	G(x,r(b) y) = G(x,y) b, \qquad \forall x,y \in X, b \in B\,.
\eeq
If $G(x,x) = 0 \iff x = 0$ we say that $G$ is definite, however we don't strictly require this property. The set, $B_G(X)$, of bounded operators on $X$ with respect to the $B$-rigging consist of maps $\mathcal{O}: X \rightarrow X$ such that\footnote{In the case that $G$ is definite, the adjoint operator $\mathcal{O}^{\dagger}$ is unique and conditions $1$ and $2$ above imply $3$.}
\begin{enumerate}
	\item There exists $k > 0$ such that $G(\mathcal{O}(x), \mathcal{O}(y)) \leq k^2 G(x,y)$ for all $x,y \in X$,
	\item There exists $\mathcal{O}^{\dagger}: X \rightarrow X$ (bounded as in point 1) such that $G(\mathcal{O}(x),y) = G(x, \mathcal{O}^{\dagger}(y))$ for all $x,y \in X$, and
	\item The operator $\mathcal{O}$ commutes with the right action of $B$, $[\mathcal{O},r(b)] = 0$ for each $b \in B$. 
\end{enumerate}
If $X$ also admits a homomorphism $\ell: A \rightarrow B_G(X)$, then it is called a $B$-rigged $A$-module. 

Let $X$ be a $B$-rigged $A$-module. Rieffel induction states that any Hilbert space representation $\rho: B \rightarrow B(H)$ can be lifted to a Hilbert space representation $\rho^{X}: A \rightarrow B(H_X)$. The Hilbert space $H_X$ is defined as follows: as a vector space it is the quotient of the tensor product $X \otimes H$ by the equivalence relation
\beq \label{Quotient for RI}
	x \otimes \rho_b(v) \sim r_b(x) \otimes v, \qquad \forall x \in X, v \in H, b \in B\,.
\eeq
One can interpret \eqref{Quotient for RI} as quotienting by the algebra $B$, and we denote this quotient space by $H_X \equiv (X \otimes H)/B$. The inner product on $H_X$ is inherited from that of $H$ and the $B$-valued inner product on $X$:
\beq \label{Inner product on H_X}
	\langle x_1 \otimes v_1, x_2 \otimes v_2 \rangle_{H_X} \equiv \langle v_1, \rho_{G(x_1,x_2)}(v_2) \rangle_H\,.
\eeq
The representation of $B$ on $H$ induces a representation of $A$ on $H_X$:
\beq
	\rho^X_a(x \otimes v) \equiv \pi_a(x) \otimes v\,.
\eeq
It is worth noting from \eqref{Inner product on H_X} that
\beq
	\langle x_1 \otimes v_1, \rho^X_a(x_2 \otimes v_2) \rangle_{H_X} = \langle x_1 \otimes v_1, \pi_a(x_2) \otimes v_2 \rangle_{H_X} = \langle v_1, \rho_{G(x_1, \pi_a(x_2))}(v_2)\rangle_H\,,
\eeq
and so we see that the original action of $B$ on $H$ is encoded implicitly in $\rho^X$ through the structure of the inner product on $H_X$. 

A useful example of Rieffel induction comes from the case where $i: B \hookrightarrow A$ is an inclusion of C$^*$-algebras. In this context, $A$ can be promoted to a $B$-rigged $A$-module by endowing it with the $B$-valued inner product induced from a conditional expectation $\mathcal{E}: A \rightarrow B$. Recall that a conditional expectation is a positive, unital, projection satisfying
\beq
	\mathcal{E}(i(b) a i(b^*)) = b \mathcal{E}(a) b^*\,.
\eeq
The map $\mathcal{E}$ induces a $B$-valued bilinear $G_{\mathcal{E}}: A \times A \rightarrow B$ given by
\beq
	G_{\mathcal{E}}(a_1, a_2) \equiv \mathcal{E}(a_1^* a_2)\,.
\eeq	
The algebra $B$ has an antirepresentation on $A$ by right multiplication in the algebra, $r_b(a) \equiv a i(b)$, such that
\beq
	G_{\mathcal{E}}(a_1, r_b(a_2)) = \mathcal{E}(a_1^* a_2 i(b)) = \mathcal{E}(a_1^* a_2) b = G_{\mathcal{E}}(a_1, a_2) b\,.
\eeq
The algebra $A$ acts on itself via the representation $\ell_{a_1}(a_2) \equiv a_1 a_2$. This representation can be regarded as a homomorphism $\ell: A \rightarrow B_{G_{\mathcal{E}}}(A)$, and thus $A$ is a $B$-rigged $A$-module as claimed. 

To implement Rieffel induction in our context, we note that the endomorphism action $\rho: \text{Fus}(\mathcal{C}) \rightarrow \text{End}(M)$ can be lifted to an action of the fusion algebra on the standard Hilbert space of the system, which we denote by $L^2(M)$.\footnote{See \cite{AliAhmad:2025oli} for a review of the standard Hilbert space from the GNS construction. Briefly, given a faithful, semifinite, normal weight $\omega$ on $M$, a dense subset of $L^2(M)$ is generated by elements $\ket{x}$ where $x \in M$ has finite expectation value under $\omega$, $\omega(x) < \infty$. The inner product on this space is induced from the weight as $\braket{x|y} \equiv \omega(x^* y)$. The action $\rho: A \rightarrow \Endvect(M)$ induces an action $\rho: A \rightarrow B(L^2(M))$ given by $\rho_a \ket{x} \equiv \ket{\rho_a(x)}$.} Given a $\text{Fus}(\mathcal{C})$-rigged $\text{Str}_{\mathcal{C}}(\mathcal{M})$-module $X$, we can then promote $\rho$ to a representation $\rho^X: \text{Str}_{\mathcal{C}}(\mathcal{M}) \rightarrow B(L^2(M)_X)$. In fact, the fusion algebra embeds into the strip algebra \cite{Jia:2024zdp}, and thus the strip algebra itself can be endowed with the structure of a $\text{Fus}(\mathcal{C})$-rigged $\text{Str}_{\mathcal{C}}(\mathcal{M})$-module by specifying a conditional expectation $\mathcal{E}: \text{Str}_{\mathcal{C}}(\mathcal{M}) \rightarrow \text{Fus}(\mathcal{C})$. In this case, the extended Hilbert space $\bigg(L^2(\text{Str}_{\mathcal{C}}(\mathcal{M})) \otimes L^2(M)\bigg)/\text{Fus}(\mathcal{C})$ is the standard Hilbert space of an algebra $M_{\mathcal{M}} \equiv (\text{Str}_{\mathcal{C}}(\mathcal{M}) \otimes M)/\text{Fus}(\mathcal{C})$, where the quotient \eqref{Quotient for RI} is lifted to the algebraic level as in \eqref{Algebraic Quotient} below. 

One may regard the algebra $M_{\mathcal{M}}$ as an extension of the system algebra by the operators in the strip algebra which do not also appear in the fusion algebra. More colloquially, we think of $M_{\mathcal{M}}$ as an extension of the system by the generators of the module category $\mathcal{M}$. Since there is a map from $M_{\mathcal{M}}$ into a dense subset of the extended Hilbert space, we can treat the map $\rho^{L^2(\text{Str}_{\mathcal{C}}(\mathcal{M}))} \equiv \rho^{\mathcal{M}}: \text{Str}_{\mathcal{C}}(\mathcal{M}) \rightarrow B(L^2(M_{\mathcal{M}}))$ as an action of $\text{Str}_{\mathcal{C}}(\mathcal{M})$ on $M_{\mathcal{M}}$.\footnote{In Appendix~\ref{App: Cat from Alg} we provide a complimentary perspective on how to determine when a given representation of the strip algebra is induced from a representation of the fusion algebra by appealing to Rieffel's notion of imprimitivity.}

\subsection{Summary}

In this section, we have explored how to induce algebraic symmetry actions from categorical ones. Our starting point was a fusion categorical action $\Phi: \mathcal{C} \rightarrow \text{End}(M)$. We showed that one can always construct an algebraic action by the fusion algebra $\text{Fus}(\mathcal{C})$, which possesses a universal C$^*$ completion, $\mathcal{L}(\mathcal{C})$, via its inclusion in the tube algebra $\text{Tub}(\mathcal{C})$. The fusion algebra can be promoted to a bialgebra, but it is not clear that it generally possesses the structure of a WHA.\footnote{Nevertheless, in many circumstances the fusion algebra has sufficient structure to define an entropic order parameter via the approach discussed in the main text. In Section~\ref{sec: examples} we provide an example of such a case in the context of TQFT. The fusion algebra can also generically be promoted to a WHA by passing through the inclusion $\text{Fus}(\mathcal{C}) \subset \text{Str}_{\mathcal{C} \boxtimes \mathcal{C}^{op}}(\mathcal{C})$.}

To obtain a weak Hopf algebraic action associated with the symmetry $\mathcal{C}$ we pass through Tannaka-Krein duality. Given any fiber functor $F: \mathcal{C} \rightarrow \text{Vec}$, the algebra of natural transformations of $F$ to itself will always satisfy the property that $\mathcal{C} = \text{Rep}(\text{End}(F))$, and can be endowed with a weak Hopf structure. As we have described, a large class of fiber functors can be formed by considering the space of semisimple, faithful module categories of $\mathcal{C}$, $\mathcal{M}$, and Frobenius algebras which represent them, $\mathcal{M} = \text{Rep}(B)$. To each (equivalence class of) pair(s) $(\mathcal{M},B)$ we obtain a different fiber functor, and in turn a different WHA $A_{\mathcal{M},B}^{\mathcal{C}}$ representing $\mathcal{C}$. This underscores the non-uniqueness of Tannaka-Krein duality for general fusion categories. For a fixed module category $\mathcal{M}$, a natural choice of Frobenius algebra presents itself as $B_{\mathcal{M}} \equiv \mathbb{C}^{\oplus |\text{Irr}(\mathcal{M})|}$, in which case $A_{\mathcal{M},B_{\mathcal{M}}}^{\mathcal{C}} = \text{Str}_{\mathcal{C}}(\mathcal{M})^*$. 

By recognizing the fusion algebra as a subalgebra of the strip algebra, we can induce the representation $\rho: \text{Fus}(\mathcal{C}) \rightarrow \text{End}(M)$ to a representation $\rho^{\mathcal{M}}: \text{Str}_{\mathcal{C}}(\mathcal{M}) \rightarrow \Endvect(M_{\mathcal{M}})$.\footnote{Of course, more general $\text{Fus}(\mathcal{C})$-rigged $\text{Str}_{\mathcal{C}}(\mathcal{M})$-modules may be used if necessary.} Here, $M_{\mathcal{M}}$ is the system algebra which is obtained by appending to the original system $M$ the generators of the module category. More specifically $M_{\mathcal{M}} = (\text{Str}_{\mathcal{C}}(\mathcal{M}) \otimes M)/\text{Fus}(\mathcal{C})$, where the quotient is under the equivalence relation
\beq \label{Algebraic Quotient}
	a \otimes \rho_f(a) \sim r_f(a) \otimes x, \qquad \forall a \in \text{Str}_{\mathcal{C}}(\mathcal{M}), f \in \text{Fus}(\mathcal{C}), x \in M\,.
\eeq 
We conclude that a fusion categorical symmetry $\Phi: \mathcal{C} \rightarrow \text{End}(M)$ gives rise to a weak Hopf algebraic symmetry $\rho^{\mathcal{M}}: \text{Str}_{\mathcal{C}}(\mathcal{M}) \rightarrow \Endvect(M_{\mathcal{M}})$ such that $\mathcal{C} = \text{Rep}(\text{Str}_{\mathcal{C}}(\mathcal{M})^*)$. This action is uniquely defined up to the choice of module category $\mathcal{M}$ and the $\text{Fus}(\mathcal{C})$-rigged structure on $L^2(\text{Str}_{\mathcal{C}}(\mathcal{M}))$. 
\section{Order Parameters for Non-Invertible Symmetries} \label{sec: order parameters}

As we have discussed in Section~\ref{sec: Prelims}, the relative entropy between a density operator $\rho$ and its invariantization under a group $G$ serves as an order parameter for detecting the breaking of the grouplike symmetry.  In this section, we will construct this entropic order parameter explicitly in the context of non-invertible symmetries encoded in finite WHAs $H$. While the definition of the entropic order parameter is the same in both cases~\cite{Casini:2020rgj,Benedetti:2024dku}, namely as a relative entropy between a state and its symmetrization, the non-invertible nature of the symmetry $H$ leads to interesting differences with the group case.  Thanks to the algorithm presented in Section~\ref{sec: algebras from categories}, this can equivalently be interpreted as a strategy for constructing order parameters probing quantum information theoretic properties of states under non-invertible categorical symmetries, provided they are encoded in a fusion category.

%In this section, we will 
%outline a broad generalization of this construction which allows for the definition of an
%order parameter detecting the breaking of non-invertible symmetries encoded in finite, WHAs $H$.

\subsection{The Analog of Group Averaging for Weak Hopf Algebras}
\label{sec: WHAaverage}

The starting point of this subsection is a WHA $(H,\mu,\eta,\Gamma,\epsilon,S)$ along with a left $H$-module algebra $A$. For our purposes, $A$ is an $H$-module algebra if it admits an action by $H$, $\rho: H \rightarrow \Endvect(A)$.\footnote{More precisely, $\rho$ must satisfy additional compatibility conditions relative to the weak Hopf structure of $H$ and the involution on $A$. For details, see \cite{nill1998weakhopfalgebrasreducible,nikshych2000finitequantumgroupoidsapplications}.}  

Relative to the discussion of Section~\ref{sec: algebras from categories}, one may interpret $\rho: H \rightarrow \Endvect(A)$ as the algebraic non-invertible symmetry action induced from a categorical symmetry action $\Phi: \mathcal{C} = \text{Rep}(H^*) \rightarrow \text{End}(M)$. The WHA $H$ and the system algebra $A$ are induced from the data of $\Phi$ along with the choice of a module category $\mathcal{M}$. That is, one may take $H = \text{Str}_{\mathcal{C}}(\mathcal{M})$ and $A = M_{\mathcal{M}}$. However, the analysis of this section is sufficiently general to apply to any WHA $H$ with left module algebra $A$. For example, in many instances one may wish to take $H = \text{Fus}(\mathcal{C})$ and $A = M$ to consider the action of the fusion algebra (when it is weak Hopf) on the non-extended system algebra $M$. The non-uniqueness of the pair $(H,A)$ in relation to the original symmetry $\Phi$ is a significant feature of categorical symmetry which differentiates it from the invertible case, and we will pay special attention to this detail in the sequel. 

In any finite WHA $H$, there exists an element $\ell \in H$ satisfying \cite{nill1998weakhopfalgebrasreducible,nikshych2000finitequantumgroupoidsapplications,etingof2017fusioncategories}
\beq \label{Left Haar Integral}
	h \ell = \epsilon_t(h) \ell , \qquad \forall h \in H\,. 
\eeq
Here, $\epsilon_t: H \rightarrow H$ is the target counital map defined in Appendix~\ref{app: WHA}. If $H$ is a genuine Hopf algebra, as opposed to a WHA, the target and source counits agree and coincide with the overall counit $\epsilon$. The element $\ell$ is called a \emph{left integral} of $H$. Finite WHAs also admit a notion of right integral
\beq
    r h = r \epsilon_s(h), \qquad \forall h \in H\,,
\eeq
where $\epsilon_s: H \rightarrow H$ is the source counital map. In fact, any finite, biconnected, semisimple WHA can be shown to admit a unique element $\Lambda \in H$ which is a left and right integral and is normalized in the sense that $\epsilon_t(\Lambda) = \mathbb{I}_H = \epsilon_s(\Lambda)$. In Appendix~\ref{app: WHA} we provide a discussion of integral theory for WHAs including expressions for the canonical left integral of any finite $H$, the unique Haar integral for C$^*$ WHAs, and (in Appendix~\ref{app: strip}) the unique Haar integral for general strip algebras. 

The usefulness of the left integral is that it provides an approach to symmetrizing elements of an algebra $A$ which is acted upon by $H$. Firstly, let us define by
\beq \label{Symmetric Subalgebra}
	A_{\rho} \equiv \{x \in A \; | \; \rho_h(x) = \rho_{\epsilon_t(h)}(x), \; \forall h \in H\}\,,
\eeq
the set of elements in $A$ which are invariant under the action of $H$ up to an action by elements in the image of $\epsilon_t$.\footnote{The image of $\epsilon_t$ is called the target base algebra and has an interpretation as defining generalized scalars for $H$; this is the sense in which $A_{\rho}$ may be regarded as a symmetric sector. In the case that $H$ is a Hopf algebra, $\epsilon_t = \epsilon$, and $A_{\rho}$ consists of all elements in $A$ which are multiplied by genuine scalars under $H$ action. The Haar integral can be chosen so that this constant is one, and thus $A_{\rho}$ is precisely the subalgebra of $A$ which is left invariant under the $H$-action.} Then, we can define the map $T_{\rho}: A \rightarrow A_{\rho}$ given by \cite{nill1998weakhopfalgebrasreducible}
\beq
	T_{\rho}(x) \equiv \rho_{\ell}(x)\,.\label{Trho}
\eeq
To see that $T_{\rho}$ lands in the invariant subalgebra is a simple application of \eqref{Left Haar Integral} and the homomorphism property of $\rho$:
\beq
    \rho_h(T_{\rho}(x)) = \rho_h \rho_\ell(x) = \rho_{h \ell}(x) = \rho_{\epsilon_t(h) \ell}(x) = \rho_{\epsilon_t(h)}(T_{\rho}(x))\,.
\eeq
Provided $T_{\rho}(\mathbb{I}_A) = c \mathbb{I}_A$ for a finite $c$, we can promote $T_{\rho}$ to a linear, unital map
\beq \label{WH Invariantization Map}
	E_{\rho}(x) \equiv \frac{1}{c} T_{\rho}(x)\,.
\eeq
When $A$ is a C$^*$-algebra, the maps $T_{\rho}$ and $E_{\rho}$ are respectively an operator valued weight and a conditional expectation \cite{nill1998weakhopfalgebrasreducible,AliAhmad:2025oli}.

A useful example to keep in mind is the case of a standard group action. Associated with a finite group $G$ we have the Hopf algebra $H = \mathcal{L}(G)$ generated by elements $\ell(g)$ along with the multiplication and comultiplication
\beq
    \mu(\ell(g) \otimes \ell(h)) \equiv \ell(gh), \qquad \delta(\ell(g)) = \ell(g) \otimes \ell(g)\,.
\eeq
The counit is given by $\epsilon(\ell(g)) = 1$ for each $g \in G$. Given a group \emph{automorphism} action $\alpha: G \rightarrow \text{Aut}(A)$, we automatically obtain an \emph{endomorphism} action of $H$ by
\beq
    \rho: H \rightarrow \text{End}(A), \qquad \rho_{\sum_{g \in G} f_g \ell(g)}(x) \equiv \sum_{g \in G} f_g \alpha_g(x)\,.
\eeq

In this case, $H$ admits a bi-integral given by $\Lambda = \sum_{g \in G} \ell(g)$. By direct computation:
\beq
    \ell(g) \Lambda = \ell(g) \sum_{h \in G} \ell(h) = \sum_{h \in G} \ell(gh) = \sum_{h \in G} \ell(h) = \epsilon(\ell(g)) \Lambda, \qquad \forall g \in G\,,
\eeq
and analogously for the right action. The invariantization map is therefore given by
\beq
    T_{\rho}(x) \equiv \rho_{\sum_{g \in G} \ell(g)}(x) = \sum_{g \in G} \alpha_g(x)\,,
\eeq
and can be normalized to the conditional expectation
\beq \label{Invariantization for group}
    E_{\rho}(x) = \frac{1}{|G|} \sum_{g \in G} \alpha_g(x)\,.
\eeq
In the case that $\alpha_g$ is implemented by a unitary representation $U: G \rightarrow U(\mathscr{H})$ so that $\alpha_g(x) = U(g)^{\dagger} x U(g)$, \eqref{Invariantization for group} precisely reproduces the invariantization map discussed in the introduction!\footnote{The density operator is conjugated as $\rho \mapsto \alpha_{g^{-1}}(\rho)$ because it transforms in the Schrodinger sense dual to the transformation of operators. This can be seen e.g. by recalling that the expectation value of an observable $x \in A$ computed in the state defined by density operator $\rho$ is given by $\text{Tr}_{\mathscr{H}}(\rho x)$.}

In the rest of the paper, unless otherwise noted, we will take the integral $\ell$ in \eqref{Trho} to be the unique Haar integral $\Lambda$ for the WHA and the corresponding conditional expectation $E_\rho$ \eqref{WH Invariantization Map} will be referred to as the Haar conditional expection.

\subsection{Entropic Order Parameters} 
A quantum state on the system algebra $A$ is a map $\psi: A \rightarrow \mathbb{C}$, where $x \mapsto \psi(x)$ is interpreted as the expectation value of the operator $x$ in the state $\psi$. For the purposes of the present note, the reader may regard $\psi$ as induced from a density operator $\rho_{\psi} \in A$ such that $\psi(x) = \text{Tr}_{\mathscr{H}}(\rho_{\psi} x)$. The relative entropy between two quantum states $\psi_1, \psi_2$ can then be written
\beq
	S(\psi_1 \mid \psi_2) = \text{Tr}_{\mathscr{H}}\bigg(\rho_{\psi_1} (\log \rho_{\psi_1} - \log \rho_{\psi_2})\bigg)\,.
\eeq
Of course, the relative entropy is well defined even when the density operators are not by appealing to spatial/modular theory. In Appendix~\ref{app: ECR} we review the definition of the relative entropy in terms of this approach which is free of ambiguities pertaining to the choice of trace or density operators.

In light of the preceding analysis, and motivated by the discussion in Section~\ref{sec: Prelims}, we propose the following as a candidate for quantifying symmetry breaking in the categorical context:
\beq
	\Delta_{E_{\rho}} S(\psi) \equiv S(\psi \mid \psi \circ E_{\rho}) \,.
\eeq
When $H = \mathcal{L}(G)$ this exactly reproduces the usual entropic order parameter. For a WHA, $\psi \circ E_{\rho}$ represents a symmetric state provided $\ell \in H$ is a bi-integral. Henceforth, we will assume we have induced $E_{\rho}$ from the Haar integral $\Lambda \in H$, which is both a left and right integral, for the purposes of producing an invariantization map. In this case
\beq
    \psi \circ E_{\rho} \circ \rho_{h}(x) = \frac{1}{c} \psi \circ \rho_{\ell h}(x) = \psi \circ E_{\rho} \circ \rho_{\epsilon_s(h)}(x)\,.
\eeq
The set of states $\omega: A \rightarrow \mathbb{C}$ such that $\omega \circ \rho_h = \omega \circ \rho_{\epsilon_s(h)}$ is the appropriate Schrodinger dual notion of invariantization relative to the Heisenberg picture \eqref{Symmetric Subalgebra}.\footnote{This is a weaker version of the weakly symmetric condition considered for mixed states \cite{Schafer-Nameki:2025fiy}. For an explicit comparison, see \eqref{rhosym} for the dual invariantization map on a density operator and its symmetry property in \eqref{weaksym}.} 

To understand the information theoretic properties of the entropic order parameter it is useful to consider the context of an inclusion of operator algebras $N \subset M$. In our case $N = A_{\rho}$ and $M = A$, but the following discussion applies more generally. Given a conditional expectation, $E: M \rightarrow N$, like \eqref{WH Invariantization Map}, we can compute a quantity called the index, $\text{Ind}(E)$, which morally measures the relative size of the algebras $M$ and $E(M)$.\footnote{See \cite{AliAhmad:2025oli} for a more technical discussion.} In \cite{Longo:2022lod} it was shown that, in the case that $N$ and $M$ are von Neumann factor algebras, the entropic order parameter $\Delta_E S(\psi) \equiv S(\psi \mid \psi \circ E)$ is upper bounded by the index
\beq
    \sup_{\psi} \Delta_E S(\psi) = \log \text{Ind}(E)\,. 
\eeq
The supremum here is taken over normal states on the algebra $M$. 

More explicitly, the index can be related to the entropic order parameter in the following way. Firstly, we can define a dual map relative to the conditional expectation, $E^{-1}: N' \rightarrow M'$, where here $M' \equiv \{\mathcal{O} \in B(H) \; | \; [\mathcal{O},x] = 0, \forall x \in M\}$ is the commutant of the algebra $M$.\footnote{The notation $E^{-1}$ is standard and doesn't refer to the inverse of the map $E$!} The map $E^{-1}$ is defined by appealing to spatial theory \cite{Kosaki1991}, and we provide a complete discussion in Appendix~\ref{app: ECR}. Kosaki demonstrated that the index of $E$ can be computed from the normalization of this map, $\text{Ind}(E) = E^{-1}(\mathbb{I})$ \cite{Kosaki1991}. Although Kosaki's original derivation concerned only the case where $N$ and $M$ are factor algebras, the Kosaki formula for the index was generalized to non-factorial algebras by Teruya in \cite{Teruya1992Index}. In the case that $M$ and $N$ are not assumed to be factors the index is not a scalar but rather an element of the center of $M$, the larger algebra.

The index can also be defined in the setting of general C$^*$-algebras by appealing to the construction of Watatani \cite{Watatani1990}. Given an inclusion of C$^*$-algebras $i: B \hookrightarrow A$, a conditional expectation $E: A \rightarrow B$ is said to be of finite index type if it admits a finite-type quasi-basis. A quasi-basis is a collection $\{u_i,v_i\}_{i \in \mathcal{I}} \subset A$ such that
\beq
    a = \sum_{i \in \mathcal{I}} u_i E(v_i a) = \sum_{i \in \mathcal{I}} E(a u_i) v_i\,, \qquad \forall a \in A\,.
\eeq
It is of finite type if the cardinality of the index set $\mathcal{I}$ is finite. Then, the index of $E$ is the central element
\beq
    \text{Ind}(E) \equiv \sum_{i \in \mathcal{I}} u_i v_i \in Z(A)\,.
\eeq

For factorial inclusions and inclusions of finite dimensional algebras the Watatani index and the Kosaki index coincide. As such we will not distinguish between these two notions of index. While we have not seen a proof in the general case, it may arise from the following observation: The quasi-basis is closely related to the notion of a Pimsner-Popa basis. Given a faithful conditional expectation $E: M \rightarrow N$, we can define an $N$-valued inner product on $M$ by $G_E(x,y) \equiv E(x^* y)$. This inner product can be diagonalized in orthonormal form by employing a generalized version of the Gram-Schmidt procedure. This results in a basis $\{\lambda_i\}_{i \in \mathcal{I}}$ such that $G_E(\lambda_i,\lambda_j) = \delta_{ij} \mathbb{I}_N$. The index is then given by $\text{Ind}(E) = \sum_{i \in \mathcal{I}} \lambda_i \lambda_i^*$. The set $\{\lambda_i,\lambda_i^*\}$ also forms a quasi-basis in the sense of Watatani.

Given a (faithful normal) state $\psi$ on $M$, one can define an associated state $\psi'$ on $N'$ by passing to the vector representative of $\psi$ on the common Hilbert space. The following identity\footnote{These so-called certainty relations were discussed in a physical context in Refs.~\cite{Casini:2020rgj,Magan:2020ake} with an emphasis on finite-dimensional algebras. Similar results hold in the infinite-dimensional case~\cite{Longo:2022lod}. In Appendix~\ref{app: ECR} we provide a derivation of the entropic certainty relation for general inclusions and discuss its relation to non-invertible symmetry breaking.} holds for any state on the bigger algebra~\cite{Xu:2018uxc, Longo:2022lod}
\begin{equation} \label{ECR}
    S_{M} (\psi \vert \psi \circ E) + S_{N'} (\psi' \vert \psi' \circ E') = \psi\bigg(\log \bigg(\text{Ind}(E)\bigg)\bigg)\,.
\end{equation}
This can be written in terms of the generalized entropic order parameter we defined as
\begin{equation}
    \Delta_E S(\psi) + \Delta_{E'} S(\psi') = \psi\bigg(\log \bigg(\text{Ind}(E)\bigg)\bigg)\,. \label{certainty}
\end{equation}

Since the order parameter is defined in terms of a relative entropy, which is non-negative, the above analysis indicates that it is bounded by zero and the expectation value of the index:
\beq \label{Bound on EOP}
    0 \leq \Delta_E S(\psi) \leq \psi\bigg(\log \bigg(\text{Ind}(E)\bigg)\bigg)\,.
\eeq 
Returning to the case of interest, the index of the conditional expectation $E_{\rho}$ is bounded 
by the quantity \cite{bohm1999weakhopfalgebrasii}
\beq \label{Upper Bound WHA}
    \mu(H) \equiv \text{dim}(H_t) |\text{Rep}(H) |\,,
\eeq
Here, $\text{dim}(H_t)$ is the complex vector space dimension of the target base algebra $H_t \equiv \epsilon_t(H)$ and
\beq
    |\text{Rep}(H)| \equiv \sum_{U \in \text{Irr}(\text{Rep}(H))} d_U^2\,,
\eeq
is the total quantum dimension of the category $\text{Rep}(H)$.\footnote{Each $d_U$ is the quantum dimension of the simple object $U$.} The quantity $\mu(H)$ coincides with the Watatani index of the conditional expectation, $\text{Ind}(E_{\rho})$ -- a fact which we derive in detail in Appendix~\ref{app: strip} for the case $H = \text{Str}_{\mathcal{C}}(\mathcal{M})$. The order parameter for weak Hopf algebraic symmetry breaking is bounded as
\beq \label{Bound for WHA}
    0 \leq \Delta_{E_{\rho}} S(\psi) \leq \log \mu(H) = \log \text{dim}(H_t) + \log |\text{Rep}(H)|\,.
\eeq
Due to the appearance of the quantity $\text{dim}(H_t)$, the upper bound on the index \eqref{Upper Bound WHA} is not a Morita invariant. That is, given a pair of WHAs $H_1,H_2$ with equivalent representation categories, $\text{Rep}(H_1) \simeq \text{Rep}(H_2)$, it will not necessarily be true that $\mu(H_1) = \mu(H_2)$. 

In the event that $H$ is a Hopf algebra the target base algebra coincides with the set of complex numbers and we find
\beq
    \mu(H) = |\text{Rep}(H)|\,,
\eeq
This quantity \emph{is} a Morita invariant. For an invertible symmetry encoded in a finite group $G$ we recover the standard result
\beq
    0 \leq \Delta_{E_{\rho}} S(\psi) \leq \log |G|\,.
\eeq

We note that the states whose associated order parameter saturate the upper bound maximally break the symmetry. Using the certainty relation \eqref{certainty} and the fact that relative entropies vanish if and only if the two states being compared are equal, we remark that the problem of saturation for the inclusion $N \subset M$ under $E$ is equivalent to the problem of invariantization for the inclusion $M' \subset N'$ under $E'$. A formal discussion of entropic order parameters, certainty relations, and maximal symmetry breaking states can be found in \ref{app: ECR}. 

\subsection{What is Broken, Category or Algebra?}

As we have discussed in Section~\ref{sec: algebras from categories}, the WHA $H$ serves as an algebraic avatar for a fusion categorical symmetry $\mathcal{C}$. However, as we have also emphasized, the duality between the category $\mathcal{C}$ and the algebra $H$ is not unique. This bears upon the following question: Given a categorical symmetry $\Phi: \mathcal{C} \rightarrow \text{End}(M)$, what is the bound on the associated order parameter? 

Actually, this question is somewhat ill-posed. Due to the aforementioned ambiguity, there is not a single unique order parameter that can be associated with the categorical symmetry. Rather, from the point of view of the present note, we get a different extended system, $M_{\mathcal{M}}$, and weak Hopf algebraic symmetry, $\text{Str}_{\mathcal{C}}(\mathcal{M})$, for each choice of module category. In the field theoretic context, the choice of module category is determined by physical considerations and thus encodes significant information. According to \eqref{Bound for WHA}, this information is quantified in the resulting order parameter. For a categorical symmetry in which the relevant module category is $\mathcal{M}$, the entropic order parameter is bounded as
\beq \label{Bound for Fusion Symmetry}
    0 \leq \Delta_{\rho{\mathcal{C},\mathcal{M}}}(\psi) \leq \log \bigg(\mu(\text{Str}_{\mathcal{C}}(\mathcal{M}))\bigg) = \log |\text{Irr}(\mathcal{M})| + \log |\mathcal{C}|\,.
\eeq
Here, we have used the notation $\rho_{\mathcal{C},\mathcal{M}}: \text{Str}_{\mathcal{C}}(\mathcal{M}) \rightarrow \Endvect(M_{\mathcal{M}})$ to emphasize that the action, and by consequence the conditional expectation $E_{\rho_{\mathcal{C},\mathcal{M}}}$, depends upon both the original symmetry $\mathcal{C}$ and the module category $\mathcal{M}$. 

This result may be somewhat surprising in light of the analogous analysis in the case of a grouplike symmetry. There, the order of the group $|G|$ coincides with the total quantum dimension of $\text{Vec}_G$. Based on this observation, one might have conjectured that the entropic order parameter for a categorical symmetry is upper bounded by $\log |\mathcal{C}|$, which would have depended only upon the category under analysis. Indeed, this is the case for Hopf algebraic symmetry actions. However, we have seen that fusion categorical symmetries generically give rise to dual algebraic symmetries encoded in WHAs. The associated bound on the entropic order parameter is dependent upon the choice of algebraic representative in this case. 

From \eqref{Upper Bound WHA}, we see that the non-uniqueness of the index is encoded through the dimension of the base algebra $H_t$. The specification of $H_t$ coincides precisely with the selection of a module category $\mathcal{M}$ in the implementation of Tannaka-Krein duality described in Section~\ref{sec: algebras from categories}.\footnote{For the strip algebra, $H_t = B_{\mathcal{M}} = \mathbb{C}^{\oplus |\text{Irr}(\mathcal{M})|}$.} This ambiguity is present even for a fusion categorical symmetry encoded in $\text{Vec}_G$. One finds that the entropic order parameter is bounded by $\log |G|$ when the categorical symmetry is translated to an algebraic symmetry by using the trivial module category $\mathcal{M} = \text{Vec}$. For a general module category, however, this will not be the case. For example, given a $\text{Vec}_G$ symmetry acting on a system in which the natural module category is $\mathcal{M} = \text{Vec}_G$ we see that the order parameter is bounded as
\beq
    0 \leq \Delta_{\rho_{\text{Vec}_G,\text{Vec}_G}}(\psi) \leq \log \bigg(\mu(\text{Str}_{\text{Vec}_G}(\text{Vec}_G))\bigg) = \log |G|^2\,.
\eeq

The ambiguity presented here underscores the exotic structure of non-invertible symmetries relative to their invertible counterpart. For example, in the invertible case one can imagine that the chosen system $M$ does not fully preserve the group symmetry $G$, but only a subgroup $K$. In this case, the entropic order parameter is the log of $\frac{\vert G \vert}{\vert K \vert }$~\cite{Casini:2020rgj}. Similarly to the group case, the choice of $(M,\gamma)$ may break the $H$ algebraic symmetry down to a subalgebra $K$. The non-uniqueness of the duality between $\mathcal{C}$ and $H$ implies an ambiguity in the symmetry breaking pattern. For instance, even if $K\subset H$ is an algebraic inclusion, their representation categories may not lie in categorical inclusion. This complicates the translation between the algebraic symmetry breaking and the breaking of the original categorical symmetry. In future work we hope to better understand the interplay between categorical and algebraic non-invertible symmetries and especially their quantum information theoretic properties.
\section{Examples} \label{sec: examples}

In this Section, we discuss a sequence of examples  with non-invertible symmetries of increasing level of sophistication,  study the corresponding entropic order parameters, and compare with the general entropic bounds derived in \eqref{Bound for WHA} and \eqref{Bound for Fusion Symmetry}.

\subsection{Toy Model with Fibonacci Symmetry} \label{sec: Fib Cat}

We start by considering a finite quantum mechanical system with Fibonacci symmetry, which could be thought of as a WHA qudit \cite{Chang_2014,Jia:2023xar}. To realize  a weak Hopf algebraic symmetry, we choose the unique indecomposable left module category $\cM=\text{Fib}$, leading to the strip algebra
\ie 
H^*={\rm Str}_{\rm Fib}({\rm Fib})\,.
\fe
In other words, the system algebra for the qudit is $H$ and the symmetry algebra is $H^*$ which acts on $H$ via the canonical action \cite{BOHM1999385,bohm1999weakhopfalgebrasii} (see Section~\ref{sec: WHAaverage} for general discussion)
\ie 
\rho_f(x)=f(x_{(1)})x_{(2)}\,,\quad \forall f\in H^*\,,~x\in H\,,
\label{canonicalHHdualaction}
\fe
where we have used the Sweedler notation of the coproduct.
The qudit Hilbert space for $H$ is the usual GNS Hilbert space (isomorphic to $H$). In the following we identify the index for the conditional expectation $E_\rho$ defined by the unique Haar element $\lambda$ in $H^*$ (equivalently the Haar measure on $H$) 
\ie 
E_\rho(x)=\lambda(x_{(1)})x_{(2)}\,,
\label{HaarCEFib}
\fe
which projects $H$ to its base algebra $H_t$, 
and compare with the relative entropy for states in the system.

Let us first recall some useful information of the Fib WHA $H$. 
As a semisimple algebra, $H$ decomposes into full matrix algebras as
\beq \label{FibHdecomp}
H\cong M_2(\mC) \oplus M_3(\mC)\,.
\eeq
This algebra is generated by a basis of matrix units which we denote by $e_{n}^{ab}$. The index $n=1,2$ corresponds to the two matrix algebras in \eqref{FibHdecomp} and $a,b = 1, ..., N_n$ with $N_1 = 2$ and $N_2 = 3$. The WHA $H$ is self-dual and contains ${\rm Fus}({\rm Fib})$ as a subalgebra
\ie 
L_1 \equiv e_1^{22}+e_2^{22}
\,,\quad L_W \equiv \xi e_1^{22}+(1-\xi) e_2^{22}
\,,
\fe
which reproduces the fusion rules
\ie 
L_W L_W=L_1+L_W\,,\quad L_1 L_W=L_W L_1=L_1\,.
\fe
Here,
\ie 
d_1=1\,,\quad d_W=\xi\,,\quad \xi={\sqrt{5}+1\over 2}\,.
\fe 
are the quantum dimensions of the irreducible objects of $\text{Fib}$ which we have labeled respectively by $1$ and $W$.

In Appendix~\ref{app: strip} we provide a complete account of the most general strip algebra using a diagrammatic calculus. In terms of these strip algebra generators, the matrix units are given explicitly by\footnote{Here the normalization of the topological junction on the boundary is such that the $\Theta$ diagram (see e.g. in \eqref{counit}) with all $W$ defects evaluates to $\xi^2$.}\footnotemark
\ie 
&e_1^{11}=
\begin{gathered}
\begin{tikzpicture}[scale=.3]
\fill [fill=gray!20] (-1.8,-1) rectangle (-1,1);
\fill [fill=gray!20] (0,-1) rectangle (.8,1);
\draw [line,dotted] (-1,-1) -- (-1,1);
\draw [line,dotted] (0,-1) -- (0,1);
\end{tikzpicture}
\end{gathered}\,,
\quad 
e_1^{12}=
\xi^{-1}\begin{gathered}
\begin{tikzpicture}[scale=.3]
\fill [fill=gray!20] (-1.8,-1) rectangle (-1,1);
\fill [fill=gray!20] (0,-1) rectangle (.8,1);
\draw [line,dotted] (-1,0) -- (-1,1);
\draw [line,dotted] (0,0) -- (0,1);
\draw [line] (-1,-1) -- (-1,0) -- (0,0) --(0,-1);
\end{tikzpicture}
\end{gathered}\,,
\quad 
e_1^{21}=
\begin{gathered}
\begin{tikzpicture}[scale=.3]
\fill [fill=gray!20] (-1.8,-1) rectangle (-1,1);
\fill [fill=gray!20] (0,-1) rectangle (.8,1);
\draw [line,dotted] (-1,0) -- (-1,-1);
\draw [line,dotted] (0,0) -- (0,-1);
\draw [line] (-1,1) -- (-1,0) -- (0,0) --(0,1);
\end{tikzpicture}
\end{gathered}\,,
\quad 
e_1^{22}=\xi^{-2}\left(
\begin{gathered}
\begin{tikzpicture}[scale=.3]
\fill [fill=gray!20] (-1.8,-1) rectangle (-1,1);
\fill [fill=gray!20] (0,-1) rectangle (.8,1);
\draw [line ] (-1,1) -- (-1,-1);
\draw [line ] (0,1) -- (0,-1);
\end{tikzpicture}
\end{gathered}
+
\begin{gathered}
\begin{tikzpicture}[scale=.3]
\fill [fill=gray!20] (-1.8,-1) rectangle (-1,1);
\fill [fill=gray!20] (0,-1) rectangle (.8,1);
\draw [line] (-1,0) -- (-1,-1);
\draw [line] (0,0) -- (0,-1);
\draw [line] (-1,1) -- (-1,0) -- (0,0) --(0,1);
\end{tikzpicture}
\end{gathered}\right)\,,
\quad 
\\
&e_2^{11}=
\begin{gathered}
\begin{tikzpicture}[scale=.3]
\fill [fill=gray!20] (-1.8,-1) rectangle (-1,1);
\fill [fill=gray!20] (0,-1) rectangle (.8,1);
\draw [line,dotted] (-1,-1) -- (-1,1);
\draw [line] (0,-1) -- (0,1);
\end{tikzpicture}
\end{gathered}\,,
\quad 
e_2^{33}=
\begin{gathered}
\begin{tikzpicture}[scale=.3]
\fill [fill=gray!20] (-1.8,-1) rectangle (-1,1);
\fill [fill=gray!20] (0,-1) rectangle (.8,1);
\draw [line] (-1,-1) -- (-1,1);
\draw [line,dotted] (0,-1) -- (0,1);
\end{tikzpicture}
\end{gathered}\,,
\quad 
e_2^{22}=\xi^{-1}
\begin{gathered}
\begin{tikzpicture}[scale=.3]
\fill [fill=gray!20] (-1.8,-1) rectangle (-1,1);
\fill [fill=gray!20] (0,-1) rectangle (.8,1);
\draw [line ] (-1,1) -- (-1,-1);
\draw [line ] (0,1) -- (0,-1);
\end{tikzpicture}
\end{gathered}
-\xi^{-2}
\begin{gathered}
\begin{tikzpicture}[scale=.3]
\fill [fill=gray!20] (-1.8,-1) rectangle (-1,1);
\fill [fill=gray!20] (0,-1) rectangle (.8,1);
\draw [line] (-1,0) -- (-1,-1);
\draw [line] (0,0) -- (0,-1);
\draw [line] (-1,1) -- (-1,0) -- (0,0) --(0,1);
\end{tikzpicture}
\end{gathered}\,,
\quad 
\\
& e_2^{12}= 
\begin{gathered}
\begin{tikzpicture}[scale=.3]
\fill [fill=gray!20] (-1.8,-1) rectangle (-1,1);
\fill [fill=gray!20] (0,-1) rectangle (.8,1);
\draw [line,dotted] (-1,1) -- (-1,0);
\draw [line] (0,1) -- (0,-1);
\draw [line] (-1,-1) -- (-1,0) -- (0,0);
\end{tikzpicture}
\end{gathered} \,,
\quad 
e_2^{21}= 
\xi^{-1}\begin{gathered}
\begin{tikzpicture}[scale=.3]
\fill [fill=gray!20] (-1.8,-1) rectangle (-1,1);
\fill [fill=gray!20] (0,-1) rectangle (.8,1);
\draw [line,dotted] (-1,-1) -- (-1,0);
\draw [line] (0,1) -- (0,-1);
\draw [line] (-1,1) -- (-1,0) -- (0,0);
\end{tikzpicture}
\end{gathered}\,,\quad 
e_2^{23}= 
\xi^{-1}\begin{gathered}
\begin{tikzpicture}[scale=.3]
\fill [fill=gray!20] (-1.8,-1) rectangle (-1,1);
\fill [fill=gray!20] (0,-1) rectangle (.8,1);
\draw [line,dotted] (0,-1) -- (0,0);
\draw [line] (-1,1) -- (-1,-1);
\draw [line] (-1,0) -- (0,0) -- (0,1);
\end{tikzpicture}
\end{gathered} \,,
\quad 
e_2^{32}= 
\begin{gathered}
\begin{tikzpicture}[scale=.3]
\fill [fill=gray!20] (-1.8,-1) rectangle (-1,1);
\fill [fill=gray!20] (0,-1) rectangle (.8,1);
\draw [line,dotted] (0,1) -- (0,0);
\draw [line] (-1,1) -- (-1,-1);
\draw [line] (-1,0) -- (0,0) -- (0,-1);
\end{tikzpicture}
\end{gathered} \,,\quad 
e_2^{13}= 
\begin{gathered}
\begin{tikzpicture}[scale=.3]
\fill [fill=gray!20] (-1.8,-1) rectangle (-1,1);
\fill [fill=gray!20] (0,-1) rectangle (.8,1);
\draw [line,dotted] (0,-1) -- (0,0);
\draw [line,dotted] (-1,1) -- (-1,0);
\draw [line] (-1,-1) -- (-1,0) -- (0,0)--(0,1);
\end{tikzpicture}
\end{gathered} \,,\quad 
e_2^{31}= 
\begin{gathered}
\begin{tikzpicture}[scale=.3]
\fill [fill=gray!20] (-1.8,-1) rectangle (-1,1);
\fill [fill=gray!20] (0,-1) rectangle (.8,1);
\draw [line,dotted] (0,1) -- (0,0);
\draw [line,dotted] (-1,-1) -- (-1,0);
\draw [line] (-1,1) -- (-1,0) -- (0,0)--(0,-1);
\end{tikzpicture}
\end{gathered} \,.
\fe 

The same strip algebra also acts on 2d CFT with Fib symmetry in the presence of boundary conditions. Here, roughly speaking, the qudit contains topological information in the sector structure of the BCFT. In that context, one can read the dotted line as the identity Cardy brane while the solid line represents the other simple boundary condition which we refer to as $W$. These correspond to the irreducible objects of the module category, which in this case is equivalent to the symmetry category $\text{Fib}$. We identify the bulk topological defects with the boundary conditions since we are dealing with the regular module category.

\footnotetext{Another useful basis for $H$ is given in terms of Jones projections. Indeed the Fib WHA here corresponds to a special case of the Temperley-Lieb algebras, denoted as $A_{1,3}$ in \cite{nikshych2002finite}, and generated by idempotents $e_{1,2,3}$ together with the identity which satisfy
\ie 
e_i e_{i\pm 1} e_i= \xi^{-2} e_i\,,
\quad 
e_i e_j=e_j e_i~{\rm for}~|i-j|\geq 2\,.
\fe
In terms of the strip algebra basis,
\ie 
&e_1=\begin{gathered}
\begin{tikzpicture}[scale=.3]
\fill [fill=gray!20] (-1.8,-1) rectangle (-1,1);
\fill [fill=gray!20] (0,-1) rectangle (.8,1);
\draw [line,dotted] (-1,-1) -- (-1,1);
\draw [line,dotted] (0,-1) -- (0,1);
\end{tikzpicture}
\end{gathered}+ \begin{gathered}
\begin{tikzpicture}[scale=.3]
\fill [fill=gray!20] (-1.8,-1) rectangle (-1,1);
\fill [fill=gray!20] (0,-1) rectangle (.8,1);
\draw [line,dotted] (-1,-1) -- (-1,1);
\draw [line] (0,-1) -- (0,1);
\end{tikzpicture}
\end{gathered}\,,
\quad 
e_3=\begin{gathered}
\begin{tikzpicture}[scale=.3]
\fill [fill=gray!20] (-1.8,-1) rectangle (-1,1);
\fill [fill=gray!20] (0,-1) rectangle (.8,1);
\draw [line,dotted] (-1,-1) -- (-1,1);
\draw [line,dotted] (0,-1) -- (0,1);
\end{tikzpicture}
\end{gathered}+
\begin{gathered}
\begin{tikzpicture}[scale=.3]
\fill [fill=gray!20] (-1.8,-1) rectangle (-1,1);
\fill [fill=gray!20] (0,-1) rectangle (.8,1);
\draw [line] (-1,-1) -- (-1,1);
\draw [line,dotted] (0,-1) -- (0,1);
\end{tikzpicture}
\end{gathered}\,,
\\
&
e_2=\xi^{-2}\left(
\begin{gathered}
\begin{tikzpicture}[scale=.3]
\fill [fill=gray!20] (-1.8,-1) rectangle (-1,1);
\fill [fill=gray!20] (0,-1) rectangle (.8,1);
\draw [line,dotted] (-1,-1) -- (-1,1);
\draw [line,dotted] (0,-1) -- (0,1);
\end{tikzpicture}
\end{gathered}
+\begin{gathered}
\begin{tikzpicture}[scale=.3]
\fill [fill=gray!20] (-1.8,-1) rectangle (-1,1);
\fill [fill=gray!20] (0,-1) rectangle (.8,1);
\draw [line,dotted] (-1,-1) -- (-1,1);
\draw [line] (0,-1) -- (0,1);
\end{tikzpicture}
\end{gathered}
+
\begin{gathered}
\begin{tikzpicture}[scale=.3]
\fill [fill=gray!20] (-1.8,-1) rectangle (-1,1);
\fill [fill=gray!20] (0,-1) rectangle (.8,1);
\draw [line,dotted] (-1,-1) -- (-1,1);
\draw [line] (0,-1) -- (0,1);
\end{tikzpicture}
\end{gathered}
+
\begin{gathered}
\begin{tikzpicture}[scale=.3]
\fill [fill=gray!20] (-1.8,-1) rectangle (-1,1);
\fill [fill=gray!20] (0,-1) rectangle (.8,1);
\draw [line] (-1,-1) -- (-1,1);
\draw [line,dotted] (0,-1) -- (0,1);
\end{tikzpicture}
\end{gathered}
+
\begin{gathered}
\begin{tikzpicture}[scale=.3]
\fill [fill=gray!20] (-1.8,-1) rectangle (-1,1);
\fill [fill=gray!20] (0,-1) rectangle (.8,1);
\draw [line ] (-1,1) -- (-1,-1);
\draw [line ] (0,1) -- (0,-1);
\end{tikzpicture}
\end{gathered}
+
\xi^{-2}\begin{gathered}
\begin{tikzpicture}[scale=.3]
\fill [fill=gray!20] (-1.8,-1) rectangle (-1,1);
\fill [fill=gray!20] (0,-1) rectangle (.8,1);
\draw [line] (-1,0) -- (-1,-1);
\draw [line] (0,0) -- (0,-1);
\draw [line] (-1,1) -- (-1,0) -- (0,0) --(0,1);
\end{tikzpicture}
\end{gathered}
+
\begin{gathered}
\begin{tikzpicture}[scale=.3]
\fill [fill=gray!20] (-1.8,-1) rectangle (-1,1);
\fill [fill=gray!20] (0,-1) rectangle (.8,1);
\draw [line,dotted] (-1,0) -- (-1,1);
\draw [line,dotted] (0,0) -- (0,1);
\draw [line] (-1,-1) -- (-1,0) -- (0,0) --(0,-1);
\end{tikzpicture}
\end{gathered}
+
\begin{gathered}
\begin{tikzpicture}[scale=.3]
\fill [fill=gray!20] (-1.8,-1) rectangle (-1,1);
\fill [fill=gray!20] (0,-1) rectangle (.8,1);
\draw [line,dotted] (-1,0) -- (-1,-1);
\draw [line,dotted] (0,0) -- (0,-1);
\draw [line] (-1,1) -- (-1,0) -- (0,0) --(0,1);
\end{tikzpicture}
\end{gathered}
+
\begin{gathered}
\begin{tikzpicture}[scale=.3]
\fill [fill=gray!20] (-1.8,-1) rectangle (-1,1);
\fill [fill=gray!20] (0,-1) rectangle (.8,1);
\draw [line,dotted] (0,-1) -- (0,0);
\draw [line,dotted] (-1,1) -- (-1,0);
\draw [line] (-1,-1) -- (-1,0) -- (0,0)--(0,1);
\end{tikzpicture}
\end{gathered} 
+
\begin{gathered}
\begin{tikzpicture}[scale=.3]
\fill [fill=gray!20] (-1.8,-1) rectangle (-1,1);
\fill [fill=gray!20] (0,-1) rectangle (.8,1);
\draw [line,dotted] (0,1) -- (0,0);
\draw [line,dotted] (-1,-1) -- (-1,0);
\draw [line] (-1,1) -- (-1,0) -- (0,0)--(0,-1);
\end{tikzpicture}
\end{gathered}
\right)
\\
&+\xi^{-3}
\left(
\begin{gathered}
\begin{tikzpicture}[scale=.3]
\fill [fill=gray!20] (-1.8,-1) rectangle (-1,1);
\fill [fill=gray!20] (0,-1) rectangle (.8,1);
\draw [line,dotted] (-1,1) -- (-1,0);
\draw [line] (0,1) -- (0,-1);
\draw [line] (-1,-1) -- (-1,0) -- (0,0);
\end{tikzpicture}
\end{gathered} +
\begin{gathered}
\begin{tikzpicture}[scale=.3]
\fill [fill=gray!20] (-1.8,-1) rectangle (-1,1);
\fill [fill=gray!20] (0,-1) rectangle (.8,1);
\draw [line,dotted] (-1,-1) -- (-1,0);
\draw [line] (0,1) -- (0,-1);
\draw [line] (-1,1) -- (-1,0) -- (0,0);
\end{tikzpicture}
\end{gathered}
+
\begin{gathered}
\begin{tikzpicture}[scale=.3]
\fill [fill=gray!20] (-1.8,-1) rectangle (-1,1);
\fill [fill=gray!20] (0,-1) rectangle (.8,1);
\draw [line,dotted] (0,-1) -- (0,0);
\draw [line] (-1,1) -- (-1,-1);
\draw [line] (-1,0) -- (0,0) -- (0,1);
\end{tikzpicture}
\end{gathered}
+
\begin{gathered}
\begin{tikzpicture}[scale=.3]
\fill [fill=gray!20] (-1.8,-1) rectangle (-1,1);
\fill [fill=gray!20] (0,-1) rectangle (.8,1);
\draw [line,dotted] (0,1) -- (0,0);
\draw [line] (-1,1) -- (-1,-1);
\draw [line] (-1,0) -- (0,0) -- (0,-1);
\end{tikzpicture}
\end{gathered} 
\right)\,.
\fe
In terms of the matrix unit basis,
\ie 
&e_1=e_1^{11}+e_2^{11}\,,\quad e_3=e_1^{11}+e_2^{33}\,,\quad 
e_2=\xi^{-2}  \sum_{n,i,j} e_n^{ij} \xi^{(-1)^{n+1}
\D_{j,2}}\,.
\fe}

%Using the observation that $\text{Fus(Fib)}$ is an included subalgebra of $H^*\cong H$ we can pass through Rieffel induction to obtain an algebraic non-invertible symmetry action of $H^*$ on an extended system which in this case is isomorphic to $H$ itself. This action $\rho: H^* \rightarrow \text{End}(H)$ is given in \eqref{canonicalHHdualaction}. 

The symmetric subalgebra under $\rho$ is $H_t = \mathbb{C}^2$ which is spanned by $q_1 = e_1^{11}+ e_2^{11}$ and $q_2 = e_1^{22} + e_2^{22}+ e_2^{33}$. In the matrix unit basis, the Haar conditional expectation $E_\rho: H \rightarrow H_t$ from \eqref{HaarCEFib} is given by
\ie 
	E(e_1^{11}) = \frac{1}{2} q_1\,,~ E(e_2^{11}) = \frac{1}{2} q_1\,,~
	E(e_1^{22}) = \frac{1}{2\xi^2} q_2\,,~E(e_2^{22}) = \frac{1}{2\xi} q_2\,,~ E(e_2^{33}) = \frac{1}{2} q_2\,,
\fe 
and vanishes elsewhere.

% By self duality it is equivalent to the canonical action of $H^* \simeq H$ on $H$ which is inherited from multiplication and the pairing of dual elements. The Haar element of $H^* \simeq H$ is given by 
% \ie 
% \Lambda={1\over 2}\sum_{i,j} e_1^{ij}\,,
% \fe
% and the associated Haar conditional expectation for this non-invertible symmetry is $E_{\rho}(h) \equiv \rho_{\Lambda}(h)$. 

According to \eqref{Bound on EOP}, the entropic order parameter for this non-invertible symmetry is naively bounded by
\beq
    S(\psi \mid \psi \circ E_{\rho}) \leq \log\bigg( 2(1 + \xi^2) \bigg).
\eeq
The quantity $2(1+\xi^2)$ is the index of the conditional expectation $E_{\rho}$ which can be read as the dimension of the target counital base algebra ($\text{dim}(H_t) = \text{dim}(\mathbb{C}^2) = 2$) multiplying the total quantum dimension of the Fibonacci category ($|\text{Fib}| = d_1^2 + d_W^2 = 1 + \xi^2$). In Appendix~\ref{app: strip} we provide a computation of this index for general strip algebras using Watatani's quasi-basis approach \cite{Watatani1990}. For completeness, we will now demonstrate how to compute this index following the Kosaki approach, since this is directly related to how the entropic certainty relation (and thus the inequality \eqref{Bound on EOP} and \eqref{Bound for WHA}) is derived (see Appendix~\ref{app: ECR}). 

In \cite{Teruya1992Index}, Kosaki's formula for the index in terms of the normalization of a dual operator valued weight is generalized to non-factorial inclusions. If $E: M \rightarrow N$ is a conditional expectation between von Neumann algebras with finite dimensional centers and $\{p_i\}_{i = 1}^n$ and $\{q_j\}_{j = 1}^m$ are complete sets of central projections for $M$ and $N$ respectively, the Kosaki index takes the following form
\beq \label{Teruya}
	\text{Ind}(E) \equiv E^{-1}(\mathbb{I}) = \sum_{i = 1}^{n} \bigg( \sum_{j = 1}^m \lambda_{ij}^{-1} \text{Ind}(E_{ij}) \bigg) p_i. 
\eeq
The quantities $\lambda_{ij}$ are determined by evaluating 
\beq
	E(p_i q_j) = \lambda_{ij} q_j\,.
    \label{lambdaij}
\eeq	
The maps $E_{ij}: M_{ij} \rightarrow N_{ij}$ are conditional expectations between factor algebras $M_{ij} \equiv p_i q_j M p_i q_j$ and $N_{ij} \equiv p_i q_j N p_i q_j$ defined by
\beq
	E_{ij}(x) \equiv \lambda_{ij}^{-1} E(x) p_i q_j. 
\eeq	

Returning to our example, we see that neither $H$ nor $H_t$ are factorial. A set of central projectors for $H$ are given by $p_1 = e_1^{11} + e_{1}^{22}$ and $p_2 = e_{2}^{11} + e_{2}^{22} + e_{2}^{33}$. The operators $q_1$ and $q_2$ are central projectors for $H_t$. It is easy to verify that $p_1 + p_2 = \mathbb{I}$ and $q_1 + q_2 = \mathbb{I}$. The products of these central projectors are all nonzero and are given by
\ie 
	&p_1 q_1 = e_1^{11}\,,~ p_1 q_2 = e_1^{22}\,,~p_2 q_1 = e_2^{11}\,,~ p_2 q_2 = e_2^{22} + e_2^{33}\,.
\fe 
The resulting factorial inclusions in this case are given by 
\ie 
M_{ij} \supset N_{ij}\,,\quad M_{ij}\equiv p_i q_j M p_i q_j\,,~N_{ij}\equiv p_i q_j N p_i q_j\,.
\fe
% \begin{flalign}
% 	&M_{11} = \text{span}\{e_{1}^{11}\} \supset N_{11} = \text{span}\{e_1^{11}\}, \nonumber \\
% 	&M_{12} = \text{span}\{e_1^{22}\} \supset N_{12} = \text{span}\{e_1^{22}\}, \nonumber \\
% 	&M_{21} = \text{span}\{e_2^{11}\} \supset N_{21} = \text{span}\{e_2^{11}\}, \nonumber \\
% 	&M_{22} = \text{span}\{e_2^{22}, e_2^{23}, e_2^{32}, e_2^{33}\} \supset N_{22} = \text{span}\{e_2^{22} + e_2^{33}\}.
% \end{flalign}
It is straightforward to compute from the definition of $E=E_\rho$ here that
\beq
	E(p_1 q_1) = \frac{1}{2} q_1\,,~ E(p_1 q_2) = \frac{1}{2 \xi^2} q_2\,,~E(p_2 q_1) = \frac{1}{2} q_1\,,~E(p_2 q_2) = {\xi\over 2} q_2\,.
\eeq
which determines by \eqref{lambdaij}
\beq	
	\lambda_{ij} = {1\over 2}\begin{pmatrix}
1& \xi^{-2} \\
1& \xi
\end{pmatrix}\,.
\eeq
The subconditional expectations are then given by
\ie 
	&E_{11}(e_1^{11}) = e_1^{11}\,,~
	E_{12}(e_1^{22}) = e_1^{22}\,,~
	E_{21}(e_2^{11}) = e_2^{11}\,,~
    \\
	&E_{22}(e_2^{22}) = \xi^{-2} (e_2^{22} + e_3^{22})\,,~ E_{22}(e_2^{33}) = \xi^{-1}(e_2^{22} + e_3^{22})\,,
\fe 
and vanish otherwise. The subconditional expectations have trivial indices $\text{Ind}(E_{ij})=1$ clearly except for $E_{22}$ which has index $\xi^3$. To show this, we first compute the operator valued inner product $G_{E_{22}}(x,y) \equiv E_{22}(x^*y)$ which has matrix elements
\beq
	G_{E_{22}} = 
    \xi^{-2}\begin{pmatrix}
        1 & 0 \\0 & \xi 
    \end{pmatrix}\otimes \begin{pmatrix}
        1 & 0 \\0 & 1
    \end{pmatrix}
 (e_2^{22} + e_2^{33})\,.
\eeq
This inner product possesses an orthonormal basis (also referred to as a Pimsner-Popa basis in this context):
\beq
	\bigg\{ \xi  e_{2}^{22}\,, \sqrt{\xi} e_{2}^{23}\,,\xi e_{2}^{32}\,, \sqrt{\xi} e_2^{33} \bigg\}\,.
\eeq 
The index of this subconditional expectation is obtained by taking the sum of $\lambda_i \lambda_i^*$ for $\lambda_i$ in this basis. By \eqref{Teruya}, the total index of the conditional expectation $E_{\rho}$ is given by
\beq
	\text{Ind}(E_{\rho}) = 2(1+\xi^2)(p_1 + p_2) = 2(1 + \xi^2) \mathbb{I}, 
\eeq
which coincides with the Watatani approach \cite{Watatani1990}. 

By an elementary direct calculation it can be shown that the state which maximizes the entropic order parameter for $H$ is given by
\beq
    \psi(h) \equiv \tau(e_{1}^{22} h), 
\eeq
with $\tau(e_n^{ab}) = \delta^{ab}$ a tracial state on the algebra. The relative entropy of this state can be computed as
\beq \label{Relative Entropy Max for H13}
    S(\psi \mid \psi \circ E_{\rho}) = \tau(e_{1}^{22} \bigg( \log(e_{1}^{22}) - \log(E^*_{\rho}(e_1^{22}))\bigg) = \log(2\xi^2),
\eeq
where $E^*_{\rho}$ is the dual of the operator valued weight which acts upon density operators and satisfies $\tau(E^*_{\rho}(x)y) = \tau(x E(y))$.\footnote{Explicitly this map is given by
\begin{flalign}
    &E^*_{\rho}(e_n^{ab}) = 0 \; a \neq b\,, \nonumber \\
    &E^*_{\rho}(e_1^{11}) = E_{\rho}^*(e_2^{11}) = \frac{1}{2}(e_1^{11} + e_2^{11})\,, \nonumber \\
    &E^*_{\rho}(e_1^{22}) = E^*_{\rho}(e_2^{22}) = E^*_{\rho}(e_2^{33}) = \frac{1}{2}\bigg(\frac{1}{\xi^2} e_1^{22} + \frac{1}{\xi} e_2^{22} + e_2^{33}\bigg)\,.
\end{flalign}
}
The relative entropy of this state is consistent with, but fails to saturate the upper bound set by the log of the index! 

This is, in fact, an expected result. The entropic order parameter is generically bounded by the log of the Kosaki index as written in \eqref{Bound on EOP}, but this bound can be strengthened in the finite dimensional setting to
\beq
    S(\psi \mid \psi \circ E) \leq \log(\text{Ind}_{\rm PP}(E))\,.
\eeq
Here, $\text{Ind}_{\rm PP}(E)$ is the Pimsner-Popa index of the conditional expectation $E$ which is the reciprocal of the smallest constant $\lambda$ such that \cite{Pimsner1986}
\beq
    E(x) \geq \lambda x\,, \qquad \forall x > 0\,.
\eeq
It is not difficult to see that $\lambda$ is equal to the minimum of $\lambda_{ij}$ over its indices which in this case is $1/2\xi^2$. Thus, for the Haar conditional expectation associated with $H$ we have $\text{Ind}_{\rm  PP}(E_{\rho}) = 2\xi^2$ agreeing with the relative entropy \eqref{Relative Entropy Max for H13}.

\subsection{Two-dimensional TQFT} \label{sec: TQFT Fus}

Going beyond the qudit example, let us now study examples of topological quantum field theories (TQFTs). Such TQFTs generally describe gapped phases of QFTs. For simplicity, we will focus on $d=2$ TQFTs with fusion category symmetry $\cal C$ generated by simple topological defects $L_i$ with the convention that $L_1=\id$ is the identity defect. Here we will also be slightly general and consider conditional expectations defined by fusion algebras which are not necessarily WHAs, which provides hints for future investigation of such more general conditional expectations (see also discussions in Section~\ref{sec: discussion}).

% To appreciate the full generality of non-invertible symmetry let us now consider extending the system by including boundary conditions. In the simplest setting, this boils down to putting the TQFT on a strip. For definiteness we will consider the fusion categorical symmetry $\cC={\rm Fib}$. We imagine this categorical symmetry acting first on the fusion algebra $\text{Fus}(\text{Fib})$. We can then pass to a complete algebraic non-invertible symmetry by following the approach outlined in Section~\ref{sec: algebras from categories}.

Indecomposable 2d TQFTs are known to be in one-to-one correspondence with indecomposable $\cal C$-module categories $\cal M$ \cite{Thorngren:2019iar,Gaiotto:2020iye,Komargodski:2020mxz,Huang:2021zvu}.\footnote{Here, the module category is being used to specify the system algebra. This is distinct from its role in determining the extended algebraic symmetry as in Section~\ref{sec: Fib Cat}.} Physically, the module category $\cal M$ labels the possible $\cal C$-symmetric gapped phases and the symmetry can be spontaneously broken. In particular, the simple objects of the module category $\cal M$ of rank $r$ correspond to a  preferred basis for the ground states $|v_a\rangle$ on $S^1$ with $a=1,2\dots,r$ that obey clustering at infinite volume. Because the theory is topological, the same basis of vacua $|v_a\rangle$ also describe boundary conditions, thus they are also referred to as Cardy branes. We will normalize them by $\la v_a |v_b\ra=\D_{ab}$.

We start by considering the case where $\cM$ is the regular module category (i.e. $\cC$ itself). This represents the case where $\cC$ undergoes a complete SSB. The system algebra of operators in this case $M = \text{span}\{\pi_i\}$ can be conveniently represented in terms of projectors $\pi_i$ for each superselection sector, 
\ie 
\pi_i \pi_j=\delta_{ij}\pi_i\,,\quad \pi_i |v_j\ra= \D_{ij} |v_j\ra\,.
\fe 
Let us also introduce the state prepared by the Euclidean path integral on a disk, namely the state corresponding to the identity operator (GNS vacuum), 
\ie 
\id =\sum_i \pi_i\,,\quad |\id \ra=\sum_i d_i |v_i\ra\,,
\fe
where $d_i$ is the quantum dimension of the topological defect $L_i$. Here $d_1=1$ and $|v_1\ra$ is also known as the identity Cardy brane (not to be confused with $|\id \ra$ above). This allows us to deduce the expectation values on $S^2$ via $\la \cO\ra_{S^2}\equiv \la \id |\cO|\id \ra$,
\ie 
\la \pi_i\ra_{S^2}=d_i^2\,,\quad \la \id  \ra_{S^2}=D\,,
\fe
where $D\equiv \sum_{i} d_i^2$ is the total quantum dimension of $\cC$. In a similar way, by cutting and gluing, the TQFT partition function on a general Riemann surface $\Sigma_g$ of genus $g$, is determined to be 
\ie 
Z_g\equiv \la \id \ra_{\Sigma_g} = \sum_i d_i^{2-2g}\,.
\fe

The $\cC$-symmetry acts on the Hilbert space and the operator algebra above via the regular representation,
\ie 
L_i |v_j\ra=\sum_k N_{ijk} |v_k\ra\,,\quad L_i|\id\ra=d_i|\id\ra\,,\quad  L_i \circ \pi_j= \sum_k N_{ijk} {d_j\over d_k}\pi_k\,.
\fe
Connecting to the general discussion in Section~\ref{sec: algebras from categories}, we note that the fusion algebra is generated by the topological defects $\text{Fus}(\mathcal{C}) = \text{span}\{L_i\}$, and the action described above is the endomorphism action of the fusion algebra on the non-extended system, $\rho: \text{Fus}(\mathcal{C}) \rightarrow \text{End}_{\textrm{v}}(M)$ given by $\rho_{L_i}(\pi_j) = L_i \circ \pi_j$.

Let us consider the following symmetry breaking faithful normal state in the $\cC$-symmetric TQFT labeled by the regular module category,
\ie 
\rho(\A_i)=\sum_i \A_i |v_i\ra \la v_i|\,,
\fe
with $\A_i>0$ and $\sum_i \A_i=1$. The symmetrized state is 
\ie
\rho_{\rm sym}={1 \over D}\sum_i d_i^2|v_i\ra \la v_i|\,,\quad 
% L_i \rho_{\rm sym}=\rho_{\rm sym} L_i\,.
\fe
Note that the symmetrization map is explicitly given by the following conditional expectation 
\ie 
E(x)={1\over D} \sum_k  d_k L_k\circ x\,,
\fe
whose adjoint,
\ie 
E^*(\pi_j)={1\over D}\sum_i d_i^2 \pi_i\,,
\label{fusEonrho}
\fe
gives the action on the density operator $\rho$. This follows from
\ie 
\tr \pi_i=1\,,\quad E(\pi_i)={d_i^2\over D} \id \,.
\fe
For the case $\cC={\rm Rep}(H)$ with Hopf algebra $H$, the above symmetrization map is of the form of our general conditional expectation \eqref{WH Invariantization Map} with normalized left Haar integral $\Lambda = \frac{1}{D} \sum_{k} d_k L_k$. However we see it is also well-defined for general fusion algebras.

The relative entropy in this case is  
\ie 
S(\rho|\rho_{\rm sym})=\tr \left(\rho (\log \rho-\log \rho_{\rm sym}) \right)
=\log D 
+\sum_i \A_i \log {\A_i\over d_i^2} 
\,,
\fe
which satisfies the inequality 
\ie 
0\leq S(\rho|\rho_{\rm sym}) \leq \log D\,.
\fe 
The lower and upper bounds are saturated respectively by the symmetric state $\rho=\rho_{\rm sym}$ and by the projection to the identity brane $\rho=|v_1\ra \la v_1|$.

The generalization to the case of $\cC$-symmetric TQFT corresponding to arbitrary module categories $\cM$ is straightforward. Let us take $\cM$ to be an indecomposable left $\cC$-module category of rank $r$. Then the system algebra are again given by projectors
\ie 
\pi_a \pi_b=\D_{ab}\pi_a\,,\quad \pi_a |v_b\ra=\D_{ab}|v_a\ra,.
\fe
The identity operator and its corresponding state becomes
\ie 
\id=\sum_a \pi_a\,,\quad |\id \ra=\sum_a  \tilde d_a |v_a\ra\,,
\fe
where $\tilde d_a$ is the quantum dimension of the Cardy branes $|v_a\ra$ and we follow the normalization in \cite{Diatlyk:2023fwf}. The corresponding total quantum dimension of the module category coincides with that of the fusion category \cite{Diatlyk:2023fwf},
\ie 
\dim \cM \equiv \sum_a \tilde d_a^2=D\,.
\fe
The $\cC$-symmetry defects $L_i$ act on the system algebra in this case via the NIM-rep $P_i$ associated with $\cM$, physically from the fusion 
 of $L_i$ with the Cardy branes (see $e.g.$ \cite{Diatlyk:2023fwf}),
\ie 
L_i\circ \pi_a=P_{iab} {\tilde d_b\over \tilde d_a}\pi_b\,.
\fe
Carrying out a similar analysis as in the regular module category case, we find that for a general state 
\ie 
\rho=\sum_{a} \A_a |v_a\ra\la v_a|\,,
\fe
its symmetrization gives
\ie 
\rho_{\rm sym}={1\over D}\sum_a \tilde d_a^2 |v_a\ra\la v_a|\,.
\fe
Consequently, the relative entropy is 
\ie 
S(\rho|\rho_{\rm sym})
=\log D 
+\sum_a \A_a \log {\A_a\over \tilde d_a^2} 
\,.
\fe
Clearly the lower bound remains the same whereas the upper bound is stronger,
\ie 
0\leq S(\rho|\rho_{\rm sym}) \leq \log D -\log \tilde d_1^2\,,
\fe
where we have ordered the Cardy branes $|v_a\ra$ such that the minimal quantum dimension $\tilde d_a$ is achieved for $a=1$ and the upper bound is saturated by the projection to this brane. Note that for $\cM$ with a single irreducible object, the inequalities collapse and $S(\rho|\rho_{\rm sym})=0$. This is expected since the TQFT becomes a $\cC$-symmetric SPT (with a non-degenerate symmetric ground state). More generally, $\tilde d_1^2$ coincides with the minimal quantum dimension of the gaugeable algebra object $A$ for symmetry $\cC$ that relates the regular TFT and the TFT labelled by the module category $\cM$ \cite{Diatlyk:2023fwf}. In the algebraic language, $\tilde d_1^2$ is the Jones index for the inclusion for the corresponding system algebras and the corresponding gaugeable algebra is also known as a Q-system \cite{Bischoff:2014xea}. 

\subsection{Conformal Field Theory on Half Line} \label{sec: ex CFT}

The examples so far concern states represented by density operators of finite rank in finite systems. Here, inspired by previous works \cite{Kusuki:2024gss,Fossati:2024ekt}, we consider entropic order parameters for states of 2d CFT in the presence of a conformal boundary condition.

 \subsubsection{Renyi Entropy on Half Line and BCFT}
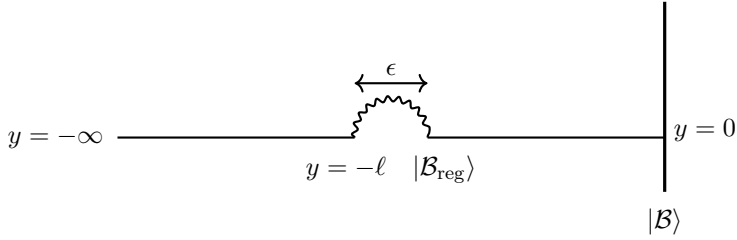
\begin{figure}[!htb]
    \centering
\begin{tikzpicture}[x=12mm,y=12mm,
  interface/.style={very thick},
  line/.style={thick},
  curly/.style={decorate,decoration={snake,aspect=0.2,segment length=4pt,amplitude=1pt}}
]
\pgfmathsetmacro{\GapLeft}{\GapCenter-\GapHalf}
\pgfmathsetmacro{\GapRight}{\GapCenter+\GapHalf}

\draw[interface] (\RightEnd,-0.6) -- (\RightEnd,1.5);
\node[scale=0.9,anchor=north] at (\RightEnd,-0.68) {$|\cB\rangle$};

\draw[line] (\LeftEnd,0) -- (\GapLeft,0);
\draw[line] (\GapRight,0) -- (\RightEnd,0);

\node[anchor=east,scale=0.9] at (\LeftEnd-0.05,0) {$y=-\infty$};
\node[anchor=west,scale=0.9] at (\RightEnd,0.08) {$y=0$};

\draw[line,curly] (\GapLeft,0) arc[start angle=180,end angle=0,radius=\GapHalf];

\draw[line,<->] (\GapCenter-.4,0.6) -- (\GapCenter+.4,0.6) ;

\node[scale=0.9] at (\GapCenter,0.75) {$\epsilon$};
\node[scale=0.9] at (\GapCenter,-0.35) {$y=-\ell\quad |\cB_{\mathrm{reg}}\rangle$};

\end{tikzpicture}
    \caption{The regulated setup for defining the density operator $\rho$ for a state on half-line with physical boundary $|\cB\ra$ and regulating boundary $|\cB_{\rm reg}\ra$.}
    \label{fig:regsetup}
\end{figure}

We consider a general non-chiral 2d CFT on the half line $y\leq 0$ with a conformal boundary condition $|\cB\ra$ and take the density operator $\rho$ on the interval $y\in [-\ell,0]$. To define $\rho$ properly, we follow the regularization procedure of \cite{Ohmori:2014eia} to factorize the Hilbert space $\cH_{\cB}$ on the half line by introducing a small regulating conformal boundary condition $|B_{\rm reg}\ra$ of size $\varepsilon$ at $y=\ell$ (see Figure~\ref{fig:regsetup}),
\ie 
\iota: \cH_{\cB} \to \cH_{ \cB_{\rm reg}} \otimes \cH_{\cB_{\rm reg} \cB}\,,
\fe
and perform the partial trace for state $|\psi\ra \in \cH_{\cB}$,
\ie \label{rhodef}
\rho=\tr_{\rm \cH_{ \cB_{\rm reg}}} \iota |\psi \ra \la \psi | \iota^\dagger\,.
\fe 

The $n$-th Renyi entropy is defined by 
\ie 
S^{(n)}(\rho)={1\over 1-n} \log {\tr_{\cH_{\cB_{\rm reg} \cB}}\rho^n\over (\tr_{\cH_{\cB_{\rm reg} \cB}} \rho)^n}\,,
\label{renyient}
\fe
and the entanglement entropy is obtained in the limit
\ie 
S(\rho)=\lim_{n\to 1}S^{(n)}(\rho)\,.
\fe

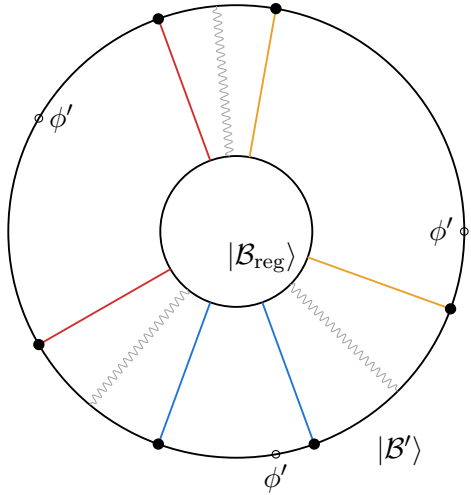
\begin{figure}[!htb]
   \centering
\begin{minipage}[c]{.45\textwidth}
\centering 
\begin{tikzpicture}[squig/.style={decorate,decoration={snake,aspect=0.2,segment length=3pt,amplitude=1.5pt}},
  bis/.style={gray!70,thin}]
  \def\Rout{3}
  \def\Rin{1}
  \def\LabelR{0.35}
 \definecolor{sheetorange}{RGB}{235,160,40}
  \definecolor{sheetred}{RGB}{214,45,45}
  \definecolor{sheetblue}{RGB}{33,117,214}

  \draw[line] (0,0) circle (\Rout);
  \draw[line] (0,0) circle (\Rin);

  \foreach \ang/\col in {340/sheetorange,80/sheetorange,
                         250/sheetblue,290/sheetblue,
                         110/sheetred, 210/sheetred}{
    \draw[line,\col] (\ang:\Rin) -- (\ang:\Rout) node [fill=black, circle, inner sep=1.5pt] {}; }
  \foreach \bisin in {95,230,315}{
    \draw[bis,squig] (\bisin:\Rin) -- (\bisin:\Rout);
  }
  \node[anchor=north] at (310:{\Rout+\LabelR}) {$|\cB'\rangle$};
  \node[anchor=east] at (340:{\Rin}) {$|\cB_{\mathrm{reg}}\rangle$};
  \draw (0:{\Rout}) circle [radius=0.05] node [left] {$\phi'$};
  \draw (280:{\Rout}) circle [radius=0.05] node [right,below] {$\phi'$};
  \draw (150:{\Rout}) circle [radius=0.05] node [right] {$\phi'$};
\end{tikzpicture} 
\end{minipage}
    \caption{The replicated annulus partition $Z_{\cB'|\cB_{\rm reg}}(q^n)[\phi']$ for $n=3$. Here the gray curly lines represent branch cuts connecting the three sheets. The colored lines represent algebra elements in the WHA symmetry that act on each sheet via the symmetrization map \eqref{rhosym}. The black dots represent (possibly non-topological) boundary changing operators on the symmetry breaking boundary $|\cB'\ra$ and the white dots represent the topological operator insertions.}
    \label{fig:entropyBCFT}
\end{figure}

For simplicity, let us consider $|\psi\ra$ to be the ground state. 
As a consequence of the enhanced conformal symmetry in 2d, the Renyi entropy is in fact computed by the annulus partition function (see Figure~\ref{fig:entropyBCFT}), 
\ie 
Z_{\cB_{\rm Reg}|\cB}(q^n)=\tr_{\cH_{\cB_{\rm reg} \cB}} \rho^n\,,
\fe
with modular parameter \cite{Ohmori:2014eia,Cardy:2016fqc}
\ie 
q =e^{-2\pi^2/W}\,,\quad W= \log{2\ell \over \varepsilon} +\cO(\varepsilon)\,,
\fe
and the entanglement Hamiltonian is identified with the open string Hamiltonian as
\ie 
\rho=q^{H^{\rm open}_{\cB_{\rm Reg}|\cB}}\,,\quad H^{\rm open}_{\cB_{\rm Reg}|\cB}=L_0-{c\over 24}\,.
\label{denopbcft}
\fe

Using the open-closed channel duality
\ie 
Z_{\cB_{\rm Reg}|\cB}(q^n)=\la \cB_{\rm reg} | \tilde q^{L_0-{c\over 24}}|\cB\ra \,,\quad \tilde q = e^{-2W}\,.
\label{openclose}
\fe
Therefore in the limit of $\ell\gg \varepsilon$, since $\tilde q\ll 1$, the close channel expression of the partition function 
\ie 
S^{(n)}(\rho)={c\over 12}{n+1\over n} \log{2\ell \over \varepsilon} + \log g_{\cB}+\log g_{\cB_{\rm reg}} + \dots\,,
\label{renyifinal}
\fe
where the first subleading corrections come from the $g$-functions of the boundaries.

\subsubsection{Entropic Order Parameter for Non-invertible Symmetry}
As in \cite{Kusuki:2024gss,Fossati:2024ekt}, we now refine the entropies discussed in the last section with symmetry information and characterize the entropic order parameter for symmetry violation, also known as entanglement asymmetry \cite{Ares:2022koq}. While the previous works focus on invertible symmetries, here we consider general non-invertible symmetries described by symmetry category $\cC$ in the bulk, and realized via a weak Hopf algebra $H={\rm Str}_\cC(\cM)$ in the presence of a boundary $|\cB\ra$. Here the module category $\cM$ encodes the multiplet of boundary conditions under $\cC$ and the Hopf algebra naturally acts on the boundary changing operators among boundary conditions in the multiplet. 

Assuming the algebraic symmetry action by $H$ is well-defined in the presence of the regulating boundary $|\cB_{\rm reg}\ra$ and the physical boundary $|\cB\ra$ (which requires including additional sectors as we have alluded to before), given the density operator $\rho$ constructed in the previous section, 
the symmetrized density operator can be identified using the 
conditional expectation defined in Section~\ref{sec: order parameters},
\ie 
\rho_{\rm sym}\equiv E^*(\rho)=V(\Lambda_{(1)})\rho\, V(S(\Lambda_{(2)})\,,
\label{rhosym}
\fe
via the adjoint action of the Haar element $\Lambda\in H$. Here we have used the Sweedler notation for the coproduct and $V(x)$ is a $*$-preserving representation of $H$ on the Hilbert space.\footnote{In \cite{nikshych2000finitequantumgroupoidsapplications}, such a representation is referred to as unitary due to its relationship with the unitary classification of a category. We have chosen not to emphasize this terminology due to the possible confusion with unitary representations in the algebraic sense. The elements $V(x)$ are \emph{not} unitary operators on the Hilbert space since they are in general not invertible. The crucial property utilized in \cite{nikshych2000finitequantumgroupoidsapplications} is that $V(x)^{\dagger} = V(x^*)$, where $\dagger$ signifies the formal adjoint with respect to the inner product of the Hilbert space.} The symmetrized density operator satisfied the following weak symmetric condition\footnote{This agrees with the symmetry condition discussed in \cite{Benini:2025lav}. There a stronger assumption was imposed, namely that the antipode is involutive, $S^2 = \id$, which restricts to weak Kac algebras and ensures that the separability idempotent is symmetric. We thank the authors of \cite{Benini:2025lav} for correspondence on this point.}
\ie 
V(h) \rho_{\rm sym}=\rho_{\rm sym} V(h) \quad \forall h\in H\,,
\label{weaksym}
\fe
and furthermore the symmetrization map \eqref{rhosym} is trace preserving and idempotent (thus qualify as a dual conditional expectation). These properties follow from the semisimplicity of the WHA $H$ which is equivalent to the existence of a separability idempotent \cite{BOHM1999385}
\ie 
q\equiv (\id \otimes S) \Delta(\Lambda)\,,  
\fe
that satisfies 
\ie 
q^2=\id \,,\quad \epsilon_s(q)=\id \,,\quad  (h\otimes \id ) q= q(\id \otimes h)\quad \forall h\in H\,.
\fe 
The above can be easily checked using the explicit formulas of Appendix~\ref{app: strip} for the WHA in the strip algebra representation. As a trivial example, if $H$ is a group algebra, \eqref{rhosym} reproduces the familiar average over the group by conjugation. 

Having defined the symmetrized density operator, 
the relative Renyi entropy that measures the amount of symmetry breaking is given by
\ie
\Delta_H S^{(n)}(\rho) ={1\over 1-n} \log {\tr_{\cH_{\cB_{\rm reg}|\cB}}\rho_{\rm sym}^n\over \tr_{\cH_{\cB_{\rm reg}|\cB}} \rho^n}\,.
\label{renyiHbreak}
\fe
We note that, although for concreteness we focus on symmetry-breaking states in the presence of a physical boundary, the symmetrization map \eqref{rhosym} applies more generally to arbitrary mixed states whenever a symmetry action is defined.

\subsubsection{$H_8$ Symmetry in the Ising$^2$ CFT with Boundary}

To be more explicit, let us consider the Ising$^2$ CFT which has a ${\rm Rep}(H_8)$ fusion category symmetry, which is a subcategory in the products of two ${\rm TY}(\mZ_2,+)$ Tambara-Yamagami fusion category symmetries. 
Here we will be explicit about the way we extend the system and identify the algebraic avatar of the category. 

We denote the $\mZ_2$ topological defects for the TY symmetries by $\eta_{i}$ and the Kramers-Wannier duality defects by $\cN_{i}$, which obey the following fusion rules
\ie 
\cN_i^2=\id +\eta_{i}\,,\quad \eta_i \cN_i=\cN_i \eta_i=\cN_i\,.
\label{isingtdls}
\fe
In terms of these, the ${\rm Rep}(H_8)$ symmetry has the simple objects,
\ie 
\id\,,~\eta_1\,,~\eta_2\,,~\eta_1\eta_2\,,~\cV\equiv \cN_1\cN_2\,,
\fe
where the only non-invertible defect is the duality defect $\cV$ of quantum dimension $d_\cV=2$.

First we take the regulating boundary to be
\ie 
|\cB_{\rm reg}\ra =|\cN\ra\,,
\fe 
which corresponds to the Kramers-Wannier duality defect $\cN$ in a single copy of Ising CFT after folding. The two copies of Ising defects in \eqref{isingtdls} can be thought of as shadows of those in this single copy to the two sides of the fold (see e.g. \cite{Choi_2023} for details). As a consequence of the fusion rules \eqref{isingtdls} and the identification $\cN=\cN_1=\cN_2$ at the fold, we see $|B\ra$ corresponds to a multiplet of rank 1 under symmetry ${\rm Rep}(H_8)$, realizing the fiber functor and thus supports naturally the action of the Hopf algebra $H_8$ on the boundary changing operators (see also \cite{Choi:2024wfm}).  

The $H_8$ C$^*$-Hopf algebra  is generated by $\{\id,a,b,z\}$ with multiplication rules (see e.g. \cite{Buerschaper:2013bju} for a review)
\ie 
a^2=b^2=\id \,,\quad z^2={1\over 2}(\id +a+b-ab)\,,\quad ab=ba\,,\quad  a z= z b\,,\quad b z = z a\,,
\label{H8mul}
\fe
and the following coproduct
\ie 
\Delta(\id)=\id \otimes \id \,,\Delta(a)=a\otimes a\,,~\Delta(b)=b\times b\,,~\Delta(z)= \frac{1}{2} (\id \otimes \id + b \otimes \id + \id \otimes a - b \otimes a ) z\otimes z\,.
\fe 
The counit and antipode are specified by
\ie 
\epsilon(a)=\epsilon(b)=\epsilon(z)=1\,,\quad 
S(a)=a\,,~S(b)=b\,,~S(z)=z\,.
\label{counitantipode}
\fe
The anti-linear involution is 
\ie 
a^*=a\,,\quad b^*=b\,,\quad z^*=z^3=z^{-1}\,.
\fe 
Finally the unique Haar element and Haar measure,
\ie 
\Lambda= {1\over 8}(\id+a+b+ab + z + z a + zb + zab)\,,\quad \lambda=\D_1\,.
\label{HaarH8}
\fe 

Next we specify the physical boundary multiplet. One can easily construct a subset of boundary conditions of the Ising$^2$ CFT from products of the Cardy boundaries $|\pm\ra,|f\ra$ of the Ising CFT, such as the following 
\ie 
|{++}\ra\,,~|{+-}\ra=\eta_2|{++}\ra\,,~|{-+}\ra=\eta_1|{++}\ra\,,~|{--}\ra=\eta_1\eta_2|{++}\ra\,,~|ff\ra=\cV|{++}\ra,,
\label{regularH8m}
\fe
which are related to one another by fusion with defects in ${\rm Rep}(H_8)$, a symptom of symmetry breaking, and mathematically described by the regular module category over ${\rm Rep}(H_8)$. 
However, the Hopf algebra $H_8$ action is not obviously defined for this boundary multiplet. Instead the natural option is to consider a generalized strip algebra in the sense of \cite{Choi:2024tri}, which may not be weak Hopf. It may still be possible to define entropic parameters for this algebra 
but it's beyond the scope of this paper (we will comment on this point in Section~\ref{sec: discussion}). 

Instead, given a starting setup with boundary conditions $|\cB\ra=|{++}\ra$ and $|\cB_{\rm reg}\ra=|\cN\ra$, we introduce a $H_8$-compatible extension, by including the entire multiplet \eqref{regularH8m} via,
\ie 
|\cB'\ra=|{++}\ra \oplus |{+-}\ra \oplus |{-+}\ra \oplus |{--}\ra \oplus 2|{ff}\ra, 
\fe
and inserting the following topological point operator,
\ie 
\phi'={1\over 8}(\id_{++}+\id_{+-}+\id_{-+}+\id_{++}+\id_{f_1f_1}+\id_{f_2f_2})\,,
\label{cloakop}
\fe 
as a linear combination of irreducible idempotents.

This effectively leads to 
the cloaking boundary condition of \cite{Brehm:2024zun} (also known as entanglement brane in \cite{Hung:2019bnq}), which is now compatible with the $H_8$ action. 
When inserted in the annulus partition function with boundaries $|\cB'\ra$ and $|\cB_{\rm reg}\ra$, the operator $\phi'$ weighs the contributions from different components of $|\cB'\ra$. 
In a similar way, the initial physical symmetry breaking boundary can be picked out by assigning
\ie 
\phi=\id_{++}\,.
\fe
Furthermore, since $|\cB'\ra$ can be identified with the identity Cardy brane fusing with the topological interface associated with half-gauging the maximal algebra object $A_{\rm max}$ in the symmetry category ${\rm Rep}(H_8)$ (which is self-dual under this gauging), we specify the topological junctions between the topological defects in ${\rm Rep}(H_8)$ and the boundary $|\cB'\ra$ via those of the algebra object $A_{\rm max}$ \cite{Diatlyk:2023fwf} (see Figure~\ref{peelrel} and discussions there). This guaranties an action of the Hopf algebra $H_8$ on the strip Hilbert space and on the corresponding density operator as defined in \eqref{rhodef}.

Having specified the extended system with boundary conditions that admit an $H_8$ action, we can proceed to calculate the relative Renyi entropy as an order parameter for the symmetry breaking \eqref{renyiHbreak}. Following the discussion  around \eqref{denopbcft}, the initial density operator for our setup  is 
\ie 
\rho=\phi\cdot  q^{H^{\rm open}_{\cB'|\cN}} \,.
\fe
Its symmetrization from \eqref{rhosym} gives,
\ie 
\rho_{\rm sym}=\phi'\cdot q^{H^{\rm open}_{\cB'|\cN}}\,.
\label{symrhoH8}
\fe
One way to see this is to note that the space of topological operators on $|\cB'\ra$ is a copy of $H_8^*$ and $\phi'={1\over 8}1_{H_8^*}$ is proportional to the identity in $H_8^*$ (i.e. the counit $\epsilon(\cdot)$ of $H_8$).  

The adjoint action in \eqref{rhosym} on the density operator $\rho$ factors through its action on $H_8^*$. The canonical adjoint action of Haar element $\Lambda$ in C$^*$-Hopf algebra $H$ on the dual Hopf algebra $H^*$ itself follows from the following identity by standard manipulations (see e.g. \cite{montgomery1993hopf}),
\ie 
f(\Lambda_{(1)} x S(\Lambda_{(2)} ))= \epsilon(x) f(\Lambda)\,,\quad \forall f \in H^*\,,~x\in H\,.
\fe
which implies 
\ie 
V(\Lambda_{(1)}) f (\cdot)V(S(\Lambda_{(2)})=f(\Lambda)\epsilon(\cdot)\,.
\fe
Applied to the specific case here, it gives
\ie 
V(\Lambda_{(1)}) \phi (\cdot)V(S(\Lambda_{(2)}) = {1\over 8} \epsilon(\cdot)\,,
\fe
as desired (the normalization on the RHS is fixed by the Haar measure on $H_8^*$) and gives \eqref{symrhoH8}.\footnote{The $H_8$ algebra is self-dual. The explicit isomorphism $x \to f_x$ for $H_8\cong H_8^*$ can be found in \cite{Buerschaper:2013bju}. Here in terms of the topological point operators on $|\cB'\ra$,  \ie
f_1=\,&\id_{++}+\id_{+-}+\id_{-+}+\id_{--}+\id_{f_1f_1}+\id_{f_2f_2}\,,
\\
f_a=\,&\id_{++}-\id_{+-}-\id_{-+}+\id_{--}+\id_{f_1f_1}-\id_{f_2f_2}\,,
\\
f_b=\,&\id_{++}-\id_{+-}-\id_{-+}+\id_{--}-\id_{f_1f_1}+\id_{f_2f_2}\,,
\\
f_{ab}=\,&\id_{++}+\id_{+-}+\id_{-+}+\id_{--}-\id_{f_1f_1}-\id_{f_2f_2}\,,
\\
f_z=\,&\id_{++}+i\id_{+-}-i\id_{-+}-\id_{--}+\id_{f_1f_2}+\id_{f_2f_1}\,.
\fe} 
We mention in passing that this is also consistent with the $E^*(\cdot)$ map on density operators \eqref{fusEonrho} discussed in the TQFT example when the symmetry is Hopf.

Consequently, the relative Renyi entropy \eqref{renyiHbreak} becomes a ratio of linear combinations of annulus partition functions with elementary boundary conditions for Ising$^2$ CFT weighted by topological operators.  
\ie 
\Delta_H S^{(n)}(\rho) = {1\over 1-n} \log {Z_{\cB'|\cN}(q^n)[\phi']\over Z_{\cB'|\cN}(q^n)[\phi]}\,.
\label{renyitoannulus}
\fe
where the $\phi'$ and $\phi$ topological point operators are each inserted $n$ terms on the replicated manifold (once in each sheet). The partition function in the denominator is simply a single irreducible annulus partition function, with the starting physical boundary $|{++}\ra$ and regulating boundary $|\cN\ra$,
\ie 
 Z_{\cB'|\cN}(q^n)[\phi]= Z_{\cB_{++}|\cN}(q^n)\,.
 \fe 
The other annulus partition function in the numerator on the RHS of \eqref{renyitoannulus} is more complicated (see Figure~\ref{fig:entropyBCFT}). Nonetheless, in the limit $\ell\gg \epsilon$, the dominant contribution comes from components of $|\cB'\ra$ that align from one sheet to another, so that there is no penalization from a non-topological boundary changing operator as is the case in \cite{Kusuki:2023bsp}. Consequently, in this limit,
\ie 
 Z_{\cB'|\cN}(q^n)[\phi']\to  {1\over 8^n} (Z_{\cB_{++}|\cN}(q^n)+Z_{\cB_{+-}|\cN}(q^n)+Z_{\cB_{-+}|\cN}(q^n)+Z_{\cB_{--}|\cN}(q^n)+2Z_{\cB_{ff}|\cN}(q^n))\,.
 \label{limitofap}
\fe 
Furthermore, using the open-closed duality \eqref{openclose}, the simplification as in \eqref{renyifinal} and taking into account the $g$-functions $g_{ff}=2g_{\pm \pm}$, we have in the limit of $\ell\gg \epsilon$, 
\ie 
\Delta_H S^{(n)}(\rho) =\log 8 - c_n \left (\varepsilon\over \ell\right)^{2\Delta^{\rm bco}_{\rm min}}+\dots\,,
\fe
where we have included the first subleading correction coming from boundary changing operators in the step \eqref{limitofap}. As explained in \cite{Kusuki:2024gss,Fossati:2024ekt}, this correction comes from the two-point function of a pair of boundary changing operators and $c_n$ is positive by reflection positivity.  On the replicated annulus, this contribution comes a boundary configuration that is almost uniformly given by an irreducible component of $|\cB'\ra$ except on a single sheet, where boundary changing operators are forced to be present when connecting to adjacent sheets. The lightest boundary changing operator on $|\cB'\ra$ has the dominant two-point function in the limit considered here \cite{Kusuki:2024gss,Fossati:2024ekt}
and $\Delta^{\rm bco}_{\rm min}$ is its conformal weight. As was observed there, this finding is consistent with the bound on entropic order parameter and saturates the bound in the strict $\ell/\varepsilon\to \infty$ limit. 

The generalization of the above consideration to cases of 2d (B)CFT with a general Hopf algebra symmetry $H$ is straightforward and the resulting relative Renyi entropy takes the expected universal form in the large size limit,
\ie 
\Delta_H S^{(n)}(\rho) =\log (\dim H) - c_n \left (\varepsilon\over \ell\right)^{2\Delta^{\rm bco}_{\rm min}}+\dots\,.
\fe
Note that the algebra dimension coincides with the quantum dimension of the corresponding category for Hopf algebras (i.e. $\dim H=|{\rm Rep}(H)|$). 

It would be interesting to generalize the discussion above to cases with strictly weak Hopf symmetry. Preliminary investigation suggests that the naive generalization produces an entropic order parameter consistent but not saturating our entropy bound. This is perhaps not too surprising since here we are studying special type of states (e.g. with conformal symmetry) and it's possible that to saturate the bound requires more exotic states. Conversely, it is also possible that the bound \eqref{Bound for WHA} can be strengthened by assuming certain local conditions on the states. We will return to this question in future work.

\section{Discussion} \label{sec: discussion}
In this work, we have provided a framework that unifies the algebraic approach to quantum theory, which is amenable to information-theoretic analyses, and symmetry structures more general than those described by a group. The two main inputs are the fusion categorical symmetry $\mathcal{C}$ and the system algebra $M$. From this data, we have shown how to induce a weak Hopf algebraic symmetry action of the strip algebra $\text{Str}_{\mathcal{C}}(\mathcal{M})$ on an extended system algebra $M_{\mathcal{M}}$. 
Crucially, the algebraic avatar of the fusion categorical symmetry is non-unique, as it depends on the choice of module category $\mathcal{M}$. 
In specific systems, such an extension $M\to M_{\cal M}$ can be realized physically by incorporating defects such as boundary conditions.
At the same time, we have also proposed an algorithm for computing entropic order parameters for symmetries described by a fusion category, which has not appeared before. This construction passes through the induction from fusion categorical to weak Hopf algebraic symmetry, and is therefore also non-unique. Rather than a bug, we argue that this ambiguity presents a feature of non-invertible/categorical symmetry which underscores its rich nature and the kinds of physical settings it can describe. 
In the following, we discuss a few exciting future directions.

\subsection*{Weak Hopf Algebras as Models for Degenerate Vacuum Structures} \label{sec: WHA Degen}
Given a fusion category $\mathcal{C}$, we showed how to obtain a family of weak Hopf algebraic symmetry actions labeled by its module categories. In the associated WHAs, the choice of module category determines the target and source counital subalgebras. As we have described, these subalgebras play a central role in defining the modified notion of `invariance', which is appropriate for systems under the action of a WHA. In particular, an observable is regarded as residing in the symmetric sector if its transformation is sensitive only to the target counital projection of a WHA element which acts upon it. In particular, the target counital subalgebra can be seen to be the trivial representation of the WHA within which it is contained (see Appendix~\ref{app: WHA} and ~\ref{app: strip}). Taken together, these observations suggest that the non-uniqueness implicit in fusion categorical symmetries (especially for the anomalous ones where the weakened WHA axioms are necessary) is in fact a useful feature for analyzing physical systems with exotic vacuum structures. In future work we plan to investigate this more carefully in the context of gauge theories and quantum gravity. 

\subsection*{Non-invertible Symmetries in Quantum Gravity}

It is typically said that there are no global symmetries in quantum gravity~\cite{banks, kallosh1995, banks2011, harlow2019}. While this arose from the context of group-like symmetries, there are indications that the absence of global symmetries in quantum gravity should be extended to more general symmetries as well~\cite{cobordism}. Thus, any symmetry $(\mathcal{C},\mathcal{M})$ must either be broken or gauged if $M_{\mathcal{M}}$ is to describe a quantum gravitational theory.

Recent results in quantum gravity have featured group-like symmetries $G$ which extend the local algebra of observables of fields $M$ (including the graviton) into a crossed product $\hat{M} = M \times_{\alpha} G$. In many cases of interest the resulting gravitational algebra acquires an intrinsic regulator associated with additional degrees of freedom that arise from quantum gravitational fluctuations. It may then be shown that the gravitational algebra $\hat{M}$ is semifinite following from Takesaki's result on the modular crossed product of Type III von Neumann factors~\cite{takesaki}. This has led to a deeper understanding of information-theoretic quantities like von Neumann entropies and relative entropies, and their realization in quantum gravity from an algebraic perspective.

To that end, it would be interesting to use our results here and elsewhere~\cite{AliAhmad:2023etg,Jensen:2023yxy,Klinger:2023auu,Klinger:2023tgi,AliAhmad:2024wja,AliAhmad:2024vdw,AliAhmad:2024eun,AliAhmad:2024saq,AliAhmad:2025oli}, to explore the interaction of quantum gravity with categorical symmetries from an algebraic perspective. For example, results in the literature indicate that replica wormholes in entropic computations may be responsible for the loss of orthogonality between superselection sectors of a theory with a group symmetry~\cite{Chen:2020ojn, Hsin:2020mfa}. From the point of view described above, this non-orthogonality may be associated with the introduction of new operators which implement a categorical symmetry intertwining between QFT sectors that are connected by gravity. Our approach may also be used to understand the process by which quantum gravity breaks or gauges a categorical symmetry in general.

It would also be important to isolate the holographic interpretation of a non-invertible symmetry by capitalizing on subregion duality~\cite{Leutheusser:2022bgi}. The non-uniqueness of the given symmetry realization, which is the degeneracy in choosing $H$ for a given fusion category $\mathcal{C}$, may make the faithful realization of global non-invertible symmetries untenable or at the very least richer. 

\subsection*{The Role of the Index}
The index of an inclusion of algebras, generalizing Jones' index for Type II$_{1}$ subfactors, appeared quite prominently in our analysis on bounding the order parameter. In quantum gravity, it has been hypothesized that the index of certain inclusions is an important physical quantity. More generally, it is known that the index can have many interpretations depending on the inclusion structure $N \subset M$, or more aptly the relevant map (here, a conditional expectation) $E: M \to N$. The intuitive description of the index $\text{Ind}(E)$ in terms of the relative size of $N$ in $M$, while a helpful heuristic, may not always be the most appropriate. 

The index has also been interpreted as a relative free energy~\cite{longo_landauers_2018}, with a direct relation to non-commutative conditional probability \cite{Longo:1996xa}. In our work, it appeared in terms of an extremization over states of a relative entropy $S(\psi | \psi \circ E)$, or as a sum of entropic order parameters for dual inclusions, $\Delta_ES(\psi) + \Delta_{E'}S(\psi')$. Local theories on spacetime subregions $\mathcal{U}$ typically have universal terms in their von Neumann entropies given by geometries quantities $G(\mathcal{U})$ like the volume or the area of the entangling surface~\cite{Gomez2024_DualityBlackHoleEvaporation,vdHeijdenVerlinde2024_operator_algebraic_bbhinformation, Leutheusser:2025zvp}. While recent work has offered different interpretations of the index in these contexts, we believe that a more systematic treatment of the role of the index as an information-theoretic quantity which \textit{depends on the choice of} the theory, the inclusion, and the spacetime subregion is an important direction to pursue in future work. 
\subsection*{Infinite-dimensional Categorical Symmetries} 

On the topic of the index, it is also of importance to understand the structure of inclusions in which the index is not finite. We have focused on WHAs due to their relationship with fusion categorical symmetries. However, a generalization of our construction can be obtained in situations allowing for less structure,
\begin{enumerate}
	\item A bialgebra $H$ admitting a right (left) invariant weight $\Lambda: H \rightarrow \mathbb{C}$ such that
	\beq
		(\text{id}_H \otimes \Lambda) \circ \Gamma(h) = \Lambda(h) \text{id}_H, \qquad \bigg((\Lambda \otimes \text{id}_H) \circ \Gamma(h) = \Lambda(h) \text{id}_H\bigg) \qquad \forall h \in H\,,\nonumber
	\eeq
	and
	\item A system algebra $A$ upon which $H$ right (left) coacts on by~$\gamma: A \rightarrow A \otimes H $ $\bigg(\gamma: A \rightarrow H \otimes A\bigg)$  such that
	\beq
		(\gamma \otimes \text{id}_H) \circ \gamma = (\text{id}_A \otimes \Gamma) \circ \gamma, \qquad \bigg( (\text{id}_H \otimes \gamma) \circ \gamma = (\Gamma \otimes \text{id}_A) \circ \gamma \bigg)\,.
	\eeq
\end{enumerate}
Given these ingredients we can always construct an invariantization map $T_{\gamma}$. The examples discussed in the present note arise from dualizing the coaction $\gamma$ into an action of the dual algebra. 

A rather general class of examples when this is the case come from the study of locally compact quantum groups \cite{VAES2001426}. As with WHAs (also known as finite quantum groupoids), locally compact quantum groups admit a duality theory, and the coaction of a locally compact quantum group can be related to an action by its dual. Thus, locally compact quantum group actions may be interpreted as a form of algebraic non-invertible symmetry. In future work we intend to investigate this case further
and especially  (a) extend it to account for infinite-dimensional weak quantum groupoids (see relevant works \cite{vandaele2012weakmultiplierhopfalgebras2,vandaele2012weakmultiplierhopfalgebras,böhm2013weakmultiplierbialgebras})  and (b) obtain a categorification of a similar form to the relationship between WHAs and fusion categories. 

It would be exciting if one can use the framework alluded to here to deal with categorical symmetries which are even more general than those encoded in fusion categories. For example, one simple relaxation is the finiteness of the number of simple objects. In the group case, this would correspond to passing from a finite to an infinite but locally compact group. Like locally compact groups, this infinite dimensional counterpart of a quantum group does not lead to an inclusion of finite index. In other words, the map $T_{\gamma}$ is generically an operator valued weight and not a conditional expectation. Nevertheless, the theory of inclusions associated with symmetrization under, and crossed products with locally compact quantum groups resembles strongly the study of depth-2 inclusions discussed in Appendix~\ref{app: ECR}. Providing a categorification of infinite index depth-2 inclusions was a goal of the work \cite{AliAhmad:2025oli} and would significantly improve our understanding of infinite dimensional non-invertible symmetries. 

\subsection*{Locality and Spacetime Symmetry} 
Starting with a single module or theory acted on by the fusion symmetry led us to the study of an inclusion of $A_{\rho} \subset A$ for the invariant subalgebra. In quantum field theory, instead of considering a single algebra one typically studies a net
\begin{equation}
    \mathcal{N}: \mathcal{U} \mapsto \mathcal{N}\left[ \mathcal{U} \right]\,,
\end{equation}
where $\mathcal{U}$ is an open spacetime region and $\mathcal{N}$ maps any such set to an algebra of observables localized in $\mathcal{U}$. It is important to note that only the relations among the algebras in local quantum field theory determine the theory as all the local algebras are isomorphic.

In accordance with typical requirements of algebraic quantum field theories, such a net is typically taken to satisfy isotony
\begin{equation}
\mathcal{U}_{1} \subseteq \mathcal{U}_{2} \implies \mathcal{N}[\mathcal{U}_{1}] \subseteq \mathcal{N}[ \mathcal{U}_{2}].
\end{equation}
and (Einstein) causality
\begin{equation} \label{Causality}
    \mathcal{U}_{1} \subset \mathcal{U}_{2}' \implies \left[ \mathcal{N}[\mathcal{U}_{1}] , \mathcal{N}[ \mathcal{U}_{2}]\right]=0\,,
\end{equation}
where the prime denotes the causal complement of a spacetime subregion.

To understand how spacetime symmetries such as the Poincaré group and notions of locality interact with a non-invertible symmetry, one would need to perform a Doplicher-Haag-Roberts analysis of the relevant superselection sectors of the theory $\mathcal{N}$. For example, a stronger notion of causality than \eqref{Causality} is Haag duality~\cite{Haag:1996hvx} in which $\mathcal{N}\left[\mathcal{U}'\right]= \mathcal{N}\left[\mathcal{U}\right]'$, e.g. the algebra assigned to the causal complement of a region $\mathcal{U}$ is equal to (rather than merely contained within) the commutant of algebra assigned to $\mathcal{U}$. The status of Haag duality is dependent upon a variety of factors pertaining to the theory, spacetime regions, and in particular symmetry. Indeed, it is well known that the failure of Haag duality signals a rich superselection sector of the theory~\cite{Bischoff:2014xea}. 

Recently, Haag duality in the context of two-dimensional conformal field theories has been linked to a notion of completeness, and modular invariance~\cite{Benedetti:2024dku}. Moreover, the usual additivity on multi-interval subregions for local quantum field theories has been argued to be related to the existence of invertible elements for the symmetry structure~\cite{Shao:2025mfj}, which suggests that locality must be modified whenever the symmetry is non-invertible from the algebraic perspective; cf.~\cite{Harlow:2025cqc} for a weakening of additivity to `disjoint additivity' which achieves this. It would be of great interest to adapt the algebraic machinery developed in this work to the analysis of nets of von Neumann algebras supporting general categorical symmetries. 

\subsection*{Non-invertible Gauge Theory and Anomalies}

In this note, we have considered the action of a WHA $H$ on a system $A$. Implicitly, we have regarded $H$ as acting as certain generalized global symmetry. Nevertheless, the inclusion of the symmetric sector $A_{\rho} \subset A$ is intimately related to the notion of gauging as represented by the crossed product inclusion $A \subset A \times_{\rho} H$. Together, these comprise a nested sequence of inclusions $A_{\rho} \subset A \subset A \times_{\rho} H$ forming a Jones triple which is described in detail in Appendix~\ref{app: ECR}. The last algebra acts naturally on a Hilbert space $L^2(A) \otimes L^2(H)$, where the standard representation of the WHA $H$ is its GNS representation with respect to the Haar integral of its dual viewed as a weight \cite{nill1998weakhopfalgebrasreducible}. This setup provides a basis for understanding theories in which a categorical symmetry is gauged.

A gauge theory with compact group $G$ typically carries the regular representation (e.g. in lattice gauge theory), which, due to compactness and the group nature of the symmetry, decomposes as in the Peter-Weyl theorem
\begin{equation}
    L^{2}(G, \mu_{\rm G}) \simeq \bigoplus_{U \in \text{Irr}(\text{Rep}(G))} U^{|U|}\,.
\end{equation}
Here, $|U|$ is the dimension of the irreducible $U$ and $\mu_{\rm G}$ is the unique Haar measure. Since any compact group is unimodular, the left  regular representation is spanned in the same way with actions related by the inverse operation $g \mapsto g^{-1}$. The kinematical algebra of observables is then $B(L^{2}(G, \mu_{\rm G}))$. The full set of compact operators on $L^2(G,\mu_G)$ can be obtained by taking a bicrossed product of the group algebra $\mathcal{L}(G)$ with its dual $\mathcal{L}(G)^* \simeq L^{\infty}(G)$. 

By analogy, we may try to construct a kinematical algebra for a general categorical symmetry $\mathcal{C}$. For fusion symmetries, this can be done by passing to a WHA representative $H$. Dualization is accomplished using the antipode $S$, where now one toggles between left and right coactions or target and source subalgebras. Viewing $H$ acting on its dual $H^{*}$, the natural generalization of $B(L^{2}(G, \mu))$ is the bicrossed product $H\sharp H^*$. This algebra is related to the Drinfeld double of $H$, $D(H)$, whose representation category is equivalent to that of the tube algebra of $\mathcal{C}$ (see e.g. \cite{Bai:2025zze}). In this sense, the double $D(H)$ is an algebraic avatar of $\mathcal{C}$ that is more intrinsically related to the category, with the dependence on the module category washing out in the doubling (at least at the level of categorical Morita equivalence). Indeed the double $D(H)$ plays an important role in constructing quantum lattice models for 3d TQFTs as center generalized gauge theory \cite{Kitaev:2011dxc,
Chang_2014,Jia:2023xar,Kawagoe:2024tgv,Jia:2024rzr} which generalize both the Kitaev model \cite{Kitaev:1997wr} and the Levin-Wen model \cite{Levin:2004mi},  unifying them in a broader setting.

It may happen that this non-invertible symmetry cannot be gauged consistently, a phenomena characterized by certain generalized anomalies. 
For invertible symmetries described by a group $G$, we have complete control over such anomalies which are classified by group cohomologies. For non-invertible symmetries, the possible consistent gaugings in 2d are fully classified by symmetric separable Frobenius algebras in the fusion category \cite{Fuchs:2002cm,Bhardwaj:2017xup,Choi:2023vgk,Diatlyk:2023fwf,Perez-Lona:2023djo,Perez-Lona:2024sds} and those in 3d are being investigated \cite{Cui:2015cbf,Carqueville:2018sld,Mulevicius:2020tgg,Mulevicius:2022gce,Carqueville:2025kqs,Heinrich:2025wkx,KNBalasubramanian:2025vum}. 
We note that there is also a different, physically motivated, notion of anomalies, as obstructions to a symmetric deformation of the system to a trivially gapped phase \cite{Chang:2018iay,Thorngren:2019iar}. While for the invertible case, the two notions of anomalies coincide, they differ for the non-invertible case \cite{Choi:2023xjw}. In any case, these anomalies encode invariants of the symmetry under continuous deformation and it would be interesting to investigate whether the algebraic perspective helps to define such a cohomology.

\subsection*{New Theories from Entropic Order Parameters}
Finally, as demonstrated in the recently developed Entanglement Bootstrap program \cite{Shi:2019mlt,Lin:2023pvl,Kim:2024suq,Li:2025czz}, 
by extremizing certain entropic quantities on a relative small set of candidate states, one can remarkably find CFT ground states and moreover reconstruct the CFT spectrum. This is possible because of the relation between the CFT Hamiltonian and entanglement Hamiltonian on local regions \cite{Cardy:2016fqc}. In particular, the relevant entropic functional, as an expectation value for the entanglement Hamiltonians, is a linear combination of entanglement entropies on three adjacent intervals (in 1+1d) and its critical points satisfy certain vector fixed point equation which is the entropic analog of usual renormalization group fixed point equation \cite{Lin:2023pvl}. Crucially the vector fixed point equation imposes nontrivial locality conditions on the state, which is behind the magic of this novel bootstrap approach. 
While this current approach has the merit of producing universal CFT bounds, it is plausible that by incorporating symmetry information, much refined bounds can be obtained and potentially new CFTs can be identified with the help of symmetry constraints. In this context, the entropic order parameter is the natural way to implement the symmetry constraints. It would be very interesting to investigate the symmetry-refined entanglement bootstrap program using the framework developed here. 

 \section*{Acknowledgements}
We thank Matteo Dell'Acqua, Corey Jones, Robert Leigh, Brandon Rayhaun, Sahand Seifnashri, and Michael Stone for many useful discussions. The work of YW was supported in part by the NSF grant PHY-2210420 and by the Simons Junior Faculty Fellows program. The work of MSK was supported by the Heising-Simons foundation ``Observable Signatures of Quantum Gravity" collaboration and the Walter Burke Institute for Theoretical Physics. 
\appendix

\section{Glossary} \label{app: gloss}

In this appendix, we list the symbols, constructions, and notations that will appear frequently in the main text. We broadly organize this section based on categorical and algebraic data.
\subsection{Category}
\begin{itemize}
    \item $\mathcal{C}$ is a category, typically taken to be a fusion category. Its Drinfeld center is $Z(\mathcal{C})$. 
    \item $\text{Irr}(\mathcal{C})$ is the set of (isomorphism classes of) simple objects of the category. The notation $\text{Irr}(S)$ for some set $S$ will also be used, denoting irreducible representations for instance.
    \item Fusion coefficients $N_{UV}^W$: Multiplicities in the tensor product decomposition $U \otimes V \cong \bigoplus_W N_{UV}^W W$.  
    \item $\mathcal{M}$: A semisimple module category for $\mathcal{C}$.  
    \item $\mathrm{End}(\mathcal{M})$: Category of endomorphisms of $\mathcal{M}$.  
    \item Specific widely-used categories:
    \begin{itemize}
        \item $\mathrm{Fib}$: Fibonacci category.  
        \item $\mathrm{Vec}_G$: Category of $G$-graded vector spaces.  
    \end{itemize}
    \item $\mathrm{Rep}(A)$: Representation category of an algebra $A$. 
\end{itemize}
\subsection{Algebra}
There are a number of important algebras that can be associated to a fusion category $\mathcal{C}$:
\begin{itemize}
    \item $\mathrm{Fus}(\mathcal{C})$: Fusion algebra formed from simple objects of $\mathcal{C}$.  
    \item $\mathrm{Tub}(\mathcal{C})$: Tube algebra of $\mathcal{C}$. 
    \item $L^{2}(\mathcal{C})$ is the GNS Hilbert space associated to the tube algebra $\text{Tub}(\mathcal{C})$. 
    \item Given a choice of module $\mathcal{M}$, one also obtains the strip algebra $\text{Str}_{\mathcal{C}}(\mathcal{M})$. 
    \item $\mathcal{L}(\mathcal{C})$: C$^*$-completion of $\mathrm{Fus}(\mathcal{C})$, obtained via embedding into the universal $C^{*}$-completion of the tube algebra $\text{Tub}(\mathcal{C})$.  
\end{itemize}
To describe quantum systems, we also have
\begin{itemize}
    \item A system algebra $M$ represented on a Hilbert space $\mathscr{H}$. 
        \item The system $M$ may not carry the action $\Phi$ of the category in the same Hilbert space, and an extended system algebra $M_{\mathcal{M}}$ is constructed such that the strip algebra $\text{Str}_{\mathcal{C}}(\mathcal{M})$ acts by endomorphisms $\rho$.
    \item The endomorphism category $\text{End}(M)$ of a given algebra $M$ carries endomorphisms and intertwiners between maps $M \to M$. 

\end{itemize}
\subsection{Weak Hopf algebras}
By Tannaka-Krein duality, any fusion category $\mathcal{C}$ can be identified as $\text{Rep}(H)$ for a non-unique choice of weak Hopf algebra $H$. Associated to weak Hopf algebras are many relevant objects
\begin{enumerate}
    \item A weak Hopf algebra $H$ is determined by $(H, \mu , \eta, \Gamma, \epsilon, S)$, where $H$ is a bi-algebra with multiplication $\mu$ and unit $\eta$,  comultiplication $\Gamma$ and counit $\epsilon$, and antipode $S: H \to H$. Its dual is denoted $H^{*}$, and is also a weak Hopf algebra with elements being linear functionals $\hat{h}: H \to \mathbb{C}$ on $H$. 
    \item Two relevant subalgebras of $H$ are the target $H_{t}$ and source $H_{s}$ weak Hopf subalgebras, which are images of counital maps $\epsilon_{t}$ and $\epsilon_{s}$ respectively. 
    \item Inside of $H$ exist integral elements obeying appropriate invariance conditions. The left, right, and (normalized, non-degenerate) Haar integrals will be denoted $\ell, r,$~and $\Lambda \in H$ respectively. 
    %\item The vector space dimension of $H$ is $|H|$. We reserve $\text{dim}(H)$ for the dimension of the representation category $\text{Rep}(H)$. 
    \item A module for $H$ is denoted as $A$, acted on by $\rho: H \to \text{End}(A)$. The invariant subalgebra under the action is $A_{\rho}$. 
    \item An integral element, say $\ell \in H$, induces an invariantization map $T_{\rho} : A \to A^{\rho}$. If $T_{\rho}$ is finite on the unit, then it can be turned into a finite-index conditional expectation $E_{\rho}$ with index $\text{Ind}(E_{\rho})$. 
\end{enumerate}
\subsection{States, Entropies, and Modular Data}
Finally, since we are interested in entropic order parameters for non-invertible symmetry breaking patterns, we provide a brief list of relevant modular data:
\begin{enumerate}
    \item A state $\psi$ is a positive linear functional on an algebra $M$. If $M$ admits a trace $\tau$, there exists a positive, self adjoint operator $\rho_{\psi} \in M$ such that $\psi(x) = \tau(\rho_{\psi} x)$. 
    \item If $\pi: M \rightarrow B(\mathscr{H})$ is a representation of $M$, $\psi$ is a state on $M$ and $\Psi$ is a state on $\pi(M)'$ we denote by $\frac{d\Psi}{d\psi}$ the spatial derivative of $\Psi$ by $\psi$. This generalizes the Radon-Nikodym derivative of measure theory and the relative modular operator of Tomita-Takesaki theory.
    \item A conditional expectation $E: M \rightarrow N$ is a completely positive, unital and homogeneous map from a $C^*$ algebra $M$ to its $C^*$ subalgebra $N$.
    \item When $E: M \rightarrow N$ is a conditional expectation between von Neumann algebras, we define the Kosaki dual operator valued weight $E^{-1}: N' \rightarrow M'$ by the spatial condition $\frac{d(\Psi \circ E^{-1})}{d\psi} = \frac{d \Psi}{d(\psi \circ E)}$. 
    \item The relative entropy between a state $\psi$ and its composition with a conditional expectation $\psi \circ E$ is denoted $\Delta_{E} S(\psi) = S(\psi| \psi \circ E)$ and is the entropic order parameter of interest in this work. 
    %\item $\Delta_S^E(\psi)$: Entropic order parameter defined by relative entropy with respect to a conditional expectation $E$.  
    %\item $T_\rho$: Invariantization map induced by an integral of a weak Hopf algebra.  
    %\item $E_\rho$: Normalized conditional expectation associated with $T_\rho$.  
    \item $\text{Ind}(E)$: Index of a conditional expectation $E$. 
    \item Quasi-basis of $E$: Collection $\{u_i, v_i\}$ such that $a = \sum_i u_i E(v_i a)$ for all $a$.  
    \item Kosaki index: Definition of index via spatial theory for von Neumann algebra inclusions. $\text{Ind}(E) = E^{-1}(\mathbb{I})$.  
    \item Watatani index: Definition of index for inclusions of C$^*$-algebras, based on quasi-bases. $\text{Ind}(E) = \sum_{i} u_i v_i$. 
    \item Pimsner-Popa index: $\text{Ind}_{\rm PP}(E)$ is another definition for the index relevant for finite dimensions which is the reciprocal of $\lambda$ where $\lambda$ is the best constant for which this inequality holds on positive elements, $E(x) \geq \lambda x$.
\end{enumerate}

\section{Fusion Categories} \label{app: FusCat}
The full machinery of category theory is not necessary for the purposes of this note, and we refer to the reader to \cite{etingof2017fusioncategories} for a comprehensive discussion. Nevertheless, for completeness, we here briefly define the properties necessary to define a fusion category. A category $\mathcal{C}$ is called semisimple if any object can be decomposed as a direct sum of irreducible (simple) objects. A category $\mathcal{C}$ is called $\mathbb{C}$-linear if $\Hom{\mathcal{C}}{X}{Y}$ is a complex vector space for every $X,Y \in \Obj{\mathcal{C}}$ and the composition is bilinear. A strict monoidal category is a category $\mathcal{C}$ together with a distinguished object $\mathbb{I}$ called the unit object and a functor $\otimes: \mathcal{C} \times \mathcal{C} \rightarrow \mathcal{C}$ called the tensor product satisfying
	\begin{enumerate}
		\item $(X \otimes Y) \otimes Z = X \otimes (Y \otimes Z), \qquad (s \otimes t) \otimes u = s \otimes (t \otimes u)$,
		\item $X \otimes \mathbb{I} = \mathbb{I} \otimes X = X, \qquad s \otimes \text{id}_{\mathbb{I}} = \text{id}_{\mathbb{I}} \otimes s = s$,
		\item $(a \otimes b) \circ (c \otimes d) = (a \circ c) \otimes (b \circ d)$. 
	\end{enumerate}
	A non-strict monoidal category possesses a tensor product functor for which the above hold only up to natural isomorphism rather than as identities. In a monoidal category, we say that a pair of objects $(O_1,O_2)$ are dual if $O_1 \otimes O_2$ contains the trivial object $\mathbb{I}$ in its decomposition by simple objects. A category is called rigid if every object possesses a dual object. A fusion category is a semisimple, $\mathbb{C}$-linear, monoidal, rigid category with finitely many simple objects.
	
\section{Weak Hopf Algebras} \label{app: WHA}

Again, we refer the reader to \cite{BOHM1999385,etingof2017fusioncategories} for a complete discussion. A weak bialgebra is a tuple $(H, \mu ,\eta, \Gamma, \epsilon)$ based on a vector space $H$. The triple $(H, \mu, \eta)$ defines a unital algebra with $\mu : H \otimes H \to H$ a multiplication and $\eta : \mathbb{C} \rightarrow H$ the unit. Likewise, $(H, \Gamma, \epsilon)$ is a co-algebra with $\Gamma : H \to H \otimes H$ a coproduct and $\epsilon : H \to \mathbb{C}$ a counit. These elements satisfy weak compatibility conditions:
\begin{enumerate}
        \item $\Gamma \circ \mu =  (\mu \otimes \mu) \circ \left( \text{id}_{H} \otimes \sigma \otimes \text{id}_{H}\right) \circ (\Gamma \otimes \Gamma)$,
        \item $\epsilon \circ \mu \circ ( \mu \otimes \text{id}_{H}) = (\epsilon \otimes \epsilon) \circ ( \mu \otimes \mu)  \left( \text{id}_{H} \otimes \Gamma \otimes \text{id}_{H}\right)= (\epsilon \otimes \epsilon) \circ ( \mu \otimes \mu)  \left( \text{id}_{H} \otimes \Gamma' \otimes \text{id}_{H}\right)$,
        \item $(\Gamma \otimes \text{id}_{H}) \circ \Gamma \circ \eta = ( \text{id}_{H} \otimes \mu \otimes \text{id}_{H})\circ (\Gamma \otimes \Gamma) \circ (\eta \otimes \eta) =  ( \text{id}_{H} \otimes \mu' \otimes \text{id}_{H})\circ (\Gamma \otimes \Gamma) \circ (\eta \otimes \eta)$, 
\end{enumerate}
    where here $\sigma: H \otimes H \rightarrow H \otimes H$ is the swap operator, and $\Gamma' \equiv \sigma \circ \Gamma$ and $\mu' \equiv \mu \circ \sigma$ are the opposite coproduct and product, respectively. 

A weak Hopf algebra $(H, \mu , \eta, \Gamma, \epsilon)$ is a weak bialgebra along a linear map $S: H \to H$ called the antipode satisfying the following properties:
\begin{enumerate}
\item $\mu \circ (\text{id}_{H} \otimes S) \circ \Gamma = (\epsilon \otimes \text{id}_{H}) \circ (\mu \otimes \text{id}_{H}) \circ (\text{id}_{H} \otimes \sigma ) \circ (\Gamma \otimes \text{id}_{H}) \circ (\eta \otimes \text{id}_{H})  $, 
\item $\mu \circ ( S \otimes \text{id}_{H} ) \circ \Gamma = (\text{id}_{H} \otimes \epsilon ) \circ ( \text{id}_{H} \otimes \mu) \circ ( \sigma  \otimes \text{id}_{H}) \circ (\text{id}_{H} \otimes \Gamma ) \circ (\text{id}_{H} \otimes \eta )  $, 
\item $S = \mu \circ ( \mu \otimes \text{id}_{H}) \circ ( S \otimes \text{id}_{H} \otimes S) \circ (\Gamma \otimes \text{id}_{H}) \circ \Gamma$. 
\end{enumerate}

The difference between an ordinary Hopf algebra and a weak Hopf algebra is well encoded in an additional pair of maps $\epsilon_t,\epsilon_s: H \rightarrow H$ which are called the \emph{target} and \emph{source} counital maps. In particular we have
\beq
    \epsilon_t(h) \equiv \mu \circ (\text{id}_H \otimes S) \circ \Gamma, \qquad \epsilon_s(h) \equiv \mu \circ (S \otimes \text{id}_H) \circ \Gamma\,,
\eeq
which are precisely the expressions appearing in the first and second compatibility conditions defining the antipode. In the event that the coproduct $\Gamma$ is unital, e.g. $\Gamma(\mathbb{I}_H) = \mathbb{I}_H \otimes \mathbb{I}_H$, the target and source counits coincide with each other and with the overall counit $\epsilon$. This will be the case if and only if $H$ is an ordinary Hopf algebra. 

\subsection{Integral Theory}

The integral theory of finite weak Hopf algebras is very rich. It is well known that unimodular groups admit a unique Haar measure $\mu_{\rm G}$ which is bi-invariant and normalized 
\begin{equation}
    \mu_{\rm G} ( g g') = \mu_{\rm G}(g)= \mu_{\rm G}(g' g) , \qquad \mu_{\rm G}(e_G)=1, \quad \forall g,g' \in G\,.
\end{equation}
For a general weak Hopf algebra, there exists a notion dual to a Haar measure which is an element $k \in H$ obeying appropriate left and/or right invariance properties. Such an element is called a left/right integral. Respectively, left and right integral elements are dual to left and right invariant measures. An element of $H$ which is both a left and a right integral \emph{and} which is suitably normalized is called a Haar integral. A Haar integral is dual to a bi-invariant, normalized measure. In general, a Haar integral will not necessarily exist for a given finite weak Hopf algebra $H$. The existence of the Haar integral hinges intimately upon the properties of semisimplicity, and non-degeneracy which we will describe below.\footnote{A standard set of examples for finite non-semisimple Hopf algebras are Taft algebras \cite{taft}. The simplest such algebra is four dimensional and also known as the Sweedler algebra $H_4$ with generators $g,x$ which satisfy $g^2=1,x^2=0,gx=-xg$. The $H_4$ algebra was recently realized via certain non-invertible symmetries on spin chains \cite{Delcamp:2024rjp}. } In particular, any finite, semisimple weak Hopf algebra will always admit a normalizable left integral, while any finite, semisimple, non-degenerate weak Hopf algebra will always admit a Haar integral.

A left (right) integral $l$ ($r$) is an element of $H$ satisfying
\begin{equation}
     x l =  \epsilon_{t}(x) l, \qquad \bigg(rx =r \epsilon_{s}(x)\bigg), \qquad \forall x \in H\,.
\end{equation}
where $\epsilon_{t}$ and $\epsilon_{s}$ are the target and source counital maps, respectively. Moreover, the left (right) integral is normalized if
\begin{equation}
    \epsilon_{t}(l) = \mathbb{I}_H, \qquad \bigg(\epsilon_{s}(r) = \mathbb{I}_H\bigg)\,.
\end{equation}
It is important to note that $l \in H$ is a left integral if and only if $S(l)$, the antipodal image of $l$ in $H$, is itself a right integral~\cite{BOHM1999385}. An element $\Lambda \in H$ which is both a normalized left and right integral is called a \emph{Haar integral} for $H$.

The dual notion to a left (right) integral $l$ ($r$) of $H$ is a left (right) invariant measure on $H$, which is an element $\mu_{l}$ $(\mu_{r})$ in $H^{*}$ satisfying
\begin{equation}
    \left( \text{id}_H \otimes \mu_{l}\right)\circ \Gamma = \left( \epsilon_{t} \otimes \mu_{l}\right) \circ \Gamma \qquad \bigg(\left(\mu_{r} \otimes \text{id}_H \right) \circ \Gamma = \left( \mu_{r} \otimes \epsilon_{s} \right) \circ \Gamma\bigg) \,,
\end{equation}
with normalization condition
\begin{equation}
    \left( \text{id}_H \otimes \mu_{l}\right) \circ \Gamma(\mathbb{I}_H) = 1 \qquad  \bigg(\left( \mu_{r} \otimes \text{id}_H \right)\circ \Gamma(\mathbb{I}_H) = 1\bigg)\,.
\end{equation}
For the Hopf algebra associated with a finite group, these conditions reproduce the defining characteristics of a left (right) invariant Haar measure. 

We introduce the standard notation
\beq
	(h_1 \rightharpoonup \phi)[h_2] \equiv \phi(h_2 h_1), \qquad (\phi \leftharpoonup h_1)[h_2] \equiv \phi(h_1 h_2)\,,
\eeq
to denote the left and right actions of $H$ on $H^*$. Given $l \in H$ and $\hat{l} \in H^*$, a pair of left integrals for $H$ and $H^*$, we say that they are dual if
\beq
	\hat{l} \rightharpoonup l = \mathbb{I}_H, \qquad l \rightharpoonup \hat{l} = \mathbb{I}_{H^*} = \epsilon\,.
\eeq
If $\hat{l}$ is non-degenerate as a map $\hat{l}: H \rightarrow \mathbb{C}$, then the dual element $l$ is uniquely determined. We will call $(l, \hat{l})$ a dual pair of left integrals. 

Given a pair of dual bases, $\{e_i\}_{i \in \mathcal{I}} \subset H$ and $\{e^i\}_{i \in \mathcal{I}} \subset H^*$, the element
\begin{align}
    \hat{\ell} = \sum_{i} e^i \leftharpoonup S^{-2}(e_i)\,,
\end{align}
can be shown to be a left integral for $H^*$. It is called the \emph{canonical} left integral of $H^*$~\cite{nikshych2000finitequantumgroupoidsapplications}. We emphasize that the canonical left integral is valid for any finite weak Hopf algebra. This element can alternatively be written\footnote{For a finite weak Hopf algebra, we actually obtain $|H_{t}|$ independent left integrals of $H^{*}$ given by
\begin{equation}
    \hat{\ell}_i(x) := \text{Tr}\left[ L_{x} \circ S^{2} \rvert_{H S(q_{i})}\right]\,.
\end{equation}
Here, $q_{i} \in H_{t}$ are the primitive central idempotents of the algebra $H_t$. By an analogous argument we also obtain $|H_{s}|$ independent right integrals.}
\beq
    \hat{\ell}(x) \equiv \text{Tr}_H\left[L_{x} \circ S^{2} \rvert_{H} \right]\,,
\eeq
where $L_x(y) \equiv xy$ is interpreted as the regular representation of $H$ on itself. If $\hat{\ell}$ is non-degenerate as a map from $H$ to $\mathbb{C}$, we say that the weak Hopf algebra $H$ is non-degenerate.

As we alluded to above, the existence and uniqueness of a normalized left integral $l \in H$ is guaranteed by the semisimplicity of $H$. Hereafter, we work in the context of a finite semisimple weak Hopf algebra.\footnote{Note that every semisimple weak Hopf algebra is automatically finite by Maschke's theorem for weak Hopf algebras~\cite{etingof2017fusioncategories}.} Moreover, it can be shown that if $H$ is a semisimple weak Hopf, then so is $H^{*}$, and thus $H$ is both semisimple and \textit{cosemisimple}. In \cite{nikshych2000finitequantumgroupoidsapplications}, it is further shown that a Haar integral for $H$ exists if and only if $\hat{\ell}$ is non-degenerate, in which case $\Lambda$ is the dual left integral of $\hat{\ell}$. Thus, whenever $H$ admits a Haar integral it is determined uniquely by the conditions
\beq \label{Haar from Canonical}
	\hat{\ell} \rightharpoonup \Lambda = \mathbb{I}_H \qquad \Lambda \rightharpoonup \hat{\ell} = \mathbb{I}_{H^*} = \epsilon\,.
\eeq

In summary, the following are necessary and sufficient conditions for the existence of a Haar integral:
\begin{enumerate}
    \item $H$ is semisimple and nondegenerate,
	\item $H$ is biconnected and semisimple,
	\item $H$ is connected, semisimple, and admits an invertible element $g \in H$ such that $S^2(h) = g h g^{-1}$ and $\text{tr}_{U}(\pi_U(g^{-1})) \neq 0$ for all $U \in \text{IrrRep}(H)$. 
\end{enumerate}
The element $g \in H$ is \emph{grouplike} in the sense that $\Gamma(g) = (g \otimes g) \Gamma(\mathbb{I}_H) = \Gamma(\mathbb{I}_H)(g \otimes g)$, and is referred to as the \emph{canonical grouplike element} of $H$. In terms of the canonical grouplike element, we can also obtain a different formulation of the canonical left integral. Let $\tau: H \rightarrow \mathbb{C}$ be a tracial map (e.g. $\tau[h_1 h_2] = \tau[h_2 h_1]$) such that
\beq
	\tau[\mathbb{I}_U] = \text{tr}_{U}( \pi_U(g^{-1})), \qquad \forall U \in \text{IrrRep}(H)\,.
\eeq	
Then, $\hat{\ell} = g \rightharpoonup \tau$. In somewhat more explicit notation:
\beq
	\hat{\ell}[h] = \sum_{i \in \mathcal{I}} e^i[g^{-1} e_i g h] = \tau[hg]\,.
\eeq	

By permuting the above arguments, we can also write down the canonical left integral for $H$ as\footnote{$\hat{S}$ is the antipodal map on $H^*$.}
\beq \label{Canonical Left Integral of H}
	\lambda \equiv \sum_{i \in \mathcal{I}} e_i \leftharpoonup \hat{S}^{-2}(e^i)\,.
\eeq
In the case that $H^*$ admits a Haar integral there will exist a grouplike element $\hat{g} \in H^*$ such that $\hat{S}^2(\phi) = \hat{g} \phi \hat{g}^{-1}$. Then,
\beq
	\lambda[\phi] = \sum_{i \in \mathcal{I}} e_i[\hat{g}^{-1} e^i \hat{g} \phi] = \hat{\tau}[\phi \hat{g}]\,,
\eeq
where $\hat{\tau}$ is a trace on $H^*$ satisfying 
\beq
	\hat{\tau}[\mathbb{I}_V] = \text{tr}_V(\pi_V(\hat{g}^{-1})), \qquad \forall V \in \text{IrrRep}(H^*)\,.
\eeq
Given the Haar integral $\Lambda \in H$ and the action $\phi \in H^* \mapsto h \hookrightarrow \phi \in H^*$ we can define a condition expectation $E_{\hookrightarrow}: H^* \rightarrow H^*_{\hookrightarrow}$ given by $\phi \mapsto \Lambda \hookrightarrow \phi$. This conditional expectation is of finite index type and possesses a quasi-basis $Q = \hat{\ell}_{(2)} \otimes \hat{S}^{-1}(\hat{\ell}_{(1)})$. Here $\hat{\ell}$ is the canonical left integral on $H^*$ and $\hat{\Gamma}(\hat{\ell}) = \hat{\ell}_{(1)} \otimes \hat{\ell}_{(2)}$ in Sweedler notation. The index of this conditional expectation is therefore
\beq
    \text{Ind}(E_{\hookrightarrow}) = \hat{\ell}_{(2)} \hat{S}^{-1}(\hat{\ell}_{(1)})\,.
\eeq

If $\rho: H \times A \rightarrow A$ is an action on a general $H$-module algebra $A$ we also obtain a conditional expectation $E_{\rho}: A \rightarrow A_{\rho}$ given by $x \mapsto \rho_{\Lambda}(x)$. For a generic $H$-module algebra we define the subalgebra $A_R \equiv \rho_H(\mathbb{I}_A) \subset A$. In this way, we obtain a homomorphism $\theta_{\rho}: H^*_R \rightarrow A_R$ such that $h \hookrightarrow \mathbb{I}_{H^*} \mapsto \rho_h(\mathbb{I}_A)$. The index of the conditional expectation $E_{\rho}$ is bounded by $\theta_{\rho}(\text{Ind}E_{\hookrightarrow})$. It is equivalent to this quantity in the event that $\rho$ is both standard and properly outer. Thus, we have the bound
\beq
    \text{Ind}E_{\rho} \leq \theta_{\rho}\bigg(\hat{\ell}_{(2)} \hat{S}^{-1}(\hat{\ell}_{(1)})\bigg)\,.
\eeq

To summarize, we will here provide component expressions for the various integral structures we have described in terms of a particular chosen basis. Let $H$ be an arbitrary finite, semisimple, non-degenerate weak Hopf algebra with dual $H^*$ and dual bases $\{e_i\}$ and $\{e^i\}$. We will use the following notations
\begin{flalign} \label{Characterization of Integral Element}
    & S^{-1}(e_i) = (S^{-1})^j_i e_j, \nonumber \\
    & \mu(e_i \otimes e_j) = \mu_{ij}^k e_k, \nonumber \\
    & \hat{S}^{-1}(e^i) = (\hat{S}^{-1})^i_j e^j, \nonumber \\
    & \hat{\Gamma}(e^i) = \hat{\Gamma}^i_{jk} e^j \otimes e^k, \nonumber \\
    & \hat{\mu}(e^i \otimes e^j) = \hat{\mu}^{ij}_k e^k, \nonumber \\
    & \mathbb{I}_H = \mathbb{I}_H^i e_i
\end{flalign}
Then, we have the following formulae:\footnote{The notation $(\hat{\mu}_{\rightharpoonup}^{-1})^{ij}_k$ deserves some explanation. These components arise when solving \eqref{Haar from Canonical} to obtain the Haar integral. In components, the action, $\rightharpoonup$, of $H^*$ on $H$ is given by $e^i \rightharpoonup e_j = \hat{\mu}^{ki}_j e_k$. We view $\hat{\mu}^{ki}_j$ as the matrix components of a linear map from $H$ to $H$ for each fixed index $i$. The notation $(\hat{\mu}^{-1}_{\rightharpoonup})^{ki}_j$ simply means that we invert the matrix $\hat{\mu}^{ki}_j$ for each fixed $i$. We include the subscript $\rightharpoonup$ to remind that we are inverting with $i$ fixed, as opposed to inverting with $k$ fixed as would be the case for the action $\leftharpoonup$. The Haar integral is obtained by inverting $\hat{\ell} \rightharpoonup \Lambda = \mathbb{I}_H$. Thus, we can write $\Lambda^i = \hat{\ell}_j (\hat{\mu}^{-1}_{\rightharpoonup})^{ij}_k \mathbb{I}_H^k$, where $\mathbb{I}_H^i$ are the components of $\mathbb{I}_H \in H$ with respect to the chosen basis.}
\begin{flalign} \label{Component Expressions}
    &\hat{\ell} = (S^{-1})^{l}_k (S^{-1})^{k}_i \mu^i_{lm} e^m, \nonumber \\
    &\Lambda = (S^{-1})^{l}_k (S^{-1})^{k}_i \mu^i_{lm} (\hat{\mu}_{\rightharpoonup}^{-1})^{pm}_q \mathbb{I}_H^q e_p, \nonumber \\
    &Q = (S^{-1})^{l}_k (S^{-1})^k_i \mu^i_{lm} \hat{\Gamma}^{m}_{pq} (\hat{S}^{-1})^p_r e^q \otimes e^r, \nonumber \\
    &\text{Ind}(E_{\rightharpoonup}) = (S^{-1})^l_k (S^{-1})^k_i \mu^i_{lm} \hat{\Gamma}^m_{pq} (\hat{S}^{-1})^p_r \hat{\mu}^{qr}_s e^s\,.
\end{flalign}
Eqn. \eqref{Component Expressions} illustrate the steps that can be undertaken in order to define the left integral, Haar integral, quasibasis and index. The coefficients defined in \eqref{Characterization of Integral Element} are structural data defining the weak Hopf algebra as represented in a particular basis and are easily obtained for any given example. In Appendix~\ref{app: strip} we evaluate these formulae for the strip algebra, $\text{Str}_{\mathcal{C}}(\mathcal{M})$. 
\subsection{C$^*$ Weak Hopf Algebras}

Every finite, C$^*$ weak Hopf algebra possesses a unique Haar integral $\Lambda \in H$. To construct this element, let us first note that the algebra $H_t$ can be regarded as the trivial representation of $H$ via the GNS construction relative to the counital map. In particular\footnote{Note that $\epsilon$ is not a faithful weight on $H$. In the GNS construction we quotient by the kernel of $\epsilon$ which is why the GNS Hilbert space of $H$ with respect to $\epsilon$ is isomorphic to $H_t$. In the case of a bonafide Hopf algebra, $H_t = H_s = \mathbb{C}$ and we see that the trivial representation is just $\mathbb{C}$.},
\beq
	\epsilon(h_1^* h_2) = \epsilon(\epsilon_t(h_1)^* \epsilon_t(h_2)), \qquad \forall h_1, h_2 \in H\,,
\eeq
and thus we can promote $H_t$ to a Hilbert space with inner product $\langle h_1^t, h_2^t \rangle_{H_t} \equiv \epsilon(h_1^t{}^* h_2^t)$. The action of $H$ on $H_t$ is given simply by $\pi_{H_t}(h_1) \epsilon_t(h_2) \equiv \epsilon_t(h_1 h_2)$. 

Although $H_t$ is the trivial representation, it is not, in general, irreducible. The irreducible decomposition of $H_t$ is given by
\beq \label{Decomposition of H_t}
	H_t = \bigoplus_{U \in Z \subset \text{IrrRep}(H)} U\,,
\eeq
where here $Z$ is the set of irreps appearing in $H_t$.\footnote{All irreps that appear in $H_t$ do so with multiplicity $1$.}

Let $\{e^t_i\}_{i \in \mathcal{I}}$ be an orthonormal basis for $H_t$ e.g. $\langle e^t_i, e^t_j \rangle_{H_t} = \delta_{ij}$. In fact, $e^s_i \equiv S(e^t_i)^*$ forms an orthonormal basis for $H_s$. Likewise, let us denote by $\{e_t^i\}$ and $\{e_s^i\}$ dual orthonormal bases for $\hat{H}_t$ and $\hat{H}_s$.\footnote{Here, we are using the notation $\hat{H}$ to refer to the dual weak Hopf algebra of $H$.} These bases may be chosen to respect the decomposition \eqref{Decomposition of H_t} such that, for example, the quantity $M^{ij} \equiv \hat{e}_t^i \hat{e}_s^j \in \hat{H}$ is zero unless $i$ and $j$ are representation indices of a common $U \in Z$. For each $U \in Z$ we can choose a dual element $E_{Uij} \in H$ such that $M^{ij}(E_{Ukl}) = \delta^i_k \delta^j_l \delta_{k \in U} \delta_{l \in U}$. With these notations in place, the Haar integral can be written 
\beq
	\Lambda = \sum_{U \in Z} \sum_{i,j  \in U} \frac{\epsilon(e^t_i)\epsilon(e^s_j)}{\epsilon_U(\mathbb{I}_H)} E_{Uij}\,.
\eeq
The elements $E_{Uij}$ can be regarded as matrix units for the trivial block of $H_U$, within the full block decomposition of $H$. The trivial block of $H_U$ is the subalgebra whereupon $\epsilon\rvert_{H_U}$ further restricts to a pure weight.\footnote{Recall that a weight on a C$^*$-algebra is called pure if it cannot be rewritten as a non-trivial convex combination of other weights.} Here, $\epsilon_U$ is the restriction of the counit to this trivial block. 

Given an action $\rho: H \rightarrow \text{End}(A)$, the condition expectation we constructed can therefore be written in the form
\beq
	E_{\rho}(x) = \sum_{U \in Z} \sum_{i,j \in U} \frac{\epsilon(e^t_i) \epsilon(e^s_j)}{\epsilon_U(\mathbb{I}_H)} \rho_{E_{Uij}}(x)\,,
\eeq
which is the generalization of the group averaging map. In particular, consider $H = \mathcal{L}(G)$ for $G$ a finite group. In this case, the target and base counital maps coincide and are equal to $\mathbb{C}$, the trivial representation which is irreducible. The only matrix unit which is present is $E_{\mathbb{C}} = \frac{1}{|G|} \sum_{g \in G} \ell(g)$ which is the projection of $\mathcal{L}(G)$ to the symmetric sector. In the case of a weak Hopf algebra, one can think of $(U,i,j)$ as labeling the distinct `vacua' with $E_{Uij}$ associated projection operators onto the appropriate symmetric sector of each vacuum. This is a useful point of view for understanding, physically, the role of the module category in Tannaka duality which fixes the target and base counital algebras and in turn the complete `vacuum' structure of the symmetry.

\section{WHA as Strip Algebras} \label{app: strip}

As we saw in the main text, the physical incarnations of WHA symmetry encountered here are naturally realized by strip algebras that act on 1+1D systems with boundaries. The axiomatic properties of WHAs have simple origins from the perspective of strip algebras. Here we give a brief review of 
the key ingredients and set up our conventions.

Given a fusion category $\cC$ and an indecomposable left module category $\cM$ over $\cC$, the strip algebra ${\rm Str}_\cC(\cM)$ is a C$^*$ bi-connected 
WHA with the following basis elements
 \ie
 v\otimes w^*=
\begin{gathered}
\begin{tikzpicture}[scale=1]
\fill [fill=gray!20] (-1.8,-1) rectangle (-1,1);
\fill [fill=gray!20] (0,-1) rectangle (.8,1);
\draw [line,-<-=.55] (-1,-1) -- (-1,0);
\draw (-1,-0.6) node [left] {$a$};
\draw [line,-<-=.55] (-1,0) -- (-1,1);
\draw (-1,0.6) node [left] {$c$};
\draw [line,->-=.55] (0,-1) -- (0,0);
\draw (0,-0.6) node [right] {$b$};
\draw [line,->-=.55] (0,0) -- (0,1);
\draw (0,0.6) node [right] {$d$};
\draw [red,line,-<-=.55] (0,0) -- (-1,0);
\draw (-0.5,0) node [below] {$L$};
\filldraw[black] (-1,0) circle (1pt) node[left] {\scriptsize $v$};
\filldraw[black] (0,0) circle (1pt) node[right] {\scriptsize $w^*$};
\end{tikzpicture}
\end{gathered} \,, \quad H= {\rm Str}_\cC(\cM)=\bigoplus_{a,b,c,d,L} V^L_{c,a}\otimes  (V^L_{d,b})^*\,,
\label{stripgens}
\fe
where $a,b,c,d$ label simple objects in $\cM$, $L$ label simple objects in $\cC$ and $v,w$ are basis vectors for the topological junction vector spaces $V^L_{c,a}$ and $V^L_{d,b}$. Whenever a WHA is given a strip algebra presentation, we refer to the basis above as its corresponding strip algebra basis. 

Physically, $\cC$ describes the topological defects in the 1+1D system $T_\cC$ (in the absence of boundaries), while $\cM$ captures the additional symmetry structure of the system in the presence of boundaries. In particular $\cM$ determines the multiplet of boundary conditions of $T_\cC$ transforming under $\cC$ by fusion. Due to the relation between boundary conditions and bulk states ($e.g.$ from the Cardy condition), $\cM$ can be thought of as a refined label for $T_\cC$.  
In the case of the regular module category $\cM=\cC$, the boundary multiplet is in one-to-one correspondence with the bulk topological defects and there is a distinguished boundary, known as the identity Cardy brane, which corresponds to the identity (trivial) topological defect. For a general module category $\cM$, a similar identification exists, between the boundary multiplet and the set of topological interfaces between the given system $T_\cC$ and its cousin $T'_{\cC'}$ after a generalized gauging by $\cM$ \cite{Diatlyk:2023fwf}. Here $\cC'$ is the dual fusion category and in this identification, we take the boundary of $T'_{\cC'}$ to be the identity Cardy brane for $\cC'$ and the boundary multiplet of $\cT_\cC$ arises from fusing the topological interfaces with this identity brane. 
We thus have the following relation,
\ie 
\begin{gathered}
\begin{tikzpicture}
\draw (-1,1) node [above] {$\cM$};
\fill [fill=gray!20] (-1.8,-1) rectangle (-1,1);
\draw [line,-<-=.55] (-1,-1) -- (-1,0);
\draw (-1,-0.4) node [left] {$a$};
\draw [line,-<-=.55] (-1,0) -- (-1,1);
\draw (-1,0.4) node [left] {$c$};
\draw [red,line,-<-=.55] (0,0) -- (-1,0);
\draw (-0.5,0) node [below] {$L$};
\filldraw[black] (-1,0) circle (1pt) node[left] {\scriptsize $v$};
\draw [red] (-1,0.8) node [right] {$T_\cC$};
\end{tikzpicture}
\end{gathered}
=\,
\begin{gathered}
\begin{tikzpicture}[scale=1]
\draw [red] (-1,0.8) node [right] {$T_\cC$};
\draw (-1,1) node [above] {$\cM$};
\draw (-1.6,1) node [above] {$\cC'$};
\fill [fill=gray!20] (-2.2,-1) rectangle (-1.6,1); 
\fill [fill=green!20] (-1.6,-1) rectangle (-1,1);
\draw (-1,-0.4) node [left] {$a$};
\draw [line,blue,-<-=.55] (-1,0) -- (-1,1);
\draw [line,blue,-<-=.55] (-1,-1) -- (-1,0);
\draw (-1,0.4) node [left] {$c$};
\draw [red,line,-<-=.55] (0,0) -- (-1,0);
\draw (-0.5,0) node [below] {$L$};
\filldraw[black] (-1,0) circle (1pt) node[left] {\scriptsize $v$};
\draw [dotted] (-1.6,-1) -- (-1.6,1);
\draw [red] (-1.68,0.81) node [right] {$T'_{\cC'}$};
\end{tikzpicture}
\end{gathered}\,,
\label{peelrel}
\fe 
where the blue lines represent the topological interfaces and the dotted line represents the identity brane for $T'_{\cC'}$.

Here we do not assume that the 1+1D theory is conformal or topological. To deduce various structure of the algebra by topological moves, we will find the relation \eqref{peelrel} useful, since the blue and red defects are topological. 

In particular, a bubble of the topological interface $a$ determines its quantum dimension $d_a$ as below,
\ie 
d_a=\begin{gathered}
\begin{tikzpicture}[scale=.7]
\begin{scope}[even odd rule]
\clip (0,0) circle (1)  (-1.4,-1.4) rectangle (1.4,1.4);
\fill [fill=green!20] (-1.4,-1.4) rectangle (1.4,1.4);
\end{scope}
\useasboundingbox (-1.3,-1.2) rectangle (1.3,1.4); 
 \draw (0,1) node [below] {$a$}; 
\draw [line,blue,->-=.27,->-=.77] (0,0) circle (1);
\end{tikzpicture}
\end{gathered}\,.
\label{qdimint}
\fe

The strip algebra generators in \eqref{stripgens} encode the action of  the topological defect $L$ with topological endpoints, implemented on the Hilbert space on a strip. The multiplication of strip algebra elements is obviously given by
\ie
\begin{gathered}
\begin{tikzpicture}[scale=1]
\fill [fill=gray!20] (-1.8,-1) rectangle (-1,1);
\fill [fill=gray!20] (0,-1) rectangle (.8,1);
\draw [line,-<-=.55] (-1,-1) -- (-1,0);
\draw (-1,-0.6) node [left] {$a_1$};
\draw [line,-<-=.55] (-1,0) -- (-1,1);
\draw (-1,0.6) node [left] {$c_1$};
\draw [line,->-=.55] (0,-1) -- (0,0);
\draw (0,-0.6) node [right] {$b_1$};
\draw [line,->-=.55] (0,0) -- (0,1);
\draw (0,0.6) node [right] {$d_1$};
\draw [red,line,-<-=.55] (0,0) -- (-1,0);
\draw (-0.5,0) node [below] {$L_1$};
\filldraw[black] (-1,0) circle (1pt) node[left] {\scriptsize $v_1$};
\filldraw[black] (0,0) circle (1pt) node[right] {\scriptsize $w_1^*$};
\end{tikzpicture}
\end{gathered} 
\times 
\begin{gathered}
\begin{tikzpicture}[scale=1]
\fill [fill=gray!20] (-1.8,-1) rectangle (-1,1);
\fill [fill=gray!20] (0,-1) rectangle (.8,1);
\draw [line,-<-=.55] (-1,-1) -- (-1,0);
\draw (-1,-0.6) node [left] {$a_2$};
\draw [line,-<-=.55] (-1,0) -- (-1,1);
\draw (-1,0.6) node [left] {$c_2$};
\draw [line,->-=.55] (0,-1) -- (0,0);
\draw (0,-0.6) node [right] {$b_2$};
\draw [line,->-=.55] (0,0) -- (0,1);
\draw (0,0.6) node [right] {$d_2$};
\draw [red,line,-<-=.55] (0,0) -- (-1,0);
\draw (-0.5,0) node [below] {$L_2$};
\filldraw[black] (-1,0) circle (1pt) node[left] {\scriptsize $v_2$};
\filldraw[black] (0,0) circle (1pt) node[right] {\scriptsize $w_2^*$};
\end{tikzpicture}
\end{gathered}
=
\begin{gathered}
\begin{tikzpicture}[scale=1]
\fill [fill=gray!20] (-1.8,-1) rectangle (-1,1);
\fill [fill=gray!20] (0,-1) rectangle (.8,1);
\draw [line,-<-=.55] (-1,-1) -- (-1,-.35);
\draw [line,-<-=.55] (-1,-.35) -- (-1,.35);
\draw [line,-<-=.55] (-1,.35) -- (-1,1);
\draw [line,->-=.55] (0,-1) -- (0,-.35);
\draw [line,->-=.55] (0,-.35) -- (0,.35);
\draw [line,->-=.55] (0,.35) -- (0,1);
\draw (-1,-0.7) node [left] {$a_2$};
\draw (-1,0.) node [left] {$a_1$};
\draw (0,0.) node [right] {$b_1$}; 
\draw (-1,0.7) node [left] {$c_1$}; 
\draw (0,-0.7) node [right] {$b_2$}; 
\draw (0,0.7) node [right] {$d_1$};
\draw [red,line,-<-=.55] (0,.35) -- (-1,.35);
\draw [red,line,-<-=.55] (0,-.35) -- (-1,-.35);
\draw (-0.5,-.35) node [below] {$L_2$};
\draw (-0.5,.35) node [above] {$L_1$};
\filldraw[black] (-1,0.35) circle (1pt) node[left] {\scriptsize $v_1$};
\filldraw[black] (0,0.35) circle (1pt) node[right] {\scriptsize $w_1^*$};
\filldraw[black] (-1,-0.35) circle (1pt) node[left] {\scriptsize $v_2$};
\filldraw[black] (0,-0.35) circle (1pt) node[right] {\scriptsize $w_2^*$};
\end{tikzpicture}
\end{gathered} \D_{a_1,c_2}\D_{b_1,d_2}\,,
\fe
and the RHS can be further reduced to a linear combination of the basis elements in \eqref{stripgens} using F-moves and the F-symbols for $\cC$ and $\cM$. The multiplicative unit is given by 
\ie 
\id =\sum_{a,b} \begin{gathered}
\begin{tikzpicture}[scale=1]
\fill [fill=gray!20] (-1.8,-1) rectangle (-1,1);
\fill [fill=gray!20] (0,-1) rectangle (.8,1);
\draw [line,-<-=.55] (-1,-1) -- (-1,1);
\draw (-1,0) node [left] {$a$};
\draw [line,->-=.55] (0,-1) -- (0,1);
\draw (0,0) node [right] {$b$}; 
\end{tikzpicture}
\end{gathered}\,. 
\fe

The co-multiplication in the strip algebra defines how the topological defect acts on a tensor product of states in the strip Hilbert space, 
\ie 
\Delta 
\left(
\begin{gathered}
\begin{tikzpicture}[scale=1]
\fill [fill=gray!20] (-1.8,-1) rectangle (-1,1);
\fill [fill=gray!20] (0,-1) rectangle (.8,1);
\draw [line,-<-=.55] (-1,-1) -- (-1,0);
\draw (-1,-0.6) node [left] {$a$};
\draw [line,-<-=.55] (-1,0) -- (-1,1);
\draw (-1,0.6) node [left] {$c$};
\draw [line,->-=.55] (0,-1) -- (0,0);
\draw (0,-0.6) node [right] {$b$};
\draw [line,->-=.55] (0,0) -- (0,1);
\draw (0,0.6) node [right] {$d$};
\draw [red,line,-<-=.55] (0,0) -- (-1,0);
\draw (-0.5,0) node [above] {$L$};
\filldraw[black] (-1,0) circle (1pt) node[left] {\scriptsize $v$};
\filldraw[black] (0,0) circle (1pt) node[right] {\scriptsize $w^*$};
\end{tikzpicture}
\end{gathered} 
\right) 
=
 \sum_{e,f,u} d_f\begin{gathered}
\begin{tikzpicture}[scale=1]
\fill [fill=gray!20] (-1.8,-1) rectangle (-1.2,1);
\fill [fill=gray!20] (-.8,-1) rectangle (-.2,1);
\fill [fill=gray!20] (0.2,-1) rectangle (.8,1); 
\draw [line, -<-=.31,-<-=.81] (-.2,-1) -- (-.2,1);
\draw [line, ->-=.33,->-=.83] (-.8,-1) -- (-.8,1);
\draw (-.5,0.6) node [] {$f$};
\draw (-.5,-.6) node [] {$e$};
\draw [line,-<-=.55] (-1.2,-1) -- (-1.2,0.2);
\draw (-1.2,-0.6) node [left] {$a$};
\draw [line,-<-=.55] (-1.2,0.2) -- (-1.2,1);
\draw (-1.2,0.6) node [left] {$c$};
\draw [line,->-=.55] (0.2,-1) -- (0.2,0.2);
\draw (0.2,-0.6) node [right] {$b$};
\draw [line,->-=.55] (0.2,0.2) -- (0.2,1);
\draw (0.2,0.6) node [right] {$d$};
\draw [red,line,->-=.65] (-1.2,0.) -- (-.8,0.);
\draw [red,line,->-=.65] (-.2,0) -- (0.2,0); 
\draw (-.95,0.2) node [] {\scriptsize $u$};
\draw (0,0.25) node [] {\scriptsize $u^*$};
\filldraw[black] (-1.2,0.) circle (1pt) node[left] {\scriptsize $v$};
\filldraw[black] (0.2,0.) circle (1pt) node[right] {\scriptsize $w^*$}; 
\filldraw[black] (-.8,0.) circle (1pt);
\filldraw[black] (-.2,0.) circle (1pt);
\end{tikzpicture}
\end{gathered}
\left(\begin{gathered}
\begin{tikzpicture}[scale=.7]
\begin{scope}[even odd rule]
\clip (0,0) circle (1)  (-1.4,-1.4) rectangle (1.4,1.4);
\fill [fill=green!20] (-1.4,-1.4) rectangle (1.4,1.4);
\end{scope}
\useasboundingbox (-1.3,-1.2) rectangle (1.3,1.4); 
 \draw (0,1) node [below] {$f$};
 \draw (0,-1) node [below=-.05] {$e$};
  \draw (0,0) node [below] {$L$};
\draw [line,blue,->-=.27,->-=.77] (0,0) circle (1);
\draw [line,->-=.53,red] (-1,0) to (1,0);
\filldraw[black] (-1,0.) circle (1pt) node[above left=-.2] {\scriptsize $u^*$};
\filldraw[black] (1,0) circle (1pt) node[below right=-.1] {\scriptsize $u$}; 
\end{tikzpicture}
\end{gathered}\right)^{-1}\,, 
\fe 
which comes from the following topological move,
\ie 
\sum_e \begin{gathered}
\begin{tikzpicture}[scale=1]
\fill [fill=gray!20] (-1.8,-1) rectangle (-1.2,1);
\fill [fill=gray!20] (0.2,-1) rectangle (.8,1); 
\draw [line,blue,->-=.29,->-=.79,fill=green!20] plot [smooth] coordinates {(-.8,-1) (-.7,-.1)  (-.3,-.1) (-.2,-1)};
\draw (-.5,-.2) node [below] {$e$};
\draw [line,-<-=.55] (-1.2,-1) -- (-1.2,0.2);
\draw (-1.2,-0.6) node [left] {$a$};
\draw [line,-<-=.55] (-1.2,0.2) -- (-1.2,1);
\draw (-1.2,0.6) node [left] {$c$};
\draw [line,->-=.55] (0.2,-1) -- (0.2,0.2);
\draw (0.2,-0.6) node [right] {$b$};
\draw [line,->-=.55] (0.2,0.2) -- (0.2,1);
\draw (0.2,0.6) node [right] {$d$};
\draw [red,line,-<-=.55] (0.2,0.2) -- (-1.2,0.2);
\draw (-0.5,0.2) node [above] {$L$};
\filldraw[black] (-1.2,0.2) circle (1pt) node[left] {\scriptsize $v$};
\filldraw[black] (0.2,0.2) circle (1pt) node[right] {\scriptsize $w^*$};
\end{tikzpicture}
\end{gathered} 
=
 \sum_{e,f,u} d_f\begin{gathered}
\begin{tikzpicture}[scale=1]
\fill [fill=gray!20] (-1.8,-1) rectangle (-1.2,1);
\fill [fill=gray!20] (0.2,-1) rectangle (.8,1); 
\draw [line,->-=.19,->-=.39,->-=.66,->-=.86,blue,fill=green!20] plot [smooth] coordinates {(-.8,-1) (-.75,.5)    (-.25,.5) (-.2,-1)};

\draw (-.5,0.65) node [below] {$f$};
\draw (-.5,-.5) node [below] {$e$};
\draw [line,-<-=.55] (-1.2,-1) -- (-1.2,0.2);
\draw (-1.2,-0.6) node [left] {$a$};
\draw [line,-<-=.55] (-1.2,0.2) -- (-1.2,1);
\draw (-1.2,0.6) node [left] {$c$};
\draw [line,->-=.55] (0.2,-1) -- (0.2,0.2);
\draw (0.2,-0.6) node [right] {$b$};
\draw [line,->-=.55] (0.2,0.2) -- (0.2,1);
\draw (0.2,0.6) node [right] {$d$};
\draw [red,line,->-=.65] (-1.2,0.) -- (-.8,0.);
\draw [red,line,->-=.65] (-.2,0) -- (0.2,0); 
\draw (-.95,0.2) node [] {\scriptsize $u$};
\draw (0,0.25) node [] {\scriptsize $u^*$};
\filldraw[black] (-1.2,0.) circle (1pt) node[left] {\scriptsize $v$};
\filldraw[black] (0.2,0.) circle (1pt) node[right] {\scriptsize $w^*$}; 
\filldraw[black] (-.8,0.) circle (1pt);
\filldraw[black] (-.2,0.) circle (1pt);
\end{tikzpicture}
\end{gathered}
\left(\begin{gathered}
\begin{tikzpicture}[scale=.7]
\begin{scope}[even odd rule]
\clip (0,0) circle (1)  (-1.4,-1.4) rectangle (1.4,1.4);
\fill [fill=green!20] (-1.4,-1.4) rectangle (1.4,1.4);
\end{scope}
\useasboundingbox (-1.3,-1.2) rectangle (1.3,1.4); 
 \draw (0,1) node [below] {$f$};
 \draw (0,-1) node [below=-.05] {$e$};
  \draw (0,0) node [below] {$L$};
\draw [line,blue,->-=.27,->-=.77] (0,0) circle (1);
\draw [line,->-=.53,red] (-1,0) to (1,0);
\filldraw[black] (-1,0.) circle (1pt) node[above left=-.2] {\scriptsize $u^*$};
\filldraw[black] (1,0) circle (1pt) node[below right=-.1] {\scriptsize $u$}; 
\end{tikzpicture}
\end{gathered}\right)^{-1}\,,
\fe 
which ensures the multiplicativity of the coproduct
\ie 
\Delta(h_1h_2)=\Delta(h_1)\Delta(h_2)\,,\quad \forall h_1,h_2\in H\,.
\fe

The 
counit is 
\ie \label{counit}
\epsilon
\left(
\begin{gathered}
\begin{tikzpicture}[scale=1]
\fill [fill=gray!20] (-1.8,-1) rectangle (-1,1);
\fill [fill=gray!20] (0,-1) rectangle (.8,1);
\draw [line,-<-=.55] (-1,-1) -- (-1,0);
\draw (-1,-0.6) node [left] {$a$};
\draw [line,-<-=.55] (-1,0) -- (-1,1);
\draw (-1,0.6) node [left] {$c$};
\draw [line,->-=.55] (0,-1) -- (0,0);
\draw (0,-0.6) node [right] {$b$};
\draw [line,->-=.55] (0,0) -- (0,1);
\draw (0,0.6) node [right] {$d$};
\draw [red,line,-<-=.55] (0,0) -- (-1,0);
\draw (-0.5,0) node [above] {$L$};
\filldraw[black] (-1,0) circle (1pt) node[left] {\scriptsize $v$};
\filldraw[black] (0,0) circle (1pt) node[right] {\scriptsize $w^*$};
\end{tikzpicture}
\end{gathered} 
\right) 
=
{\D_{a,b} \D_{c,d}  \over d_d}
\begin{gathered}
\begin{tikzpicture}[scale=.7]
\begin{scope}[even odd rule]
\clip (0,0) circle (1)  (-1.4,-1.4) rectangle (1.4,1.4);
\fill [fill=green!20] (-1.4,-1.4) rectangle (1.4,1.4);
\end{scope}
\useasboundingbox (-1.3,-1.2) rectangle (1.3,1.4); 
 \draw (0,1) node [below] {$c$};
 \draw (0,-1) node [below=-.05] {$a$};
  \draw (0,0) node [below] {$L$};
\draw [line,blue,->-=.27,->-=.77] (0,0) circle (1);
\draw [line,->-=.53,red] (-1,0) to (1,0);
\filldraw[black] (-1,0.) circle (1pt) node[above left=-.1] {\scriptsize $v$};
\filldraw[black] (1,0) circle (1pt) node[below right=-.15]  {\scriptsize $w^*$}; 
\end{tikzpicture}
\end{gathered}\,,
\fe 
which in CFT (and TQFT) captures how the algebra element act on the ground states in the strip Hilbert space. 

The antipode $S$, is a linear algebra anti-homomorphism, defined as in 
\ie
S
\left(
\begin{gathered}
\begin{tikzpicture}[scale=1]
\fill [fill=gray!20] (-1.8,-1) rectangle (-1,1);
\fill [fill=gray!20] (0,-1) rectangle (.8,1);
\draw [line,-<-=.55] (-1,-1) -- (-1,0);
\draw (-1,-0.6) node [left] {$a$};
\draw [line,-<-=.55] (-1,0) -- (-1,1);
\draw (-1,0.6) node [left] {$c$};
\draw [line,->-=.55] (0,-1) -- (0,0);
\draw (0,-0.6) node [right] {$b$};
\draw [line,->-=.55] (0,0) -- (0,1);
\draw (0,0.6) node [right] {$d$};
\draw [red,line,-<-=.55] (0,0) -- (-1,0);
\draw (-0.5,0) node [above] {$L$};
\filldraw[black] (-1,0) circle (1pt) node[left] {\scriptsize $v$};
\filldraw[black] (0,0) circle (1pt) node[right] {\scriptsize $w^*$};
\end{tikzpicture}
\end{gathered} 
\right) 
={d_a\over d_c}\,
\begin{gathered}
\begin{tikzpicture}[scale=1]
\fill [fill=gray!20] (-1.8,-1) rectangle (-1,1);
\fill [fill=gray!20] (0,-1) rectangle (.8,1);
\draw [line,-<-=.55] (-1,-1) -- (-1,0);
\draw (-1,-0.6) node [left] {$d$};
\draw [line,-<-=.55] (-1,0) -- (-1,1);
\draw (-1,0.6) node [left] {$b$};
\draw [line,->-=.55] (0,-1) -- (0,0);
\draw (0,-0.6) node [right] {$c$};
\draw [line,->-=.55] (0,0) -- (0,1);
\draw (0,0.6) node [right] {$a$};
\draw [red,line,->-=.55] (0,0) -- (-1,0);
\draw (-0.5,0) node [above] {$L$};
\filldraw[black] (-1,0) circle (1pt) node[left] {\scriptsize $w^*$};
\filldraw[black] (0,0) circle (1pt) node[right] {\scriptsize $v$};
\end{tikzpicture}
\end{gathered} \,,
\label{antipode}
\fe
which is essential to define dual representations for the WHA. 

The star involution is an anti-linear algebra anti-homomorphism, which physically realizes Hermitian conjugation (which involves
time-reversal in the Euclidean signature),
\ie
\left(
\begin{gathered}
\begin{tikzpicture}[scale=1]
\fill [fill=gray!20] (-1.8,-1) rectangle (-1,1);
\fill [fill=gray!20] (0,-1) rectangle (.8,1);
\draw [line,-<-=.55] (-1,-1) -- (-1,0);
\draw (-1,-0.6) node [left] {$a$};
\draw [line,-<-=.55] (-1,0) -- (-1,1);
\draw (-1,0.6) node [left] {$c$};
\draw [line,->-=.55] (0,-1) -- (0,0);
\draw (0,-0.6) node [right] {$b$};
\draw [line,->-=.55] (0,0) -- (0,1);
\draw (0,0.6) node [right] {$d$};
\draw [red,line,-<-=.55] (0,0) -- (-1,0);
\draw (-0.5,0) node [above] {$L$};
\filldraw[black] (-1,0) circle (1pt) node[left] {\scriptsize $v$};
\filldraw[black] (0,0) circle (1pt) node[right] {\scriptsize $w^*$};
\end{tikzpicture}
\end{gathered} 
\right)^*
=
\sqrt{
d_a d_b\over d_c d_d
}\,\begin{gathered}
\begin{tikzpicture}[scale=1]
\fill [fill=gray!20] (-1.8,-1) rectangle (-1,1);
\fill [fill=gray!20] (0,-1) rectangle (.8,1);
\draw [line,-<-=.55] (-1,-1) -- (-1,0);
\draw (-1,-0.6) node [left] {$c$};
\draw [line,-<-=.55] (-1,0) -- (-1,1);
\draw (-1,0.6) node [left] {$a$};
\draw [line,->-=.55] (0,-1) -- (0,0);
\draw (0,-0.6) node [right] {$d$};
\draw [line,->-=.55] (0,0) -- (0,1);
\draw (0,0.6) node [right] {$b$};
\draw [red,line,->-=.55] (0,0) -- (-1,0);
\draw (-0.5,0) node [above] {$L$};
\filldraw[black] (-1,0) circle (1pt) node[left] {\scriptsize $v^*$};
\filldraw[black] (0,0) circle (1pt) node[right] {\scriptsize $w$};
\end{tikzpicture}
\end{gathered} \,.
\fe
The coefficient on the RHS is compatible with the counit $\epsilon(h^*)=\overline{\epsilon(h)}$, which is needed to define the adjoint action on the Hilbert space. 

We note that $S\circ *$ is another involution (ensured by the coefficient on the RHS in \eqref{antipode}) that is 
an anti-linear algebra homomorphism, and  corresponds to the CPT symmetry in relativistic systems. 

For weak Kac algebras, the antipode is also an involution. In general, $S^2$ is not identity but a nontrivial inner automorphism,
\ie 
S^2(h)=g h g^{-1}\,,
\fe
where $g$ is the canonical group-like element of the WHA defined in the strip algebra basis by
\ie
g=\sum_{a,b} {d_b\over d_a}\begin{gathered}
\begin{tikzpicture}[scale=1]
\fill [fill=gray!20] (-1.8,-1) rectangle (-1,1);
\fill [fill=gray!20] (0,-1) rectangle (.8,1);
\draw [line,-<-=.55] (-1,-1) -- (-1,1);
\draw (-1,0) node [left] {$a$};
\draw [line,->-=.55] (0,-1) -- (0,1);
\draw (0,0) node [right] {$b$}; 
\end{tikzpicture}
\end{gathered} \,,
\fe 
which obviously satisfies $\Delta(g)=\Delta(\id) g\otimes g$,
and in the strip algebra basis, $S^2$ acts by a scalar multiplication of infinite order in general,
\ie 
S^2 \left(
\begin{gathered}
\begin{tikzpicture}[scale=1]
\fill [fill=gray!20] (-1.8,-1) rectangle (-1,1);
\fill [fill=gray!20] (0,-1) rectangle (.8,1);
\draw [line,-<-=.55] (-1,-1) -- (-1,0);
\draw (-1,-0.6) node [left] {$a$};
\draw [line,-<-=.55] (-1,0) -- (-1,1);
\draw (-1,0.6) node [left] {$c$};
\draw [line,->-=.55] (0,-1) -- (0,0);
\draw (0,-0.6) node [right] {$b$};
\draw [line,->-=.55] (0,0) -- (0,1);
\draw (0,0.6) node [right] {$d$};
\draw [red,line,-<-=.55] (0,0) -- (-1,0);
\draw (-0.5,0) node [above] {$L$};
\filldraw[black] (-1,0) circle (1pt) node[left] {\scriptsize $v$};
\filldraw[black] (0,0) circle (1pt) node[right] {\scriptsize $w^*$};
\end{tikzpicture}
\end{gathered} 
\right)  = {d_a d_d\over d_b d_c}  
\begin{gathered}
\begin{tikzpicture}[scale=1]
\fill [fill=gray!20] (-1.8,-1) rectangle (-1,1);
\fill [fill=gray!20] (0,-1) rectangle (.8,1);
\draw [line,-<-=.55] (-1,-1) -- (-1,0);
\draw (-1,-0.6) node [left] {$a$};
\draw [line,-<-=.55] (-1,0) -- (-1,1);
\draw (-1,0.6) node [left] {$c$};
\draw [line,->-=.55] (0,-1) -- (0,0);
\draw (0,-0.6) node [right] {$b$};
\draw [line,->-=.55] (0,0) -- (0,1);
\draw (0,0.6) node [right] {$d$};
\draw [red,line,-<-=.55] (0,0) -- (-1,0);
\draw (-0.5,0) node [above] {$L$};
\filldraw[black] (-1,0) circle (1pt) node[left] {\scriptsize $v$};
\filldraw[black] (0,0) circle (1pt) node[right] {\scriptsize $w^*$};
\end{tikzpicture}
\end{gathered} \,.
\fe

The WHA is a Hopf algebra if and only if any of the four following conditions are satisfied
\ie 
\epsilon(h_1 h_2)=\epsilon(h_1)\epsilon(h_2)\,,~ \Delta(\id )=\id\otimes \id \,,~S(h_{(1)}) h_{(2)}=\epsilon(h)\,,~ h_{(1)}S(h_{(2)}) =\epsilon(h)\,.
\fe
In the strip algebra basis, these conditions are equivalent to $\cM$ having a single simple object (i.e. ${\rm rk} \cM=1$), namely $\cM$ corresponds to a fiber functor of the fusion category $\cC$.  

More generally, we have nontrivial target and source counital maps
\ie 
\epsilon_t(h)=h_{(1)}S(h_{(2)})=\epsilon(\id_{(1)} h) \id_{(2)}\,,\quad 
\epsilon_s(h)=S(h_{(1)})h_{(2)}=\id_{(1)} \epsilon(h \id_{(2)}  ) \,,
\fe
whose images are the target and source counital subalgebras $H_t$ and $H_s$ of $H$. If $H$ is Hopf, $H_t=H_s=\mC$ is trivial and the above counital maps coincides with the WHA counit. More generally, $\dim H_t =\dim H_s ={\rm rk} \cM$ and captures the multiple channels for the identity defect to act in the WHA.

In the strip algebra basis, the target and source counital subalgebras have the following bases,
\ie 
e^t_a= \sum_{b} \begin{gathered}
\begin{tikzpicture}[scale=1]
\fill [fill=gray!20] (-1.8,-1) rectangle (-1,1);
\fill [fill=gray!20] (0,-1) rectangle (.8,1);
\draw [line,-<-=.55] (-1,-1) -- (-1,1);
\draw (-1,0) node [left] {$a$};
\draw [line,->-=.55] (0,-1) -- (0,1);
\draw (0,0) node [right] {$b$}; 
\end{tikzpicture}
\end{gathered} \,,\quad 
e^s_b= \sum_{a} \begin{gathered}
\begin{tikzpicture}[scale=1]
\fill [fill=gray!20] (-1.8,-1) rectangle (-1,1);
\fill [fill=gray!20] (0,-1) rectangle (.8,1);
\draw [line,-<-=.55] (-1,-1) -- (-1,1);
\draw (-1,0) node [left] {$a$};
\draw [line,->-=.55] (0,-1) -- (0,1);
\draw (0,0) node [right] {$b$}; 
\end{tikzpicture}
\end{gathered}\,,
\fe
which enter in the important projection below needed for defining the tensor product for the representations of the WHA,
\ie 
\Delta(\id)=\id_{(1)}\otimes \id_{(2)}=\sum_a e_a^s \otimes e_a^t\,.
\fe

The target and source counital maps are 
\ie 
\epsilon_t
\left(\begin{gathered}
\begin{tikzpicture}[scale=1]
\fill [fill=gray!20] (-1.8,-1) rectangle (-1,1);
\fill [fill=gray!20] (0,-1) rectangle (.8,1);
\draw [line,-<-=.55] (-1,-1) -- (-1,0);
\draw (-1,-0.6) node [left] {$a$};
\draw [line,-<-=.55] (-1,0) -- (-1,1);
\draw (-1,0.6) node [left] {$c$};
\draw [line,->-=.55] (0,-1) -- (0,0);
\draw (0,-0.6) node [right] {$b$};
\draw [line,->-=.55] (0,0) -- (0,1);
\draw (0,0.6) node [right] {$d$};
\draw [red,line,-<-=.55] (0,0) -- (-1,0);
\draw (-0.5,0) node [above] {$L$};
\filldraw[black] (-1,0) circle (1pt) node[left] {\scriptsize $v$};
\filldraw[black] (0,0) circle (1pt) node[right] {\scriptsize $w^*$};
\end{tikzpicture}
\end{gathered} 
\right)
=&\epsilon
\left(\begin{gathered}
\begin{tikzpicture}[scale=1]
\fill [fill=gray!20] (-1.8,-1) rectangle (-1,1);
\fill [fill=gray!20] (0,-1) rectangle (.8,1);
\draw [line,-<-=.55] (-1,-1) -- (-1,0);
\draw (-1,-0.6) node [left] {$a$};
\draw [line,-<-=.55] (-1,0) -- (-1,1);
\draw (-1,0.6) node [left] {$c$};
\draw [line,->-=.55] (0,-1) -- (0,0);
\draw (0,-0.6) node [right] {$b$};
\draw [line,->-=.55] (0,0) -- (0,1);
\draw (0,0.6) node [right] {$d$};
\draw [red,line,-<-=.55] (0,0) -- (-1,0);
\draw (-0.5,0) node [above] {$L$};
\filldraw[black] (-1,0) circle (1pt) node[left] {\scriptsize $v$};
\filldraw[black] (0,0) circle (1pt) node[right] {\scriptsize $w^*$};
\end{tikzpicture}
\end{gathered} 
\right) e^t_d\,,
\\
\epsilon_s
\left(\begin{gathered}
\begin{tikzpicture}[scale=1]
\fill [fill=gray!20] (-1.8,-1) rectangle (-1,1);
\fill [fill=gray!20] (0,-1) rectangle (.8,1);
\draw [line,-<-=.55] (-1,-1) -- (-1,0);
\draw (-1,-0.6) node [left] {$a$};
\draw [line,-<-=.55] (-1,0) -- (-1,1);
\draw (-1,0.6) node [left] {$c$};
\draw [line,->-=.55] (0,-1) -- (0,0);
\draw (0,-0.6) node [right] {$b$};
\draw [line,->-=.55] (0,0) -- (0,1);
\draw (0,0.6) node [right] {$d$};
\draw [red,line,-<-=.55] (0,0) -- (-1,0);
\draw (-0.5,0) node [above] {$L$};
\filldraw[black] (-1,0) circle (1pt) node[left] {\scriptsize $v$};
\filldraw[black] (0,0) circle (1pt) node[right] {\scriptsize $w^*$};
\end{tikzpicture}
\end{gathered} 
\right)
=&\epsilon
\left(\begin{gathered}
\begin{tikzpicture}[scale=1]
\fill [fill=gray!20] (-1.8,-1) rectangle (-1,1);
\fill [fill=gray!20] (0,-1) rectangle (.8,1);
\draw [line,-<-=.55] (-1,-1) -- (-1,0);
\draw (-1,-0.6) node [left] {$a$};
\draw [line,-<-=.55] (-1,0) -- (-1,1);
\draw (-1,0.6) node [left] {$c$};
\draw [line,->-=.55] (0,-1) -- (0,0);
\draw (0,-0.6) node [right] {$b$};
\draw [line,->-=.55] (0,0) -- (0,1);
\draw (0,0.6) node [right] {$d$};
\draw [red,line,-<-=.55] (0,0) -- (-1,0);
\draw (-0.5,0) node [above] {$L$};
\filldraw[black] (-1,0) circle (1pt) node[left] {\scriptsize $v$};
\filldraw[black] (0,0) circle (1pt) node[right] {\scriptsize $w^*$};
\end{tikzpicture}
\end{gathered} 
\right) e^s_a\,.
\fe 
We mention in passing that the target and source counital subalgebras are not WHA but together they generate the minimal WHA in $H$ given below,
\ie 
H_{\rm min}={\rm span}_{a,b}\left\{
\begin{gathered}
\begin{tikzpicture}[scale=1]
\fill [fill=gray!20] (-1.8,-1) rectangle (-1,1);
\fill [fill=gray!20] (0,-1) rectangle (.8,1);
\draw [line,-<-=.55] (-1,-1) -- (-1,1);
\draw (-1,0) node [left] {$a$};
\draw [line,->-=.55] (0,-1) -- (0,1);
\draw (0,0) node [right] {$b$}; 
\end{tikzpicture}
\end{gathered}
\right\}\,.
\fe

The unique Haar element of the strip algebra is 
\ie 
\Lambda=
{1\over {\rm rk}\cM}\sum_{a,b,L,v} {d_L\over d_a}
\left(\begin{gathered}
\begin{tikzpicture}[scale=.7]
\begin{scope}[even odd rule]
\clip (0,0) circle (1)  (-1.4,-1.4) rectangle (1.4,1.4);
\fill [fill=green!20] (-1.4,-1.4) rectangle (1.4,1.4);
\end{scope}
\useasboundingbox (-1.3,-1.2) rectangle (1.3,1.4); 
 \draw (0,1) node [below] {$b$};
 \draw (0,-1) node [below=-.05] {$a$};
  \draw (0,0) node [below] {$L$};
\draw [line,->-=.27,->-=.77] (0,0) circle (1);
\draw [line,->-=.53,red] (-1,0) to (1,0);
\filldraw[black] (-1,0.) circle (1pt) node[above left=-.1] {\scriptsize $v$};
\filldraw[black] (1,0) circle (1pt) node[below right=-.15]  {\scriptsize $v^*$}; 
\end{tikzpicture}
\end{gathered}\right)^{-1}
\begin{gathered}
\begin{tikzpicture}[scale=1]
\fill [fill=gray!20] (-1.8,-1) rectangle (-1,1);
\fill [fill=gray!20] (0,-1) rectangle (.8,1);
\draw [line,-<-=.55] (-1,-1) -- (-1,0);
\draw (-1,-0.6) node [left] {$a$};
\draw [line,-<-=.55] (-1,0) -- (-1,1);
\draw (-1,0.6) node [left] {$b$};
\draw [line,->-=.55] (0,-1) -- (0,0);
\draw (0,-0.6) node [right] {$a$};
\draw [line,->-=.55] (0,0) -- (0,1);
\draw (0,0.6) node [right] {$b$};
\draw [red,line,-<-=.55] (0,0) -- (-1,0);
\draw (-0.5,0) node [above] {$L$};
\filldraw[black] (-1,0) circle (1pt) node[left] {\scriptsize $v$};
\filldraw[black] (0,0) circle (1pt) node[right] {\scriptsize $v^*$}  ;
\end{tikzpicture}
\end{gathered} 
=
{1\over {\rm rk}\cM}\sum_{a,b} {1\over d_a}
\begin{gathered}
\begin{tikzpicture}[scale=1]
\fill [fill=gray!20] (-1.8,-1) rectangle (-1.4,1);
\fill [fill=green!20] (-1.4,-1) rectangle (-1,1);
\fill [fill=green!20] (-1,-.2) rectangle (0,.2);
\fill [fill=green!20] (0,-1) rectangle (.4,1);
\fill [fill=gray!20] (0.4,-1) rectangle (.8,1);
\draw [dotted] (-1.4,-1) -- (-1.4,1);
\draw [dotted] (.4,-1) -- (.4,1);
\draw [line,->-=.20,->-=.55,->-=.85,blue] (-1,1) -- (-1,0.2) -- (0,0.2) -- (0,1);
\draw (-.5,0.6) node [left] {$b$}; 
\draw [line,-<-=.20,-<-=.55,-<-=.85,blue] (-1,-1) -- (-1,-0.2) -- (0,-0.2) -- (0,-1);
\draw (-.5,-0.6) node [right] {$a$};   
\end{tikzpicture}
\end{gathered}\,, 
\fe
where in the second equality, we have used \eqref{peelrel}. This last expression makes it easy to check the following properties of the Haar element:
\ie 
h\Lambda=\epsilon_t(h)\Lambda\,,\quad \Lambda h=\Lambda \epsilon_s(h)
\fe 
as a two sided integral, the normalization satisfies 
\ie 
\epsilon_s(\Lambda)=\epsilon_t(\Lambda)=\id \,,
\fe
and finally,
\ie 
\Lambda=\Lambda^2=S(\Lambda)=\Lambda^*\,,\quad S(\Lambda_{(1)})\Lambda_{(2)}=\Lambda_{(1)}S(\Lambda_{(2)})=\id \,.
%\epsilon_L(e_b^L)=e_b^L 
\fe
In the case $\cM$ is rank one, thus corresponding to a fiber functor, and $H$ is Hopf,
\ie
\Lambda={1\over D}\sum_{L,v} d_L \begin{gathered}
\begin{tikzpicture}[scale=1]
\fill [fill=gray!20] (-1.8,-1) rectangle (-1,1);
\fill [fill=gray!20] (0,-1) rectangle (.8,1);
\draw [line,-<-=.55] (-1,-1) -- (-1,0);
\draw (-1,-0.6) node [left] { };
\draw [line,-<-=.55] (-1,0) -- (-1,1);
\draw (-1,0.6) node [left] { };
\draw [line,->-=.55] (0,-1) -- (0,0);
\draw (0,-0.6) node [right] { };
\draw [line,->-=.55] (0,0) -- (0,1);
\draw (0,0.6) node [right] { };
\draw [red,line,-<-=.55] (0,0) -- (-1,0);
\draw (-0.5,0) node [above] {$L$};
\filldraw[black] (-1,0) circle (1pt) node[left] {\scriptsize $v$};
\filldraw[black] (0,0) circle (1pt) node[right] {\scriptsize $v^*$}  ;
\end{tikzpicture}
\end{gathered}  \,,
\fe
where we take the junction vector basis $v$ to be orthonormal.

The Haar measure for $H$, or equivalently the Haar element of the dual WHA $\hat H$, is 
\ie 
\lambda 
\left(\begin{gathered}
\begin{tikzpicture}[scale=1]
\fill [fill=gray!20] (-1.8,-1) rectangle (-1,1);
\fill [fill=gray!20] (0,-1) rectangle (.8,1);
\draw [line,-<-=.55] (-1,-1) -- (-1,0);
\draw (-1,-0.6) node [left] {$a$};
\draw [line,-<-=.55] (-1,0) -- (-1,1);
\draw (-1,0.6) node [left] {$c$};
\draw [line,->-=.55] (0,-1) -- (0,0);
\draw (0,-0.6) node [right] {$b$};
\draw [line,->-=.55] (0,0) -- (0,1);
\draw (0,0.6) node [right] {$d$};
\draw [red,line,-<-=.55] (0,0) -- (-1,0);
\draw (-0.5,0) node [above] {$L$};
\filldraw[black] (-1,0) circle (1pt) node[left] {\scriptsize $v$};
\filldraw[black] (0,0) circle (1pt) node[right] {\scriptsize $w^*$};
\draw (-1,-.8) node [above] {};
\end{tikzpicture}
\end{gathered} 
\right)
={1\over {\rm rk} \cM}{\D_{a,c}\D_{b,d}  \over      d_c d_d}
\,\begin{gathered}
\begin{tikzpicture}[scale=1]
\filldraw [fill=green!20,line,-<-=.27
] (-1.4,0) circle (.4);
\filldraw [fill=green!20,line,-<-=.27] (.4,0) circle (.4);
\draw (-1.2,0.65) node [left] {$c$};
\draw (0.2,0.65) node [right] {$d$};
\draw [red,line,-<-=.55] (0,0) -- (-1,0);
\draw (-0.5,0) node [above] {$L$};
\filldraw[blue] (-1,0) circle (1pt) node[left] {\scriptsize $v$};
\filldraw[blue] (0,0) circle (1pt) node[right] {\scriptsize $w^*$};
\end{tikzpicture}
\end{gathered} 
\,.
\fe
Note that $\lambda$ is nonzero only for elements of $H_{\rm min}$.  This is the unique positive functional on $H$ that satisfies
\ie 
h_{(1)}\lambda(h_{(2)})=\epsilon_t(h_{(1)})\lambda(h_{(2)})\,,\quad \lambda \circ S=\lambda \,,\quad \lambda\circ \epsilon_t=\epsilon
\fe

 The corresponding Haar conditional expectation, defined by \ie  
E(h)\equiv \lambda \rightharpoonup h=\lambda(h_{(2)})h_{(1)}\,,\quad \forall h\in H\,,
\fe 
is,
\ie 
E\left(\begin{gathered}
\begin{tikzpicture}[scale=1]
\fill [fill=gray!20] (-1.8,-1) rectangle (-1,1);
\fill [fill=gray!20] (0,-1) rectangle (.8,1);
\draw [line,-<-=.55] (-1,-1) -- (-1,0);
\draw (-1,-0.6) node [left] {$a$};
\draw [line,-<-=.55] (-1,0) -- (-1,1);
\draw (-1,0.6) node [left] {$c$};
\draw [line,->-=.55] (0,-1) -- (0,0);
\draw (0,-0.6) node [right] {$b$};
\draw [line,->-=.55] (0,0) -- (0,1);
\draw (0,0.6) node [right] {$d$};
\draw [red,line,-<-=.55] (0,0) -- (-1,0);
\draw (-0.5,0) node [above] {$L$};
\filldraw[black] (-1,0) circle (1pt) node[left] {\scriptsize $v$};
\filldraw[black] (0,0) circle (1pt) node[right] {\scriptsize $w^*$};
\end{tikzpicture}
\end{gathered} 
\right)
={1\over {\rm rk} \cM}{\D_{a,c}\D_{b,d}  \over      d_c d_d}
\,\begin{gathered}
\begin{tikzpicture}[scale=1]
\filldraw [fill=green!20,line,-<-=.27
] (-1.4,0) circle (.4);
\filldraw [fill=green!20,line,-<-=.27] (.4,0) circle (.4);
\draw (-1.2,0.65) node [left] {$c$};
\draw (0.2,0.65) node [right] {$d$};
\draw [red,line,-<-=.55] (0,0) -- (-1,0);
\draw (-0.5,0) node [above] {$L$};
\filldraw[blue] (-1,0) circle (1pt) node[left] {\scriptsize $v$};
\filldraw[blue] (0,0) circle (1pt) node[right] {\scriptsize $w^*$};
\draw (-1,-.8) node [above] {};
\end{tikzpicture}
\end{gathered} \,e_a^t\,,
\fe 
which projects the WHA $H$ to its left counital subalgebra $H_t$ by a simple topological move.\footnote{This move is implemented by pretending the boundaries are topological (more precisely via \eqref{peelrel}) and bending them into a dumbbell which can be further reduced.}

The image of the Haar element under the  Haar conditional expectation is a positive element \cite{BOHM1999385,bohm1999weakhopfalgebrasii},
\ie 
E(\Lambda)=\lambda \rightharpoonup \Lambda
={1\over ({\rm rk}\cM)^2} \sum_{a} {1\over d_a^2} e_a^t \,, 
\fe
and  we have the following invertible elements $g_t\in H_t$ and $g_s\in H_s$\,,
\ie 
g_t=(E(\Lambda))^{1\over 2}={1\over {\rm rk}\cM} \sum_{a} {1\over d_a} e_a^t \,, \quad g_s=S(g_t)={1\over {\rm rk}\cM}\sum_{a} {1\over d_a} e_a^s \,,
\fe
which factorizes the canonical group element by
\ie 
g=g_t g_s^{-1}\,.
\fe

A quasi-basis for $E(\cdot)$ is defined by $\sum_A a_A\otimes b_A\in H\otimes H$ where $A$ runs over the strip algebra basis such that
\ie 
\sum_A a_A E( b_A h)=h\,,\quad \forall h \in H\,.
\fe
From \cite{bohm1999weakhopfalgebrasii}, the quasi-basis is determined by $g_s$ and the Haar element $\Lambda$,
\ie 
\sum_A a_A\otimes b_A = S(\Lambda_{(1)})\otimes g_s^{-2}\Lambda_{(2)}
\fe
we have, taking $a_A$ to be given by the strip algebra basis,
\ie 
&a_A=\begin{gathered}
\begin{tikzpicture}[scale=1]
\fill [fill=gray!20] (-1.8,-1) rectangle (-1,1);
\fill [fill=gray!20] (0,-1) rectangle (.8,1);
\draw [line,-<-=.55] (-1,-1) -- (-1,0);
\draw (-1,-0.6) node [left] {$f$};
\draw [line,-<-=.55] (-1,0) -- (-1,1);
\draw (-1,0.6) node [left] {$e$};
\draw [line,->-=.55] (0,-1) -- (0,0);
\draw (0,-0.6) node [right] {$b$};
\draw [line,->-=.55] (0,0) -- (0,1);
\draw (0,0.6) node [right] {$a$};
\draw [red,line,->-=.55] (0,0) -- (-1,0);
\draw (-0.5,0) node [above] {$L$};
\filldraw[black] (-1,0) circle (1pt) node[left] {\scriptsize $w^*$};
\filldraw[black] (0,0) circle (1pt) node[right] {\scriptsize $v$};
\end{tikzpicture}
\end{gathered} \,,
\quad 
b_A
={ {\rm rk}\cM} \sqrt{d_f d_b \over d_e d_a} \begin{gathered}
\begin{tikzpicture}[scale=1]
\fill [fill=gray!20] (-1.8,-1) rectangle (-1,1);
\fill [fill=gray!20] (0,-1) rectangle (.8,1);
\draw [line,-<-=.55] (-1,-1) -- (-1,0);
\draw (-1,-0.6) node [left] {$e$};
\draw [line,-<-=.55] (-1,0) -- (-1,1);
\draw (-1,0.6) node [left] {$f$};
\draw [line,->-=.55] (0,-1) -- (0,0);
\draw (0,-0.6) node [right] {$a$};
\draw [line,->-=.55] (0,0) -- (0,1);
\draw (0,0.6) node [right] {$b$};
\draw [red,line,-<-=.55] (0,0) -- (-1,0);
\draw (-0.5,0) node [above] {$L$};
\filldraw[black] (-1,0) circle (1pt) node[left] {\scriptsize $w$};
\filldraw[black] (0,0) circle (1pt) node[right] {\scriptsize $v^*$};
\end{tikzpicture}
\end{gathered} \,,
\fe
then
\ie 
\sum_A a_A b_A=\epsilon_s(g_s^{-2} \Lambda)= {\rm rk} \cM \cdot D \id \,,
\fe
where $D=\sum_L d_L^2$ is the total quantum dimension of $\cC$ and thus
\ie 
{\rm Ind} (E)=  {\rm rk \cM}\cdot D\,.
\fe

The following left non-degenerate integral
\ie 
\ell=
{ {\rm rk}\cM}\sum_{a,b} d_a
\begin{gathered}
\begin{tikzpicture}[scale=1]
\fill [fill=gray!20] (-1.8,-1) rectangle (-1.4,1);
\fill [fill=green!20] (-1.4,-1) rectangle (-1,1);
\fill [fill=green!20] (-1,-.2) rectangle (0,.2);
\fill [fill=green!20] (0,-1) rectangle (.4,1);
\fill [fill=gray!20] (0.4,-1) rectangle (.8,1);
\draw [dotted] (-1.4,-1) -- (-1.4,1);
\draw [dotted] (.4,-1) -- (.4,1);
\draw [line,->-=.20,->-=.55,->-=.85,blue] (-1,1) -- (-1,0.2) -- (0,0.2) -- (0,1);
\draw (-.5,0.6) node [left] {$b$}; 
\draw [line,-<-=.20,-<-=.55,-<-=.85,blue] (-1,-1) -- (-1,-0.2) -- (0,-0.2) -- (0,-1);
\draw (-.5,-0.6) node [right] {$a$};   
\end{tikzpicture}
\end{gathered}\,, 
\fe
satisfies
\ie 
E(\ell)=\lambda \rightharpoonup \ell =1\,,\quad   \ell\rightharpoonup \lambda =\epsilon\,,
\fe
so that $(\ell,\lambda)$ defines  a pair of dual
integrals.

\section{Entropic Order Parameters from the Algebraic Viewpoint} \label{app: ECR}

In the main text we stated the so-called \emph{entropic certainty relation} \eqref{ECR}. By non-negativity of the relative entropy this relation also implies a bound on the entropic order parameters for the inclusions $N \subset M$ and $M' \subset N'$ \eqref{Bound on EOP}. In this section we will explore the entropic certainly relation more closely with a specific focus on states that saturate the inequality for the entropic order parameter. In particular, we will uncover an intimate relationship between the state which maximizes the entropic order parameter of the inclusion $N \subset M$ and the state that minimizes the entropic order parameter of the inclusion $M' \subset N'$. This provides a recipe for identifying states that contain the most information about the `difference' between the algebras $N$ and $M$. From the non-invertible symmetry perspective this is related to the duality theory for weak Hopf algebras. 

\subsection{Entropic Certainty Relation for Reducible Inclusions} 

To begin, let us derive the entropic certainly relation. Although this result has appeared elsewhere, the following derivation is unique in that it allows for general inclusions of von Neumann algebras with non-trivial centers, and conditional expectations that needn't be minimal. 

Let $N \subset M$ be an inclusion of von Neumann algebras which are not required to be factorial. Take $\pi: M \rightarrow B(L^2(M))$ to be a standard representation of $M$. Here, the notation $L^2(M)$ refers to the GNS Hilbert space of $M$ with respect to any faithful, semifinite, normal state. Letting this state be given by $\varphi \in S(M)$, there is an associated vector $\ket{\varphi} \in L^2(M)$ such that $\bra{\varphi} \pi(x) \ket{\varphi} = \varphi(x)$. The vector $\ket{\varphi}$ is both cyclic and separating for the algebra $M$ and thus a dense set of states can be generated by acting on this state, $\ket{x} \equiv \pi(x) \ket{\varphi}$. The inner product on this dense set is induced from the state as $\braket{x|y} = \bra{\varphi} \pi(x^*) \pi(y) \ket{\varphi} = \varphi(x^* y)$. 

The inclusion $N \subset M$ is dual to an inclusion of their commutant algebras, $\pi(M)' \subset \pi(N)'$. To see that $\pi(M)' \subset \pi(N)'$ defines a good algebraic inclusion, note that every operator in $B(L^2(M))$ which commutes with all operators in $\pi(M)$ must also commute with all operators in $\pi(N)$ since $\pi(N) \subset \pi(M)$, however the converse is obviously not true. The duality between the pair inclusions is established by considering the Jones basic construction. Firstly, we must specify a conditional expectation $E: M \rightarrow N$. Then, there exists a natural extension of the algebra\footnote{To be precise, this extension is obtained by taking the von Neumann union of $M$ with the Jones projection of $E$: $M_1 = M \vee e_E$.} $M$, called the basic extension, such that we obtain the sequence of nested inclusions
\beq
	N \subset M \subset M_1\,.
\eeq
The inclusion $M \subset M_1$ admits a conditional expectation $\mathcal{E}: M_1 \rightarrow M$ whose index agrees with that of $E$. The inclusion $M \subset M_1$ is anti-isomorphic to the inclusion $\pi(M)' \subset \pi(N)'$ via the modular conjugation;  this extends the well known fact that $\pi(M)' = J \pi(M) J$ by the observation that $\pi(N)' = J M_1 J$. 

The duality between the conditional expectations $E$ and $\mathcal{E}$, as well as the equality of their indices, is established by appealing to Kosaki's theory of dual operator valued weights. Let $r: M \rightarrow B(H)$ be a representation of $M$ acting on a Hilbert space $H$ that needn't be standard. Given a normal state $\psi$ on $M$ and a normal faithful state $\Psi$ on $r(M)'$ we can define a positive, self adjoint operator $\frac{d\Psi}{d\psi}$ on $H$ called the spatial derivative of $\Psi$ with respect to $\psi$.\footnote{For a complete review of the spatial derivative see \cite{AliAhmad:2025oli}.} The spatial derivative generalizes the relative modular operator to the setting of non-standard representations. If $\xi \in H$ we will denote by $\omega_{\xi}$ the state on $B(H)$ induced by $\xi$ and the inner product, e.g. $\omega_{\xi}(\mathcal{O}) \equiv \langle \xi, \mathcal{O}(\xi) \rangle_{H}$. The restriction of $\omega_{\xi}$ to $r(M)' \subset B(H)$ therefore defines a state on $\pi(M)'$. For such a state we can define the relative entropy
\beq \label{Relative entropy}
	S_{r(M)}(\xi \parallel \psi) \equiv \langle \xi, \log \frac{d \omega_{\xi} \rvert_{r(M)'}}{d\psi} \xi \rangle_H\,.
\eeq
If $H = L^2(M)$ and $\omega_{\xi}\rvert_{r(M)} = \eta$ defines a state on the algebra $M$ then $\frac{d \omega_{\xi}\rvert_{r(M)'}}{d\psi} = \Delta_{\eta \mid \psi}$ is Araki's relative modular operator. By consequence we have that $S_{r(M)}(\xi \parallel \psi) = S(\eta \parallel \psi)$ where the right hand side is Araki's relative entropy \cite{araki1975relative}
\beq
	S(\eta \parallel \psi) = \bra{\eta} \log \Delta_{\eta \mid \psi} \ket{\eta}\,.
\eeq

The Kosaki dual of the conditional expectation $E$ is an operator valued weight $E^{-1}: \pi(N)' \rightarrow \pi(M)'$ satisfying the condition
\beq \label{Kosaki's equation}
	\frac{d (\Psi \circ E^{-1})}{d\psi} = \frac{d \Psi}{d(\psi \circ E)}, \qquad \forall \; \Psi \in P(\pi(M)'), \; \psi \in P(N)\,.
\eeq
The normalization of the map $E^{-1}$ defines the Kosaki index of $E$ as \cite{Teruya1992Index}
\beq
	E^{-1}(\mathbb{I}) = \text{Ind}(E) \in Z(M)\,.
\eeq
The index is an invertible central operator and thus we can define a normalized operator valued weight (hence a conditional expectation) $E' \equiv \text{Ind}(E)^{-1} E^{-1}$. This is the Kosaki dual \emph{conditional expectation} relative to $E$. The conditional expectation $E': \pi(N)' \rightarrow \pi(M)'$ is anti-isomorphic to the conditional expectation $\mathcal{E}$ via the modular conjugation. In particular, $\mathcal{E}(x) = J E'(JxJ) J$.  

Appealing to \eqref{Kosaki's equation} we see that the conditional expectations $E$ and $E'$ are related by
\beq
	\pi(\text{Ind}(E)) \frac{d(\Psi \circ E')}{d \psi} = \frac{d \Psi}{d(\psi \circ E)}, \qquad \forall \; \Psi \in P(\pi(M)'), \; \psi \in P(N)\,.
\eeq
Let $\ket{\xi} \in L^2(M)$ and denote by $\omega_{\ket{\xi}}\rvert_{\pi(N)}$ and $\omega_{\ket{\xi}}\rvert_{\pi(M)'}$ the states induced by restricting the expectation values of this vector state to the appropriate subalgebras of $B(L^2(N))$. We then find
\beq
	\pi(\text{Ind}(E)) \frac{d(\omega_{\ket{\xi}}\rvert_{\pi(M)'} \circ E')}{d \omega_{\ket{\xi}}\rvert_{\pi(N)}} =  \frac{d \omega_{\ket{\xi}}\rvert_{\pi(M)'}}{d( \omega_{\ket{\xi}}\rvert_{\pi(N)} \circ E)}\,.
\eeq
Taking logs of both sides and then computing the expectation value with respect to the state $\omega_{\ket{\xi}}$ we obtain
\begin{flalign} \label{Entropic Certainty 1}
	\bra{\xi} \log \frac{d \omega_{\ket{\xi}}\rvert_{\pi(M)'}}{d( \omega_{\ket{\xi}}\rvert_{\pi(N)} \circ E)} \ket{\xi} &= \bra{\xi} \log \frac{d(\omega_{\ket{\xi}}\rvert_{\pi(M)'} \circ E')}{d \omega_{\ket{\xi}}\rvert_{\pi(N)}} \ket{\xi} + \bra{\xi} \pi(\log \text{Ind}(E)) \ket{\xi} \nonumber \\
	& -\bra{\xi} \log \frac{d \omega_{\ket{\xi}} \rvert_{\pi(N)}}{d(\omega_{\ket{\xi}} \rvert_{\pi(M)'} \circ E')} \ket{\xi} + \bra{\xi} \pi(\log \text{Ind}(E)) \ket{\xi}\,.
\end{flalign}
This equation can be rearranged to produce the entropic certainty relation
\beq
	S_{\pi(M)}( \ket{\xi} \parallel \omega_{\ket{\xi}}\rvert_{\pi(N)} \circ E) + S_{\pi(N)'}(\ket{\xi} \parallel \omega_{\ket{\xi}}\rvert_{\pi(M)'} \circ E') = \bra{\xi} \pi(\log \text{Ind}(E)) \ket{\xi}\,.
\eeq

Suppose that $\ket{\xi} = \ket{\psi}$ is the vector representative of a state $\psi$ on $M$. Then we can write
\beq
	S(\psi \parallel \psi \circ E) + S_{\pi(N)'}(\ket{\psi} \parallel \omega_{\ket{\psi}} \circ E') = \psi(\log \text{Ind}(E))\,.
\eeq
Invoking the anti-isomorphisms $M_1 = J \pi(N)' J$ and $\pi(M) = J \pi(M)' J$ we can further identify
\beq
    S_{\pi(N)'}(\ket{\psi} \parallel \omega_{\ket{\psi}} \circ E') = S_{M_1}(\psi_1 \parallel \psi_1 \circ \mathcal{E})\,,
\eeq
where $\psi_1$ is the state on $M_1$ induced from the vector representative $\ket{\psi}$. As a result we obtain the more familiar looking equality
\beq
    S_M(\psi \parallel \psi \circ E) + S_{M_1}(\psi_1 \parallel \psi_1 \circ \mathcal{E}) = \psi( \log \text{Ind}(E))\,.
\eeq
The second term is non-negative and thus we obtain a bound
\beq \label{Bound again}
	S(\psi \parallel \psi \circ E) \leq \psi(\log \text{Ind}(E))\,.
\eeq
This bound will be saturated if the relative entropy $S_{M_1}(\psi_1 \parallel \psi_1 \circ \mathcal{E})$ vanishes. This will occur if and only if $\psi_1 \circ \mathcal{E} = \psi_1$. Thus, we conclude that the entropic order parameter associated with the conditional expectation $E$ will be saturated if and only if there exists a state on $M_1$ with vector representative $\ket{\psi} \in L^2(M)$ which is invariant under the Kosaki dual conditional expectation. Although the existence of a state on $M_1$ which is invariant under $\mathcal{E}$ is guaranteed by the idempotence of $\mathcal{E}$, it may be the case that no such state possesses a vector representative in $L^2(M)$.

When such a state exists, the bound on the entropic order parameter of the conditional expectation $E$ is realized by the state $\omega_{\ket{\psi}}\rvert_{\pi(M)}$. In general, even if an invariant state does not exist and thus the entropic order parameter does not saturate the bound \eqref{Bound again}, we still observe that the state which contains the most information about the conditional expectation $E$ (maximizes the entropic order parameter for $E$) contains the least information about the conditional expectation $\mathcal{E}$ (minimizes the entropic order parameter for $\mathcal{E}$) and visa-versa.

\subsection{Depth-2 Inclusions and Non-invertible Symmetry Breaking} 

The duality between $E$ and $\mathcal{E}$ at the level of their entropic order parameters realizes a particularly physical interpretation in the non-invertible symmetry context. For a general conditional expectation $E: M \rightarrow N$ the basic construction can be iterated indefinitely to produce a tower of inclusions
\beq
	N \subset M \subset M_1 \subset M_2 \subset \,...
\eeq
where each subsequent inclusion admits a conditional expectation Kosaki dual to the previous one. For a special class of inclusions, however, all of the relevant information in the tower of inclusions is specified in the sequence $N \subset M \subset M_1$. Such inclusions are called \emph{depth-2} because they depend only upon a two sequence Jones tower.\footnote{This description is really only heuristic. For a more rigorous discussion of inclusions of depth-2 we refer the reader to \cite{nill1998weakhopfalgebrasreducible}.} In \cite{nill1998weakhopfalgebrasreducible} it is shown that any finite index, depth-2 inclusion $E: M \rightarrow N$ gives rise to a Jones tower such that $H \equiv N' \cap M_1$ and $\hat{H} \equiv M' \cap M_2$ are a dual pair of weak Hopf algebras acting regularly on $M$ and $M_1$. In the language of this note that means we obtain a pair of algebraic non-invertible symmetry actions $\rho: H \rightarrow \Endvect(M)$ and $\hat{\rho}: \hat{H} \rightarrow \Endvect(M_1)$. 

The algebra $N$ is the symmetric subalgebra under the action of $\rho$ as we have described in \eqref{Symmetric Subalgebra}. The conditional expectation $E$ may then be identified with the Haar conditional expectation \eqref{WH Invariantization Map}. The algebra $M_1$ is the crossed product of the algebra $M$ by the action $\rho$, which we denote by $M \times_{\rho} H$. The crossed product is based on the algebra $M \otimes H)/H_t$ where the quotient is related to the equivalence relation
\beq
	x \rho_{h_t}(\mathbb{I}_M) \otimes h \sim x \otimes h_t h, \qquad \forall h_t \in H_t\,.
\eeq
We shall denote a general element in the crossed product by $x \times_{\rho} h$ which can be regarded as a standard representative of the equivalence class of elements under the above relation. The product in the crossed product is given by
\beq
	(x \times_{\rho} h) (x \times_{\rho} h') \equiv x \rho_{h_{(1)}}(x') \times_{\rho} h_{(2)} h'\,,
\eeq
and the involution by
\beq
	(x \times_{\rho} h)^* \equiv (\mathbb{I}_M \times_{\rho} h^*)(x^* \times_{\rho} \mathbb{I}_H) = \rho_{h_{(1)}^*}(x^*) \times_{\rho} h_{(2)}^*\,.
\eeq	

The inclusion $M \subset M \times_{\rho} H$ comes equipped with a canonical conditional expectation $\mathcal{E}: M \times_{\rho} H \rightarrow M$ which can be regarded as the weak Hopf algebra generalization of Haagerup's operator valued weight for crossed products by locally compact groups \cite{HaagerupOVW}. This conditional expectation is dual to $E$ through the basic construction e.g. it is anti-isomorphic to the Kosaki dual of the Haar conditional expectation. The crossed product $M \times_{\rho} H$ naturally admits a representation of the dual algebra $\hat{H}$ as
\beq
	\hat{\rho}_{\hat{h}}(x \times_{\rho} h) \equiv x \times_{\rho} (\hat{h} \rightharpoonup h)\,,
\eeq
with $\rightharpoonup$ the canonical action of $\hat{H}$ on $H$. The symmetric subalgebra of $M \times_{\rho} H$ under the action $\hat{\rho}$ is isomorphic to $M$ \cite{nill1998weakhopfalgebrasreducible}. In this sense we can regard $\mathcal{E}$ as the Haar conditional expectation associated with the weak Hopf algebra $\hat{H}$ and action $\hat{\rho}$. 

To proceed in the Jones tower we simply continue taking crossed products. The algebra $M_2 \equiv M_1 \times_{\hat{\rho}} \hat{H} = (M \times_{\rho} H) \times_{\hat{\rho}} \hat{H}$ is an alternating double crossed product of the algebra $M$. The result of this double crossed product is simply to amplify the space -- $M_2 = M \otimes \text{End}(H)$ -- which can be viewed as a far reaching generalization of Takai duality \cite{TAKAI197525}.\footnote{The bicrossed product of $H$ and its dual $\hat{H}$ is often referred to as a quantum double and denoted by $D(H)$. This can be viewed as an algebraic analog of the Drinfeld center discussed in \ref{sec: algebras from categories} around eqn. \eqref{Tube, Double, Center}.} Consequently, each subsequent step up the Jones tower repeats the cycle of crossing first with an action $\rho$ of $H$ and then with an action $\hat{\rho}$ of $\hat{H}$. This is the heuristic characterization of a depth-2 inclusion. 

Armed with these insights, the relation between entropic order parameters for the dual inclusions $N \subset M$ and $\pi(M)' \subset \pi(N)'$ becomes more concrete. Let $N \subset M \subset M \times_{\rho} H$ be a depth-2 inclusion associated with the non-invertible symmetry action $\rho: H \rightarrow \Endvect(M)$. The conditional expectation $E: M \rightarrow N$ encodes the symmetrization map associated with this action. The entropic order parameter $\Delta_E S(\psi)$ measures the asymmetry of the initial state $\psi$ relative to the action by $H$. Likewise, the Kosaki dual conditional expectation $\mathcal{E}: M \times_{\rho} H \rightarrow M$ encodes the symmetrization map associated with the action $\hat{\rho}: \hat{H} \rightarrow \Endvect(M \times_{\rho} H)$, and the entropic order parameter $\Delta_{\mathcal{E}} S(\psi)$ measures the asymmetry of the initial state $\psi$ relative to the action by $\hat{H}$.

From the entropic certainly relation \eqref{Entropic Certainty 1} and our discussion of its saturation conditions we recognize that the state which maximizes the entropic order parameter for the symmetry $H$ minimizes the entropic order parameter for $\hat{H}$. In other words, the state which is most symmetric with respect to the non-invertible symmetry $H$ is least symmetric with respect to the non-invertible symmetry $\hat{H}$. The `amount of symmetry' in the dual provides a strengthening to the bound \eqref{Bound again}:
\beq
    \max_{\ket{\psi} \in L^2(M)} S_M(\psi \parallel \psi \circ E) = \psi(\log \text{Ind}(E)) - \min_{\ket{\psi} \in L^2(M)} S_{M \times_{\rho} H}(\psi_1 \parallel \psi_1 \circ \mathcal{E})\,. 
\eeq

\section{Imprimitivity: From Strip to Fusion} \label{App: Cat from Alg}

In the main text we have described how to derive an algebraic symmetry action $\rho: \text{Str}_{\mathcal{C}}(\mathcal{M}) \rightarrow \Endvect(M_{\mathcal{M}})$ starting from a categorical symmetry action $\Phi: \mathcal{C} \rightarrow \text{End}(M)$ and a module category $\mathcal{M}$. In this section, we will briefly describe how the reverse problem can be addressed by invoking the notion of imprimitivity. 

To obtain a representation of $\text{Str}_{\mathcal{C}}(\mathcal{M})$ acting on the extended system $M_{\mathcal{M}}$, we started with the canonical representation of the fusion algebra defined by $\Phi$ and invoked Rieffel induction. Suppose instead that we began with a representation of $\text{Str}_{\mathcal{C}}(\mathcal{M})$. Under what circumstances can we conclude that our representation is equivalent to \emph{some} representation induced from the fusion algebra? This is the question of imprimitivity. Stated more generally, given a pair of C$^*$-algebra $A$ and $B$ and a represntation of $A$, $\tilde{\rho}: A \rightarrow B(\tilde{H})$, we ask when there exists a $B$-rigged $A$-module $X$ and a Hilbert space representation $\rho: B \rightarrow B(H)$ such that $\tilde{\rho} \simeq \rho^X$. 

The problem of imprimitivity is solved by introducing the notion of an imprimitivity bimodule. Given an pair of C$^*$-algebras $C$ and $D$, a $C-D$ imprimitivity bimodule ${}_C X_D$ is a vector space $X$ which is simultaneously a left $C$-rigged and a right $D$-rigged space. That is, $X$ admits a left representation of $C$, $\ell^C: C \rightarrow B(X)$, a right representation of $D$, $r^D: D \rightarrow B(X)$, as well as $C$ and $D$ valued inner products $G_C: X \times X \rightarrow C$ and $G_D: X \times X \rightarrow D$ such that
\begin{enumerate}
	\item $\ell^C_{G_C(x,y)}(z) = r^D_{G_D(y,z)}(x), \; \forall x,y,z \in X$,
	\item $G_D(\ell^C)c(x),\ell^C_c(x)) \leq \norm{c}^2 G_D(x,x), \; \forall x \in X, c \in C$,
	\item $G_C(r^D_d(x),r^D_d(x)) \leq \norm{d}^2 G_C(x,x), \; \forall x \in X, d \in D$. 
\end{enumerate}

Given a $C-D$ imprimitivity bimodule ${}_C X_D$, we automatically obtain a $D-C$ imprimitivty bimodule, ${}_D \overline{X}_D$, by dualization. The dual imprimitivity bimodule is based on the vector space $X$ but with representations and algebra valued inner products obtained via involution as
\beq
	\overline{\ell}^C_c(x) \equiv \ell^C_{c^*}(x), \overline{r}^D_d(x) \equiv r^D_{d^*}(x), \qquad \overline{G}_C(x,y) \equiv G_C(x,y)^*, \overline{G}_D(x,y) \equiv G_D(x,y)^*\,.
\eeq
It is straightforward to check that $\overline{\ell}^C$ is a right representation and $\overline{r}^D$ a left. Likewise, the adjoint inner products satisfy all the same compatibility conditions as the original ones. 

The existence of a $C-D$ imprimitivity bimodule therefore implies that $X$ is both a $C$-rigged $D$-module and a $D$-rigged $C$-module. Thus, by Rieffel's construction, $X$ induces a functor from the category of modules of $D$ to the category of modules of $C$, and $\overline{X}$ induces a functor from the category of modules of $D$ to the category of modules of $C$. Specifically, $F_X(V) \equiv X \otimes V/D$ and $F_{\overline{X}}(W) \equiv X \otimes W/C$ when $V$ and $W$ are respectively representations of $D$ and $C$. Rieffel's inversion theorem tells us that these functors are inverses of each other. That is $F_X \circ F_{\overline{X}}(W) \simeq W$ and $F_{\overline{X}} \circ F_X(V) \simeq V$. Consequently, we see that the existence of an imprimitivity bimodule between two C$^*$-algebras actually implies that they share the same representation categories and thus are strongly Morita equivalent. 

In the general context of Rieffel induction, we don't expect our algebras to be strongly Morita equivalent. Thus, the existence of an imprimitivity bimodule is too strong to answer the question of imprimitivity. Still, we can use it as a tool in the following way. Given any $B$-rigged space, $X$, we can define a C$^*$-algebra, $E(X)$, which, roughly speaking, indexes all of the possible representations which can be induced from $B$ via $X$. This algebra is called the imprimitivity algebra of $X$. The imprimitivity algebra of $X$ is simply the linear span of all operators from $X$ to $X$ of the form
\beq
	T_{x,y}(z) \equiv r_{G_B(y,z)}(x), \qquad x,y \in X. 
\eeq
If we use an abstract `bra-ket' notation in which $\bra{x} \ket{y} \equiv G_B(x,y)$, the above operator can be written in the form $T_{x,y} \sim \ket{x} \bra{y}$. That is, $T_{x,y}$ should be intepreted as a $B$-valued generalization of a rank-1 projection on a Hilbert space. The imprimitivity algebra is a two -sided ideal inside the full C$^*$-algebra $B_{G_B}(X)$. 

By definition, $X$ is a $B_G(X)$-rigged module, and thus it inherits the structure of a $E(X)$-rigged module from the inclusion $E(X) \subset B_G(X)$. The $E(X)$ valued inner product is given by $G_{E(X)}(x,y) \equiv T_{x,y}$. In this way, \emph{any} $B$-rigged space $X$ is automatically made into a $E(X)-B$ imprimitivity bimodule. The set of representations of $E(X)$ constitute the set of all representations which can be induced from $B$ via $X$. 

At last, we can state Rieffel's imprimitivity theorem. A representation $\tilde{\rho}: A \rightarrow B(\tilde{H})$ is equivalent to a representation induced from $\rho: B \rightarrow B(H)$ via the $B$-rigged space $X$ if and only if $\tilde{H}$ can be made into a representation of $E(X)$ compatible with $\tilde{\rho}$. If this is the case, the strong Morita equivalence of $B$ and $E(X)$ implies the existence of \emph{some} $(\rho,H)$ such that $F_X(H) \simeq \tilde{H}$. To determine whether $\tilde{\rho}$ is a representation induced from $\rho$ for \emph{any} $B$-rigged space, we can scan through all of the associated imprimitivity spaces. 

Returning to our original context, given an algebraic symmetry action $\tilde{\rho}: \text{Str}_{\mathcal{C}}(\mathcal{M}) \rightarrow \Endvect(M_{\mathcal{M}})$, the above provides an algorithm for determining a representation $\rho: \text{Fus}(\mathcal{C}) \rightarrow \text{End}_{\textrm{v}}(M)$ such that $\tilde{\rho} = \rho^X$, provided such a representation exists. In this case, we can use the representation $\rho$ to reconstruct the categorical symmetry action starting from an algebraic one, constituting an inverse to the construction of Section~\ref{sec: algebras from categories}.

\bibliographystyle{JHEP}

\providecommand{\href}[2]{#2}\begingroup\raggedright\endgroup

\end{document}